\documentclass[11pt, a4paper]{article}

\usepackage[margin=1in]{geometry}

\usepackage[a-2u]{pdfx}

\usepackage[utf8]{inputenc}

\usepackage[english, czech]{babel}
\usepackage{lmodern}
\usepackage[T1]{fontenc}
\usepackage{textcomp}
\usepackage{hyperref}

\usepackage{amsmath}
\usepackage{amsbsy}
\usepackage{amsfonts}
\usepackage{amsthm}
\usepackage{bm}
\usepackage{mathrsfs}
\usepackage{physics}
\usepackage{accents}
\usepackage{cancel}
\usepackage{mathtools}
\usepackage{braket}
\usepackage{tensor}

\usepackage{bbding}
\usepackage{enumerate} 
\usepackage{booktabs}
\usepackage{pdfpages}
\usepackage{float}
\usepackage{xcolor}
\usepackage{bm}
\usepackage{graphicx}

\newcommand*{\bdot}[1]{\accentset{\mbox{\large\bfseries .}}{#1}}

\newcommand*{\closedsqrt}[1][\hphantom{3}]{%
	\def\DHLindex{#1}\mathpalette\DHLhksqrt}
\def\DHLhksqrt#1#2{%
	\setbox0=\hbox{$#1\sqrt[\DHLindex]{#2\,}$}\dimen0=\ht0
	\advance\dimen0-0.2\ht0
	\setbox2=\hbox{\vrule height\ht0 depth -\dimen0}%
	{\box0\lower0.4pt\box2}}

\usepackage{authblk}

\title{Canonical Quantization of Spherically Symmetric Einstein-Scalar Field Solutions}
\author{Ji\v r\'i \v Cern\'y}
\affil{Charles University, Faculty  of  Mathematics  and  Physics, Prague, Czech Republic.}
\date{}

\begin{document}
	\maketitle
	
\begin{abstract}
	We investigate canonical quantization of a general spherically symmetric spacetimes with a massless scalar-field source and examine the associated constraint algebra. The spacetimes are quantized using Dirac’s quantization method for constrained systems, yielding a set of Wheeler-DeWitt equations. A general solution for some of these equations, specifically the momentum constraint, is found and discussed. A complete solution to the whole set of equations (namely the Hamiltonian constraint) remains yet to be found, however. A full solution can be recovered for a static version of these spacetimes.
\end{abstract}

\section{Introduction}

Canonical quantum gravity is based on the Hamiltonian formulation of general relativity. This formulation was developed by Richard Arnowitt, Stanley Deser and Charles Misner by rewriting the Einstein's field equations in Hamilton's canonical formalism \cite{ADM-1} - \cite{ADM-12}. The starting point for the Hamiltonian formulation of general relativity is its Lagrangian description via Hamilton's variational principle $\delta S = 0$. The action $S = \int_{\mathcal{M}} \mathcal{L}(x) \, d^{4}x$ is given via the Einstein-Hilbert Lagrangian density
\begin{align}
	\label{Einstein-Hilbert Lagrangian density with sources}
	\mathcal{L} 
	= \frac{\closedsqrt{- g}}{2 \kappa} \left(R - 2 \Lambda\right)
	+ \mathcal{L}_{M}
	\, .
\end{align}
where $R$ is the scalar curvature of the spacetime manifold $\mathcal{M}$, $g$ is the metric determinant, $\Lambda$ the cosmological constant, $\kappa = 8 \pi G c^{-4}$, and $\mathcal{L}_{M} = \closedsqrt{- g} \, \mathscr{L}_M$ is the Lagrangian density of the matter source, where $\mathscr{L}_M$ contains the source fields. Varying the action with respect to the spacetime metric $g_{\mu\nu}$, and setting the result to zero yields (up to boundary terms) Einstein's vacuum gravitational equations $R^{\mu \nu} - \frac{1}{2} R \, g^{\mu \nu} + \Lambda \, g^{\mu \nu} = \kappa \, T^{\mu \nu}$ with stress-energy tensor $T^{\mu \nu} \equiv \frac{2}{\closedsqrt{-g}} \frac{\delta \mathcal{L}_{M}}{\delta g_{\mu\nu}}$. In the Hamiltonian formalism, we transition from the Lagrangian description and express the action via the corresponding Hamiltonian. In order to make this transition, the Hamiltonian formalism requires a distinct notion of time. As we see below, this can be achieved for globally hyperbolic manifolds by defining a foliation function that splits the manifold into a sequence of space-like hypersurfaces. Not all spacetime can be foliated in this manner.

General relativity formulated in this canonical formalism on phase space can be subjected to the program of canonical quantization, proposed by Paul Dirac \cite{dirac-1}, and further developed by Peter Bergmann and James Anderson \cite{anderson-bergmann}. In later years, John Wheeler \cite{wheeler} proposed the idea of describing quantum states of geometry with a wave functional $\Psi(\bm{q})$ parametrized by a 3-metric $\bm{q}$. In the same time, Bryce DeWitt \cite{dewitt-1} derived a second order functional differential equation (now called the Wheeler-DeWitt equation) for Wheeler's wave functional. 

Spacetimes that can be foliated are of two kinds, minisuperspace and midisuperspace. A minisuperspace model is described by a metric for which the canonical variables do not depend on points on $\Sigma$, for example, FLRW spacetimes studied by James Hartle and Stephen Hawking \cite{hartle-hawking}. A midisuperspace model is described by a metric for which the canonical variables dependent on $\Sigma$. 
Quantization of midisuperspace spacetime was done by Karel Kucha\v r \cite{kuchar} for the Schwarzschild black hole using his method of reduced phase space quantization. This method is stim based on ADm, but is diffrent from Wheeler and DeWitt's approach. An attempt to use reduced phase space quantization for spacetimes with a massless scalaer field was later made by Joseph Romano \cite{romano}. A successfull solution to the Wheeler-DeWitt equation for the Schwarzschild spacetime was found by Masakatsu Kenmoku, Hiroto Kubotani, Eiichi Takasugi and Yuki Yamazaki \cite{kenmoku-kubotani-takasugi-yamazaki}.

\subsection{3+1 Decomposition}
\label{3+1 Decomposition}

Consider some (4-dimensional) manifold $\mathcal{M}$ described by some general coordinates $x = \{x^{\mu}\}_{\mu=1}^{4}$, and some metric $g_{\mu\nu}(x)$ defined on the manifold. A manifold $\mathcal{M}$ is globally hyperbolic iff it admits existence of Cauchy hypersurfaces, that is, space-like hypersurfaces $\Sigma \subset \mathcal{M}$ whose domain of dependence covers the whole manifold $\mathcal{M}$. On globally hyperbolic manifolds it is always possible to define a smooth function (scalar field) $t = t(x) \in \mathbb{R}$ such that the hypersurfaces on which $t=const.$ are Cauchy hypersurfaces \cite{geroch}. The foliation function $t(x)$ is called foliation time $t$. For a fixed foliation time $t = t_0$, the set $\Sigma(t_0) = \{ p(x) \in \mathcal{M} \, | \, t(x) = t_0\}$ is a space-like hypersurface called a foliation hypersurface $\Sigma$. A globally hypergolic manifold $\mathcal{M}$ can then be foliated with space-like hypersurfaces $\Sigma(t)$ such that $\mathcal{M} = \bigcup_{t \in \mathbb{R}} \Sigma(t)$, and is thus topologically isomorphic to $\mathbb{R} \times \Sigma$. From now on we choose some globally hyperbolic manifold $\mathcal{M}$ to work on, a metric $g_{\mu\nu}$, and a torsionless, metric-annihilating (Levi-Civita) covariant derivative $\nabla_{\mu} = \partial_{\mu} + \bm{\Gamma}_{\mu}$ where $\partial_{\mu}$ is the canonical derivative associated with the coordinate system $x$. In instances where some tensors or their products need to be symmetrized, the symmetrization will be denoted by round brackets around the affected indices, e.g. $A_{\mu} B_{\nu} + A_{\nu} B_{\mu} = 2 A_{(\mu} B_{\nu)}$. Naturally, Einstein's summation convention is used throughout the text.

Gradient $\nabla_{\mu} t = \partial_{\mu} t(x)$ of the foliation function is perpendicular to the hypersurface $t(x) = const.$, and so to $\Sigma(t)$. Its normalization $N \equiv ( - g^{\mu\nu} \nabla_{\mu} t \, \nabla_{\nu} t)^{-\frac{1}{2}}$ is a function called "lapse". The (covariant) vector $n_{\mu} = - N \nabla_{\mu} t$ is then normal to the hypersurface $t = const.$. It has time-like normalization $n^{\alpha} n_{\alpha} = -1$, corresponding to the fact that these vectors are normal to space-like hypersurfaces $\Sigma(t)$. A projector to the normal direction, which the 1-dimensional space, is the negative tensor product $- n^{\mu} n_{\nu}$ of two normals. Tangential projection to the foliation hypersurface $\Sigma(t)$ is given by the symmetric tensor $q^{\mu}_{\nu} = \delta^{\mu}_{\nu} + n^{\mu} n_{\nu}$ as such tensor satisfies $q^{\mu}_{\alpha} q^{\alpha}_{\nu} = q^{\mu}_{\nu}$, meaning it is a projector, and $q^{\mu}_{\nu} n_{\mu} = q^{\mu}_{\nu} n^{\nu} = 0$, showing that it projects to the space that is perpendicular to the normal, i.e. to the foliation hypersurface $\Sigma$. Normal projections of tangent tensors indices and tangential projections of normal indices are all zero. All contractions between a tangentially projected index and an unprojected index, the tangent index acts as the tangential projector. Similarly, all contractions between normal indices and tangent indices are always zero. As a result, the metric on $\mathcal{M}$ can be decomposed into it tangential and normal parts:
\begin{align}
	\label{metric on manifold via metric on Sigma}
	g_{\mu\nu} = q_{\mu\nu} - n_{\mu} n_{\nu}
	\, , \quad 
	g^{\mu\nu} =  q^{\mu\nu} - n^{\mu} n^{\nu}
	\, , \quad 
	\delta^{\mu}_{\nu} =  q^{\mu}_{\nu} - n^{\mu} n_{\nu} 
	\, .
\end{align}
where the symmetric tensors $q_{\mu\nu}$ and $q^{\bm{\mu\nu}}$ are metric and its inverse on $\Sigma$, and tensors $- n_{\mu} n_{\nu}$ and $- n^{\mu} n^{\nu}$ function as a metric and its inverse on the (1-dimensional) normal space to $\Sigma$. The identity tensor $\delta^{\mu}_{\nu}$ in (\ref{metric on manifold via metric on Sigma}) is especially useful for decomposing tensors from $\mathcal{M}$ into their tangential and normal constituents.

To describe how the points on $\Sigma(t)$ shift with the foliation time $t(x)$, we first introduce time-flow vector $t^{\mu}$ implicitly by the formula $t^{\mu} \nabla_{\mu} t = t^{\mu} \partial_{\mu} t = 1$. Explicitly, components of the time-flow vector are $t^{\alpha}(x) = \partial x^{\alpha} / \partial t$. The time-flow vector is therefore tangent to the flow given by the foliation time $t$ at the point $x$. The contraction between the time-flow vector and the normal gives the lapse function: $n_{\mu} t^{\mu} = - N$. The shift vector $N^{\mu}$ is a $\Sigma$-tangential projection of the time-flow vector: $N^{\mu} \equiv q^{\mu}_{\nu} t^{\nu} 
= t^{\mu} - N n^{\mu}$. From its definition, the shift vector is tangent to $\Sigma$ and thus perpendicular to the normal. Physically speaking, the shift vector describes how the position of a point $x \in \Sigma(t)$ changes between $\Sigma(t)$ and $\Sigma(t + dt)$. 

Instead of some general coordinate system $\{x^{\mu}\}_{\mu=1}^{4}$ we have been using, we introduce a new set of, so called adapted, coordinates $\{y^{\mu}\}_{\mu=0}^{4}$ that will respect the foliation on our globally hyperbolic manifold. To do that, we set $y^{\mu} = (t, y^{i})$ where the first adapted coordinate $y^0$ is equal to the foliation time $t$, and $\{y^{i}\}_{i=1}^{3}$ are some general (spatial) coordinates on $\Sigma$. 
Because our manifold is $\mathcal{M} = \bigcup_{t \in \mathbb{R}} \Sigma(t)$, we are not losing any additional generality by this choice of coordinates. The advantage of adapted coordinates is that they drastically simplify objects that are either fully normal, or fully tangent to $\Sigma$, for instance, components of the normal covector are $n_{\mu}(y) = -N \delta^{0}_{\mu}$. The spacial components of $g_{\mu\nu}$ are thus $g_{ij}(y) = q_{ij}(y) - n_{i}(y) n_{j}(y) = q_{ij}(y)$. For the time-flow vector we have $t^{\mu}(y) = {\partial y^{\mu}} / {\partial t} = \delta^{\mu}_{0}$, and so the shift becomes $N^{\mu}(y) = \delta^{\mu}_{i} N^{i}$. The infinitesimal line element $ds^{2} = g_{\mu\nu} dx^{\mu} dx^{\nu}$ on $\mathcal{M}$ in adapted coordinates splits into
\begin{align}
	\label{square of lenght in adapted coordinates}
	ds^{2}
	= \left( - N^{2} + q_{ij} N^{i} N^{j} \right)dt^{2} 
	+ 2 \, q_{ij} N^{i} dt \, dy^{j} + q_{ij} \, dy^{i} dy^{j}
	\, .
\end{align}
A line element on $\Sigma(t)$, where $t$ is constant, is thus $ds^{2}|_{\Sigma} = q_{ij} \, dy^{i} dy^{j}$, meaning that $q_{ab}$ is the metric of the foliation hypersurfaces $\Sigma$. As such, there exists an inverse metric $q^{ab}$ for which $q_{jk} q^{ki} = \delta^{i}_{j}$.
Returning back to the full metric $g_{\mu\nu}(y)$, its components are 
\begin{align}
	\label{metric matrix in adapted coordinates}
	g_{\mu\nu}(y)
	= \left(
	\begin{matrix}
		N_{k} N^{k} - N^{2}, & N_{j} \\
		N_{i}, & q_{ij} \\
	\end{matrix} 
	\right)
	\, ,
\end{align}
where $N_{i} = q_{ik} N^{k}$. Inverting this matrix yields the inverse metric
\begin{align}
	\label{inverse metric matrix in adapted coordinates}
	g^{\mu \nu}(y)
	= 
	\frac{1}{N^{2}}
	\left(
	\begin{matrix}
		- 1, & N^{j} \\
		N^{i}, & N^{2} q^{ij} - N^{i} N^{j} \\
	\end{matrix} 
	\right)
	\, .
\end{align}
Regarding the metric determinants $g \equiv \det(g_{\mu\nu})$ and $q \equiv \det(q_{ij})$, applying Cramer's rule $(A^{-1})_{ji} = \det(A)^{-1} \cdot \text{minor}_{ij} (A)$ on the $00$-th element of $g_{\mu\nu}$ gives $g^{0 0} = g q$. The metric determinant therefore decomposes as $g = - N^2 q$, and for metric density: $\closedsqrt{-g} = N \closedsqrt{q}$.

Connection ${^{(\parallel)}\!}\nabla_{\mu}$ on $\Sigma$ is obtained as the full tangential projection of the connection $\nabla_{\mu}$ on $\mathcal{M}$: ${^{(\parallel)}\!}\nabla_{\mu} T^{\alpha\ldots}_{\beta\ldots} \equiv q^{\nu}_{\mu} q^{\alpha}_{\gamma} \ldots q^{\delta}_{\beta} \ldots \nabla_{\nu} T^{\gamma\ldots}_{\delta\ldots}$. In adapted coordinates, the spacial part of this connection can be written as ${^{(\parallel)}\!}\nabla_{i} = \partial_{i} + \bm{\Gamma}_{i}$ where $\partial_{i}$ is the canonical covariant derivative associated with the spatial coordinates $y^i$ and $\Gamma^{a}_{ib} = \frac{1}{2} q^{ak} \left( \partial_{i} q_{bk} + \partial_{b} q_{ik} - \partial_{k} \, q_{ib} \right)$ are the Christoffel symbols. This covariant derivative ${^{(\parallel)}\!}\nabla_{i}$ annihilates the spatial metric $q_{ij}$, and is Levi-Civita connection on $\Sigma$.

The foliation hypersurfaces $\Sigma$ are characterized by their intrinsic curvature ${^{(\parallel)}\!}R$, but also by the extrinsic curvature $K_{\mu\nu} \equiv q^{\alpha}_{\mu} q^{\beta}_{\nu} \nabla_{\alpha} n_{\beta} = {^{(\parallel)}\!}\nabla_{\mu} n_{\nu}$, which describes how they are embedded in the manifold $\mathcal{M}$. 
The extrinsic curvature can be alternatively expressed as a Lie derivative of the tangential metric along the normal field, and for there also as Lie derivatives with respect to the time-flow vector and the shift vector: 
$K_{\mu\nu} = \frac{1}{2} \mathsterling_{\bm{n}} \, q_{\mu\nu} = \frac{1}{2 N} (\mathsterling_{\bm{t}} \, q_{\mu\nu} - \mathsterling_{\bm{N}} \, q_{\mu\nu})$. Although $K_{\mu\nu}$ is fully tangent to $\Sigma$, the two Lie derivatives are generally not. Without loss of generality, we can thus project $K_{\mu\nu}$ onto $\Sigma$. Since $n_{\mu}(y) = \delta^{0}_{\mu}$ in adapted coordinates, the extrinsic curvature (as a fully tangential tensor) is determined solely by its spatial part $K_{ab}$. Projection $q^{\alpha}_{\mu} q^{\beta}_{\nu} \mathsterling_{\bm{t}} \, q_{\alpha\beta}$ of the time-flow Lie derivative then becomes $\delta^{\mu}_{a} \delta^{\nu}_{b} \partial_{t} q_{\mu\nu}(y)
= \partial_{t} q_{ab}(y)$. Altogether, the extrinsic curvature is
\begin{align}
	\label{extrinsic curvature in adapted coordinates}
	K_{ab} 
	= \frac{1}{2 N} \left( \partial_{t} q_{ab} - {^{(\parallel)}\!}\nabla_{(a} N_{b)} \right) 
	\, .
\end{align}
Decomposition of the intrinsic curvature $R$ of the whole manifold $\mathcal{M}$ into normal and tangential parts is given by Gauss–Codazzi equations. In our case we have
\begin{align}
	\label{scalar curvature decomposition}
	R = 
	{^{(\parallel)}\!}R + K_{ab} \, K^{ab} - K^{2} 
	- 2 \, \nabla_{\mu} \left( a^{\mu} - n^{\mu} \, K \right)
	\, ,
\end{align}
where ${^{(\parallel)}\!}R$ is the scalar curvature of $\Sigma$, $K = K^{a}_{a}$ is the extrinsic curvature trace (scalar), and $a^{\mu} \equiv n^{\alpha} \nabla_{\alpha} n^{\mu}$ is the "acceleration" vector field for the normals. Since $a^{\mu} n_{\nu}$, this normal acceleration is perpendicular to the normal field and tangent to $\Sigma$.

\subsection{ADM Formalism}
\label{ADM Formalism}

In this section we summarize basic concepts of ADM formulation \cite{ADM-recap} of general relativity in vacuum case. From now on, we will work with some sets of adapted coordinates. Different sets of adapted coordinates can be then denoted variously, for instance $x = (t, \bm{x})$ where $\bm{x} = \{x^{i}\}_{i=1}^{3}$ are some arbitrary (spacial) coordinates on $\Sigma$. The action, which is generally given via Lagrangian density as integral $S = \int_{\mathcal{M}} \mathcal{L}(x) \, d^{4}x$ over the full manifold, can be rewritten as integration over $\Sigma$ and the foliation time $t$ separately: $S = \int_{\mathbb{R}} \int_{\Sigma} \mathcal{L}(t, \bm{x})d^3 x \, dt$. Alongside the Lagrangian density, we have the associated Lagrangian as the spatial integral $L = \int_{\Sigma} \mathcal{L} \, d^3 x$, and the action $S = \int_{\mathbb{R}} L(t) dt$. The procedures below are then performed on some general $\Sigma$ first.

The vacuum Lagrangian density is $\mathcal{L} = \frac{N \closedsqrt{q}}{2 \kappa} R$ with the decomposed scalar curvature (\ref{scalar curvature decomposition}). The last term in (\ref{scalar curvature decomposition}) is in the form of a total covariant divergence and can be converted to a surface integral over the boundary of $\mathcal{M}$. This, and other boundary terms that we will see later, all depend on our choice of boundary and can be adjusted or eliminated with appropriate boundary conditions. One such condition posed at the Lagrangian level is variation with fixed ends, i.e. requirement that $\delta g_{\mu\nu} |_{\partial\mathcal{M}} = 0$, which after decomposition of metric translates to $\delta \phi_{A} |_{\partial\mathcal{M}} = 0$ for $\phi_{a} \in \{q_{ab},\, N,\, N^{a}\}$. This condition eliminates some boundary terms, but is generally not sufficient for all the terms. Boundary terms do not affect the local field equations of the theory, i.e. equation on some $\Omega \subset \mathcal{M}$. Proper investigation of boundary terms was can be found in \cite{bojowald} and goes beyond the scope of this text. As will be discussed later, the boundary terms will have no effect for out choice of spacetime we will be focus on later. For all those reasons we will, by default, omit such boundary terms, and formally understand this as either setting the boundary conditions appropriately so that the boundary contributions are eliminated, or considering only compact manifolds (manifolds without boundary). Still, to mark the places where boundary terms were dropped from the equations we will use the "interior equality" $\overset{\text{int.}}{=}$, which is a regular equality on the interior of the respective set. We thus write the vacuum action as
\begin{align}
	\label{Einstein-Hilbert action without boundary terms}
	S\left[ q_{ab}, N, N^{a} \right]
	\overset{\text{int.}}{=} \int_{\mathbb{R}} \int_{\Sigma} 
	\frac{N \closedsqrt{q}}{2 \kappa} \left( {^{(\parallel)}\!}R + K_{ab} \, K^{ab} - K^{2} \right) d^3 x \, dt
	\, .
\end{align}
Legendre transformations of the Lagrangian $L(t)$ yield the momenta
\begin{align}
	\label{canonical momenta}
	p^{ab} \equiv \frac{\delta L}{\delta \bdot{q}_{a b}} 
	= \frac{\closedsqrt{q}}{2 \kappa} \left( K^{ab} - q^{ab} K \right) 
	\, , \quad 
	P \equiv \frac{\delta L}{\delta \bdot{N}} = 0 
	\, , \quad
	{P}_{a} \equiv \frac{\delta L}{\delta \bdot{N}^{a}} = 0 
	\, .
\end{align}
The resulting Hamiltonian $H \equiv \int_{\Sigma} \mathcal{H} \, d^3 x$ has the Hamiltonian density 
\begin{align}
	\label{vacuum Hamiltonian via sumer-Hamiltonian and supermomentum}
	H \,\overset{\text{int.}}{=}\, \int_{\Sigma} \left( N \, \mathcal{H}_{\perp} + N^{a} \, \mathcal{H}_{a} + \bdot{N} P + \bdot{N}^{a} P_{a} \right) d^3 x 
	\, ,
\end{align}
where the terms proportional to lapse are collectively called super-Hamiltonian $\mathcal{H}_{\perp}$, and the term contracted with the shift vector is called super-momentum $\mathcal{H}_{a}$:
\begin{align}
	\label{vacuum super-Hamiltonian}
	\mathcal{H}_{\perp} 
	&= \frac{2 \kappa}{\closedsqrt{q}} \left( p_{ab} \, p^{ab} - \frac{1}{2} (p_{a}^{a})^{2} \right) 
	- \frac{\closedsqrt{q}}{2 \kappa} \, {^{(\parallel)}\!}R 
	\, , 
	\\
	\label{vacuum super-momentum}
	\mathcal{H}_{a} 
	&= - 2 \, q_{ab} \, {^{(\parallel)}\!}\nabla_{c} \, p^{cb} 
	\, .
\end{align}
We see in (\ref{canonical momenta}) that the lapse and shift momenta are implicit, primary constraints $P \approx 0$ and $P_{a} \approx 0$. The lapse and shift velocities $\bdot{N}$ and $\bdot{N}^{a}$ then play a role of Lagrange multipliers. Denoting collectively $\phi_{A} = (q_{ab},\, N,\, N^{a})$ the canonical coordinates and $\pi^{A} = (p^{ab},\, P,\, P^{a})$ the canonical momenta, Hamilton's canonical equations are 
\begin{align}
	\label{Hamilton's canonical equations}
	\bdot{\pi}^{A} = - \frac{\delta H}{\delta \phi_{A}} 
	\, , \quad  
	\bdot{\phi}_{A} = \frac{\delta H}{\delta \pi^{A}} 
	\, .
\end{align}
The functional derivatives of one function/field $F(x)$ with respect to other $G(y)$, where $x = (x^{0}, \bm{x})$ and $y = (y^{0}, \bm{y})$, are in the context of the canonical formalism always taken on the same foliation hypersurface $\Sigma(t)$ with $t = x^0 = y^0$:
\begin{align*}
	\frac{\delta F(\bm{x})}{\delta G(\bm{y})} 
	= \left.\frac{\delta F(x)}{\delta G(y)}\right|_{\Sigma(x^0 = y^0)}
	\, .
\end{align*}
For this reason, we will omit the explicit reference to the foliation time $t$ and only denote the spatial coordinates, understanding that the dependence on $t$ still exists. The notation $F(\bm{x})$ should be understood as an abbreviation for $F(x^0, \bm{x})$ when used on $\Sigma(x^0)$.

The Poisson brackets on the phase space of spanned by the canonical variables $(\phi_{I}, \pi^{J})$ between two functions/functionals $F[\phi_{I}, \pi^{J}]$ and $G[\phi_{I}, \pi^{J}]$ are 
\begin{align}
	\label{Poisson brackets - definition}
	\left\{F,\, G \right\} 
	\equiv \sum\limits_A \int\limits_{\Sigma} 
	\left( \frac{\delta F}{\delta \phi_{A}(x)} \, \frac{\delta G}{\delta \pi^{A}(x)} 
	- \frac{\delta G}{\delta \phi_{A}(x)} \, \frac{\delta F}{\delta \pi^{A}(x)} \right) d^3 x  
	\, .
\end{align}
Derivatives $\bdot{F}$ of phase space functions/functional $F[\phi_{A}, \pi^{B}]$ with respect to the foliation time $t$ are then expressed as $\bdot{F} = \left\{ F, H \right\}$. The canonical Poisson brackets are those between the canonical variables:
\begin{align}
	\label{canonical Poisson brackets}
	\left\{ q_{ab}(\bm{x}), p^{ij}(\bm{y}) \right\} 
	= \delta^{i}_{(a} \delta^{j}_{b)} \, {\delta}^{(3)}(\bm{x} - \bm{y})
	\, ,
\end{align}
$\left\{ N(\bm{x}), P(\bm{y}) \right\} = {\delta}^{(3)}(\bm{x} - \bm{y})$, and $\left\{ N^{a}(\bm{x}), P_{i}(\bm{y}) \right\} = \delta^{a}_{i} \, {\delta}^{(3)}(\bm{x} - \bm{y})$, while Poisson brackets of all the other combinations of canonical variables are zero. Returning back to the constraints $P \approx 0$ and $P_{a} \approx 0$ in (\ref{canonical momenta}), the consistency condition for every and all constraints on a phase space demand that the constraints do not evolve, or evolve into another constraint. For us this means the requirement that $\bdot{P} \approx 0$ and $\bdot{P}_{a} \approx 0$. From Hamilton's equations we get $\bdot{P} = \mathcal{H}_{\perp}$ and $\bdot{P}_{a} = \mathcal{H}_{a}$. Our system can thus only be consistent with two additional constraints 
\begin{align}
	\label{Hamiltonian and momentum contraint}
	\mathcal{H}_{\perp} \approx 0  
	\, , \quad
	\mathcal{H}_{a} \approx 0  
	\, ,
\end{align}
called the Hamiltonian constraint and the momentum (diffeomorphism) constraint respectively. This also shows that the lapse $N$ and shift $N^{a}$ are Lagrange multipliers. The entire Hamiltonian density is then a constraint $\mathcal{H} \approx 0$. Furthermore, the super-Hamiltonian and super-momentum satisfy the following constraint algebra \cite{adm-algebra}: 
\begin{align}
	\label{constraint algebra - 2 super-Hamiltonians}
	\left\{ \mathcal{H}_{\perp}(\bm{x}), \mathcal{H}_{\perp}(\bm{y}) \right\} 
	&= \left[ q^{ab}(\bm{x}) \, \mathcal{H}_{b}(\bm{x}) - q^{ab}(\bm{y}) \, \mathcal{H}_{b}(\bm{y}) \right] \partial_{x^a} {\delta}^{(3)}(\bm{x} - \bm{y}) \, ,
	\\
	\label{constraint algebra - super-Hamiltonian and supermomentum}
	\left\{ \mathcal{H}_{a}(\bm{x}), \mathcal{H}_{\perp}(\bm{y}) \right\} 
	&= \mathcal{H}_{\perp}(\bm{x}) \, \partial_{x^a} {\delta}^{(3)}(\bm{x} - \bm{y}) \, , 
	\\
	\label{constraint algebra - 2 supermomenta}
	\left\{ \mathcal{H}_{a}(\bm{x}), \mathcal{H}_{b}(\bm{y}) \right\} 
	&= \mathcal{H}_{b}(\bm{x}) \, \partial_{x^a} {\delta}^{(3)}(\bm{x} - \bm{y}) 
	- \mathcal{H}_{a}(\bm{y}) \, \partial_{x^b} {\delta}^{(3)}(\bm{x} - \bm{y}) 
	\, . 
\end{align}
This constraint algebra is closed as all the Poisson brackets of two constraints are also constraints, meaning $\bdot{\mathcal{H}}_{\perp} = \{\mathcal{H}_{\perp}, H \} \approx 0$ and $\bdot{\mathcal{H}}_{a} = \{\mathcal{H}_{a}, H \} \approx 0$ are satisfied naturally on their own. Constraints whose Poisson brackets vanish on the constraint surface are called 1st class. Our system thus contains only 1st class constraints. The Lagrange multipliers $\bdot{N}$ and $\bdot{N}^a$ are completely arbitrary functions. This mean the evolution of phase space functionals $F[N, N^a, P, P_a]$ cannot be determined from the system, and all the system's natural dynamics happens only on the constraint hypersurface $P_N \approx 0$, $P_a \approx 0$. The original phase space $( q_{ab}, N^a, N, p^{ab}, P, P_{a} )$ is then effectively reduced to the space of the dynamical phase-space coordinates $( q_{ab}, p^{ab})$, and the system dynamics is determined by Arnowitt–Deser–Misner (ADM) Hamiltonian
\begin{align}
	\label{ADM Hamiltonian}
	H = \int_{\Sigma} \left( N \, \mathcal{H}_{\perp} + N^{a} \, \mathcal{H}_{a} \right) d^3 x 
	\, ,
\end{align}
and the action of our system on the reduced phase space $( q_{ab}, p^{ab} )$ is the ADM action 
\begin{align}
	\label{ADM action}
	S[q_{ab},\, p^{ab};\, N^{a},\, N] = \int_{\mathbb{R}} \int_{\Sigma} \left( p^{ab} \, \bdot{q}_{ab} - N \mathcal{H}_{\perp} - N^{a} \mathcal{H}_{a} \right) d^3 y \, dt 
	\, . 
\end{align}

When performing canonical quantization, we work on the space $\text{Riem}(\Sigma)$ of all (physical or unphysical) spatial metrics defined on the space-like hypersurfaces $\Sigma$. The metrics that are mutually related via diffeomorphisms, i.e. that are diffeomorphically equivalent, describe the same geometry of $\Sigma$. As a configuration space of our canonical theory we thus chose the space of all spatial geometries $\text{Riem}(\Sigma)/\text{Diff}(\Sigma)$, where $\text{Diff}(\Sigma)$ is the set (group) of all diffeomorphisms on $\Sigma$. This space $\text{Riem}(\Sigma)/\text{Diff}(\Sigma)$ of all geometries on $\Sigma$ (i.e. of all diffeomorphically non-equivalent spatial metrics of $\Sigma$) is called the superspace of $\Sigma$. On the space $\text{Riem}(\Sigma)$ of all $q_{ab}(x)$ there exists a metric called the DeWitt supermetric
\begin{align}
	\label{DeWitt supermetric}
	G^{abcd} 
	= \frac{\closedsqrt{q}}{2} \left( q^{ac} q^{bd} + q^{ad} q^{bc} - 2 \, q^{ab} q^{cd} \right) 
	\, ,
\end{align}
and its inverse 
\begin{align}
	\label{DeWitt inverse supermetric}
	G_{abcd} 
	= \frac{1}{2\closedsqrt{q}} \big( q_{ac} \, q_{bd} + q_{ad} \, q_{bc} - q_{ab} \, q_{cd} \big) 
	\, ,
\end{align}
These two metrics are inverse in the sense of their contraction $G^{abcd} \, G_{klcd} = \delta^{a}_{(k} \, \delta^{b}_{l)}$. The supermetric allows us to formally simplify several expressions we have encountered, for instance, the super-Hamiltonian has concise form
\begin{align}
	\label{super-Hamiltonian via supermetric}
	\mathcal{H}_{\perp} 
	= 2 \kappa \, G_{abcd} \, p^{ab} p^{cd} - \frac{\closedsqrt{q}}{2 \kappa} \, {^{(\parallel)}\!}R  
	\, .
\end{align}
As should be apparent, the supermetric (as it is defined above) is quite complicated object; it is a 4th degree tensors with symmetries $G^{abij} = G^{(ab)(ij)} = G^{ijab}$. However, in practice one has a smaller-than-general set of momenta $\{ p^{ab} \}_{a,b} = \{ p^{A} \}_{A}$. The effective (relative to the symmetries of our system) form of the inverse supermetric from the momentum part of the super-Hamiltonian: $G_{abcd} \, p^{ab} p^{cd} = G_{AB} \, p^{A} p^{B}$ where $G_{AB}$ can be now treated as a common matrix.

\subsection{Wheeler-DeWitt Equations}
\label{Wheeler-DeWitt Equations}

Dirac's canonical quantization program is a procedure that allows us to quantize the Hamiltonian and momentum constraints directly, in the canonical variables $q_{ab}$ a $p^{ab}$, without the need to solve the constraints first. In general, states of a quantum system are represented by vectors $\ket{\psi}$ from a Hilbert space $\mathscr{H}$, and quantum mechanical operators are linear maps $\widehat{A}: \mathscr{H} \rightarrow \mathscr{H}, \, \ket{\psi} \mapsto \widehat{A}\ket{\psi}$ on these vectors. Consider a phase space $(q^I, \, p_{J})$, and some functions $A$ and $B$ on this phase space. The standard quantization procedure assigns to these functions corresponding Hermitian\footnote{
	Quantum operators corresponding to classical phase space functions are required to be Hermitian so that they have real eigenvalues and thus can represent real physical measurements.} 
operators $\widehat{A}$ and $\widehat{B}$ that represent observable quantities given by $A$ and $B$ is the classical case. The classical Poisson bracket $\left\lbrace A , \, B \right\rbrace$ of the phase space functions is then replaced by a commutator $(i \hbar)^{-1} [ \widehat{A}, \, \widehat{B}] = (i \hbar)^{-1} (\widehat{A} \widehat{B} - \widehat{B} \widehat{A})$ of the corresponding quantum operators. The canonical commutators for the operators $(\widehat{q}^{\,I}, \, \widehat{p}_{J})$ corresponding to the phase space coordinates are set to $[ \widehat{q}^{\,I} , \, \widehat{p}_{J} ] = i \hbar \, \delta^I_J$ and $[ \widehat{q}^{\,I} , \, \widehat{q}^{\,J} ] = [ \widehat{p}_{I} , \, \widehat{p}_{J} ] = 0$ in accordance with the classical case. A special case is that of the system's Hamiltonian $H(q^I, \, p_{J})$. From classical formalism we know that $\bdot{F} = \left\{ F,\, H \right\}$ for phase space functions $F(q^I, \, p_{J})$. Applying the quantization to Poisson brackets yields the Heisenberg equation 
$i\hbar \, \partial_t \widehat{F} = [\widehat{F}, \, \widehat{H}]$. 

The process of assigning an operator to a phase space function is not always simple. Consider a phase space function $F(q^A, \, p_{B})$. The corresponding operator $\widehat{F}$ is formally obtained by substitution $\widehat{F} = F(\widehat{q}^{\,A}, \, \widehat{p}_{B})$. The potential problem arises when $F$ contains mutual products of the canonical variables $q^A$ and $p_{B}$ as they commute in the classical system, but their quantum operator versions generally do not commute, and it is not always clear how to order them. Swapping corresponding coordinate and momentum operators yields a term proportional to $\hbar$. Commutator of 2 operators $\widehat{F}$ and $\widehat{G}$ is then equivalent to the Poisson bracket of their classical functions $F(q^A, \, p_{B})$ and $G(q^A, \, p_{B})$ with substituted coordinate and momentum operators up to the terms of order $\mathcal{O}(\hbar^2)$. For system with constraints, Dirac's quantization procedure should be used. In this procedure, instead of assigning the operator commutator to the Poisson bracket, it is assigned to the (more complicated) Dirac bracket, which also incorporates the system constraints. However, in cases where the system only contains 1st class constraints (which is our case), Dirac brackets effectively reduce to Poisson brackets. Let us thus consider a system with only 1st class constraints $\mathcal{C}_\alpha(q^A, \, p_{B}) = 0$. In consistency with the classical case, we demand that the corresponding operators  $\widehat{\mathcal{C}}_{\alpha} = \mathcal{C}_{\alpha}(\widehat{q}^{\,A}, \, \widehat{p}_{B})$ satisfy $\widehat{\mathcal{C}}_{\alpha} \ket{\psi} = 0$ for all physical states $\ket{\psi}$ of the system, restricting the Hilbert space to its "physical" part. Another consistency condition is to require the physical states to always evolve into another physical states. Evolution of quantum states is given by the Schrödinger equation $i\hbar \, \partial_t \ket{\psi} = \widehat{H} \ket{\psi}$. Physical states $\ket{\psi}$ remain physical if $\widehat{\mathcal{C}}_{\alpha} \widehat{H} \ket{\psi} = 0$ holds for all the constraints. Equivalently we can write $[\widehat{\mathcal{C}}_{\alpha}, \, \widehat{H}] \ket{\psi} = 0$, meaning that on physical states, $\widehat{\mathcal{C}}_{\alpha}$ and $\widehat{H}$ should commute.

Our system is a phase space with canonical coordinates $q_{ab}$ and the associated canonical momenta $p^{ab}$. The functions $q_{ab}(y)$ are components of the metric $q_{\bm{ab}}$ on the foliation hypersurface $\Sigma$, expressed in some set of adapted coordinates $y$. Quantum sates $\ket{\Psi}$ of this system will be represented by wave functionals $\Psi[q_{ij}]$ parameterized by metrics on $\Sigma$. Such choice is commonly called metric representation. The space of all such wave functionals (parametrized by both physical and non-physical metrics) is the representation space $\mathscr{F} = \{ \Psi[\bm{q}];\, \bm{q} \in \text{Riem}(\Sigma) \}$. The representation space $\mathscr{F}$ is generally not a Hilbert space. The canonical commutators are
\begin{align}
	\label{canonical commutators}
	[\widehat{q}_{ab}(\bm{x}), \, \widehat{p}^{\,ij}(\bm{y})] 
	=  i \hbar \, \delta^{i}_{(a} \delta^{j}_{b)} {\delta}^{(3)}(\bm{x} - \bm{y}) 
	\, , \quad
	[\widehat{q}_{ab},\, \widehat{q}_{ij}] = [\widehat{p}^{\,ab},\, \widehat{p}^{\,ij}] = 0
	\, .
\end{align}
The metric and momentum operators are prescribed to act on functionals $\Psi[q_{ij}]$ as
\begin{align}
	\label{metric and momentum operators action on functionals in metric representation} 
	\widehat{q}_{ab} \Psi = q_{ab} \Psi
	\, , \quad
	\widehat{p}^{\, ab} \Psi = - i \hbar \, \frac{\delta \Psi}{\delta q_{ab}} 
	\, .
\end{align}
The same could be done for the lapse function $N$ and the shift vector $N^a$ as they are also formally canonical variables on the full system, but because the lapce and shift momenta are constraints, we would obtain conditions $\widehat{P} \Psi = - i \hbar \frac{\delta \Psi}{\delta N} = 0$ and $\widehat{P}_{a} \Psi = - i \hbar \frac{\delta \Psi}{\delta N^{a}} = 0$ for the functional $\Psi[q_{ij},\, N, \, N^a]$, implying that such $\Psi$ does not depend on the lapse and shift. The functionals $\Psi$ therefore depend only on the metric and we can restrict our analysis only on phase space $( q_{ab},\, p^{ij})$ without loss of any physical part of the representation space.

For the operators of super-Hamiltonian $\widehat{\mathcal{H}}_{\perp} = \mathcal{H}_{\perp}\left(\widehat{q}_{ij},\, \widehat{p}^{\, ij} \right)$ and super-momentum $\widehat{\mathcal{H}}_{a} = \mathcal{H}_{a}\left(\widehat{q}_{ij},\, \widehat{p}^{\, ij} \right)$ we have constraint conditions 
\begin{align}
	\label{quantum constraint conditions}
	\widehat{\mathcal{H}}_{\perp} \Psi = 0
	\, , \quad
	\widehat{\mathcal{H}}_{a} \Psi = 0
	\, ,
\end{align}
restricting the representation space to the space of physical metrics. The Hamiltonian operator is given by the ADM Hamiltonian 
\begin{align}
	\label{ADM Hamiltonian operator}
	\widehat{H}
	= \int_{\Sigma} \left( N \, \widehat{\mathcal{H}}_{\perp} + N^{a} \, \widehat{\mathcal{H}}_{a} \right) d^3 y 
	\, , 
\end{align}
where the $N$ and shift $N^{a}$ are just functions. The Schrödinger equation $- i\hbar \, \partial_{t} \Psi = \widehat{H} \Psi = 0$ for states that satisfy constraint condition (\ref{quantum constraint conditions}) implies that the physical states $\Psi$ does not explicitly depend on the foliation time $t$. In metric representation, the constraint condition (\ref{quantum constraint conditions}), when expressed in the conventional qp-ordering (i.e. first act the momenta operators and then the metric), have the explicit form of so called Wheeler-DeWitt equations\footnote{
	Strictly speaking, these consist of 4 equations for each point on $\Sigma$, so "$4\times \infty^3$" in total.}
\begin{align}
	\label{Wheeler-DeWitt equations - Hamiltonian constraint} 
	\widehat{\mathcal{H}}_{\perp} \Psi
	&= - 2 \kappa \hbar^2 \, G_{abcd} \, \frac{\delta^2 \Psi}{\delta q_{ab} \, \delta q_{cd}} 
	- \frac{\closedsqrt{q}}{2 \kappa} \, {^{(\parallel)}\!}R \, \Psi
	= 0
	\, ,
	\\
	\label{Wheeler-DeWitt equations - momentum constraint}
	\widehat{\mathcal{H}}_{a} \Psi
	&= 2 i \hbar \, {^{(\parallel)}\!}\nabla_{c} \left( q_{ab} \, \frac{\delta \Psi}{\delta q_{cb}} \right) 
	= 0
	\, ,
\end{align}
originally formulated by John Wheeler \cite{wheeler} and Bryce DeWitt \cite{dewitt-1}.
Wheeler-DeWitt equations restrict how the wave functionals $\Psi[q_{ij}]$ can depend on the metric variables $q_{ij}$. the space $\mathscr{F}_{WDW}$ of all solutions to the Wheeler-DeWitt equations is a subset $\mathscr{F}_{WDW} \subset \mathscr{F}$ of the full representation space $\mathscr{F}$. The space $\mathscr{F}_{phys.}$ of all physical wave functionals (those that describe realistic states of spacetime) is the subset $\mathscr{F}_{phys.} \subseteq \mathscr{F}_{WDW}$ as the Wheeler-DeWitt equations might generally not pose sufficient restrictions on $\mathscr{F}$. The space $\mathscr{F}_{WDW}$ thus might not be Hilbert space, but $\mathscr{F}_{phys.}$ is a candidate to be the Hilbert space of all physical states of the spacetime.
On such Hilbert space, one can introduce a scalar product of two wave functionals. One option is to consider formal definition
\begin{align}
	\label{scalar product of wave functionals} 
	\braket{\Psi_{1}|\Psi_{2}}
	= \int_{\text{Riem}(\Sigma)} \Psi^{*}_{1}[q_{ij}] \, \Psi_{2}[q_{ij}] \, \mathcal{D}\mu[q_{ij}]
	\, ,
\end{align}
where $\mathcal{D}\mu[q_{ij}]$ is a measure on $\mathscr{F}$. In general, no such Lebesgue measure exists on $\mathscr{F}$, making this definition strictly formal. However, a well defined scalar product might be possible to define on the space $\mathscr{F}_{WDW}$. Another problem is the question of wheter all the constraint operators $\widehat{\mathcal{C}}_{\perp}$ and $\widehat{\mathcal{C}}_{a}$ should be Hermitean or not. In quantum theory, it is natural to demand Hermiticity, but, as stated above, the representation space $\mathscr{F}$ is not a Hilbert space. It is only the space $\mathscr{F}_{phys.}$ that should have the structure of a Hilbert space. For more in dept discussion on this topic see \cite{isham}.

\subsection{Boundary Terms}
\label{Boundary Terms}

In this subsection we briefly examine the boundary terms which we ignored in the previous text. More detailed analysis of all the boundary terms can be found in \cite{bojowald}, \cite{brotz}, and \cite{kuchar}.

The first boundary term is the York-Gibbons-Hawking (YGH) term, which comes from the Euler-Lagrange variational calculus:
\begin{align*}
	B_{YGH} 
	= \int_{\partial\mathcal{M}} s^{\alpha} g^{\mu\nu} \left( \nabla_{\mu} \delta g_{\alpha\nu} 
	- \nabla_{\alpha} \delta g_{\mu\nu} \right) \closedsqrt{-g} \, d^4 x 
	\, ,
\end{align*}
with some boundary normal $s_{\mu}$. The fixed-ends variation condition $\delta g_{\mu\nu}|_{\partial\mathcal{M}} = 0$ simplifies the YGH term a bit, but does not fully eliminate it. The second term comes from the total covariant divergence at the end of scalar curvature decomposition (\ref{scalar curvature decomposition}): 
\begin{align*}
	B_R 
	= - \kappa^{-1} \int_{\partial\mathcal{M}} 
	s_{\mu} \left( a^{\mu} - n^{\mu} K \right) N \closedsqrt{q} \, d^4 x 
	\, , 
\end{align*}
The third and last term comes from moving the shift vector from under the covariant derivative so that the Hamiltonian can be written in the form (\ref{vacuum Hamiltonian via sumer-Hamiltonian and supermomentum}):
\begin{align*}
	B_H 
	= 2 \int_{\partial\Sigma} s_{a} \, p^{ab} \, N_{b} \, d^3 x 
	\, . 
\end{align*}
Our time-like foliated spacetime is topologically $\mathcal{M} \simeq \mathbb{R}\times\Sigma$. We therefore assume its boundary to be time-like and have a topological structure $\partial\mathcal{M} \simeq \mathbb{R} \times \partial\Sigma$. In case of formally finite, non-asymptotic boundary, one can define boundary-adapted coordinates $(t, z, l^n)$, where $z(x)$ is a scalar function describing the boundary via constraint $z(x) = \textit{const.}$, and $l^n = (l^1, l^2)$ are 2 spatial coordinates. Normal to the boundary is $s_{\mu} = M \, \partial_{\mu} z(x)$ with some normalization factor $M$ chosen such that $s_{\mu} s^{\mu} = 1$. The boundary normal is tangent to $\Sigma$, and so $s_{\mu} n^{\mu} = 0$. The induced metric on the boundary $\partial\Sigma$ is $b_{\mu\nu} = q_{\mu\nu} - s_{\mu}s_{\nu}$, that is $b_{\mu\nu} = g_{\mu\nu} + n_{\mu}n_{\nu} - s_{\mu}s_{\nu}$, with metric determinant $b \equiv \det(b_{\mu\nu})$. Additionally, we also define the extrinsic curvature of the boundary ${^{(\partial\Sigma)}\!}K_{\mu\nu} \equiv b^{\alpha}_{\mu} b^{\beta}_{\nu} \nabla_{\alpha} s_{\beta}$, and more importantly, its trace ${^{(\partial\Sigma)}\!}K = b^{\mu\nu} \nabla_{\mu} s_{\nu} = q^{ab} \, {^{(\parallel)}\!}\nabla_{a} s_{b}$. Under these assumptions, all 3 boundary terms can be collected together to form 
\begin{align}
	\label{finite boundary Hamiltonian contribution}
	H_{\partial\Sigma}[N,\, N^{a}]
	= \int_{\partial\Sigma} 
	\bigg(\frac{1}{\kappa} \, N \, {^{(\partial\Sigma)}\!}K 
	- N^{a} \, \frac{2}{\closedsqrt{q}} \, q_{ab} \, p^{bc} s_{c} \bigg) \closedsqrt{b} 
	\, d^2 l
	\, .
\end{align}
The contribution to the action is
\begin{align}
	\label{finite boundary action contribution}
	S_{\partial\Sigma}[N,\, N^{a}]
	= \int_{\mathbb{R}} H_{\partial\Sigma} \, dt 
    \, .
\end{align}
Varying this part of the action with respect to the lapse and shift according to the variation principle $\delta_{N,N^a} S_{\partial\Sigma} = 0$ gives 2 conditions ${^{(\partial\Sigma)}\!}K|_{\partial\Sigma} = 0$ and $(q_{ab} \, p^{bc} s_{c}) |_{\partial\Sigma} = 0$. However, these conditions are generally not satisfied. There are several ways to fix this inconsistency. One method described in \cite{bojowald} suggests re-normalizing the extrinsic curvature, spatial metric and momenta, and the boundary normals. The contribution (\ref{finite boundary Hamiltonian contribution}) can be re-normalized with functions ${^{(\partial\Sigma)}}\overline{K}$, $\overline{p}^{ab}$, and $ \overline{s}_{a}$ related to a referential metric $\overline{g}_{\mu\nu}$. The re-normalized boundary contribution 
\begin{align}
	\label{re-normalized finite boundary action contribution}
    H^{norm.}_{\partial\Sigma}[N,\, N^{a}] 
    = &- \frac{1}{\kappa} \int_{\partial\Sigma} 
    N \left( {^{(\partial\Sigma)}\!}K \closedsqrt{b} 
    - {^{(\partial\Sigma)}}\overline{K} \closedsqrt{\overline{b}} \right) d^2 l
    + \nonumber
    \\
    & + 2 \int_{\partial\Sigma} N^{a} \left( 
    \frac{\closedsqrt{b}}{\closedsqrt{q}} \, q_{ab} \, p^{bc} s_{c} 
    - \frac{\closedsqrt{\overline{b}}}{\closedsqrt{\overline{q}}} \, \overline{q}_{ab} \, \overline{p}^{bc} \, \overline{s}_{c} \right) d^2 l
    \, . 
\end{align}
The choice of a specific referential metric $\overline{g}_{\mu\nu}$ (or $\overline{q}_{ab}$) depends of the nature of the studied problem. In case of the Minkowski metric on a spherical boundary of radius $R$, flotiated with $N=1$ and $N^{a}=0$, is $H_{\partial\Sigma} = - 8\pi\kappa^{-1} R$, despite being a flat spacetime. For general folation $N$ and $N^{a}$ of the Minkowski metric we take the Minkowski metric in foliation $N=1$ and $N^{a}=0$ as the referential metric, which then gives the correct result $H^{norm.}_{\partial\Sigma}[1,0] = 0$. For general spacetimes, if one chooses the Minkowski metric as referential, non-zero values of $H^{norm.}_{\partial\Sigma}$ would imply energy deviations the flat Minkowski spacetime.

Another method, presented in \cite{brotz}, demands that $\delta N|_{\partial\Sigma} = 0$ and $\delta N^{a}|_{\partial\Sigma} = 0$ parameterize the lapse and shift at the boundary with time derivatives $\bdot{\tau}$ and $\bdot{\tau}^{a}$ of some functions $\tau$ and ${\tau}^{a}$, respectively:
\begin{align}
    \label{bondary parametrization of lapse and shift}
    N|_{\partial\Sigma} = \bdot{\tau}
    \, , \quad
    N^{a}|_{\partial\Sigma} = \bdot{\tau}^{a}
    \, .
\end{align}
The variational principle $\delta S_{\partial\Sigma} = 0$ then yields 2 conservation equations 
\begin{align}
\label{conservation equations for finite boundary terms}   
\frac{d}{dt} \left( {^{(\partial\Sigma)}\!}K \closedsqrt{b} \right)\Big|_{\partial\Sigma}   
= 0
\, , \quad 
\frac{d}{dt} \left( \frac{2}{\closedsqrt{q}} \, q_{ab} \, p^{bc} s_{c} \, \closedsqrt{b} \right)\bigg|_{\partial\Sigma} 
= 0
\, .
\end{align}
for the boundary terms. The canonical momenta  
\begin{align*}
P^{(\partial\Sigma)} 
&= \frac{\delta S_{\partial\Sigma}}{\delta \bdot{\tau}} 
= \frac{1}{\kappa} \, {^{(\partial\Sigma)}\!}K \closedsqrt{b} 
\, , 
\\
P^{(\partial\Sigma)}_{a}
&= \frac{\delta S_{\partial\Sigma}}{\delta \bdot{\tau}^{a}} 
= \frac{2}{\closedsqrt{q}} \, q_{ab} \, p^{bc} s_{c} \closedsqrt{b}
\end{align*}
are completely independent of $\bdot{\tau}$ and $\bdot{\tau}^{a}$ and cannot be therefore inverted, raising 2 new (so called external) constraints 
\begin{equation}
\begin{aligned}
\label{boundary lapce and shift constraints}
\mathcal{C}^{(\partial\Sigma)}_{\perp} 
&\equiv P^{(\partial\Sigma)} - \frac{1}{\kappa} \, {^{(\partial\Sigma)}\!}K \closedsqrt{b} \, \approx 0
\, , 
\\
\mathcal{C}^{(\partial\Sigma)}_{a} 
&\equiv P^{(\partial\Sigma)}_{a} - \frac{2}{\closedsqrt{q}} \, q_{ab} \, p^{bc} s_{c} \closedsqrt{b} \, \approx 0 
\, .
\end{aligned}
\end{equation}
Constraints $\mathcal{C}^{(\partial\Sigma)}$ and $\mathcal{C}^{(\partial\Sigma)}_{a}$ exist only on the boundary and do not directly affect other constraints in the interior. Consistency conditions for these constraints are satisfied since the action part $S_{\partial\Sigma}$ does not depend on $\tau$ and ${\tau}^{a}$, and so from Hamilton equations $\bdot{P}^{(\partial\Sigma)} = 0$ and $\bdot{P}^{(\partial\Sigma)}_{a} = 0$, which in combination with (\ref{conservation equations for finite boundary terms}) shows the constraints are indeed conserved. These new constraints are thus first class, and ought to be included in the action:
\begin{align}
	\label{finite boundary action contribution with parameterized lapse and shift}
    &S_{\partial\Sigma}\left[\tau,\, \tau^{a},\, P^{(\partial\Sigma)},\, P^{(\partial\Sigma)}_{a};\, N,\, N_{a}\right]
    = \nonumber
    \\
    & \hspace{+4em} = \int_{\mathbb{R}} \int_{\partial\Sigma} 
    \left(P^{(\partial\Sigma)} \, \bdot{\tau} + P^{(\partial\Sigma)}_{a} \, \bdot{\tau}^{a} 
    - N \mathcal{C}^{(\partial\Sigma)}_{\perp} - N^{a} \mathcal{C}^{(\partial\Sigma)}_{a} \right) d^2 l \,dt
    \, .
\end{align}

In case of asymptotic boundary, it is more convenient to obtain all boundary terms directly by varying the ADM action (\ref{ADM action}) with respect to the canonical metric and momenta. Performing the variations yields 4 boundary terms
\begin{align}
	\label{asymptotic boundary terms - NR}
	B_{R}^{(N)} 
	&= \frac{1}{2 \kappa} \int_{\partial\Sigma} 
	s_{a} \, \partial_{b} N \left( q^{am} q^{bn} - q^{ab} q^{mn} \right) 
	\delta q_{mn} \closedsqrt{b} \, d^2 l 
	\, ,
	\\
	\label{asymptotic boundary terms - qR}
	B_{R}^{(q)} 
	&= - \frac{1}{2 \kappa} \int_{\partial\Sigma} 
	s_{a} N \left( q^{am} q^{bn} - q^{ab} q^{mn} \right) 
	{^{(\parallel)}\!}\nabla_{b} \delta q_{mn} \, \closedsqrt{b} \, d^2 l 
	\, ,
	\\
	\label{asymptotic boundary terms - qNdp}
	B_{N\nabla p}^{(q)} 
	&= - \int_{\partial\Sigma} 
	\frac{N_{b}}{\closedsqrt{q}} \left( p^{cb} q^{ia} - p^{ac} q^{ib} \right) 
	\Big( s_{a} \delta q_{ic} + s_{c} \delta q_{ia} - s_{i} \delta q_{ac} \Big) 
	\closedsqrt{b} \, d^2 l 
	\, ,
	\\
	\label{asymptotic boundary terms - pNdp}
	B_{N\nabla p}^{(p)} 
	&= - \, 2 \int_{\partial\Sigma} 
	\frac{s_{a} N_{b}}{\closedsqrt{q}} \, \delta p^{ab} \, \closedsqrt{b} \, d^2 l 
	\, . 
\end{align}
An important case are asymptotically flat spacetimes. Taking $r$ as a radial distance from the origin,  an asymptotically flat metric has to satisfy fall-off condition $g_{\mu\nu} \sim \eta_{\mu\nu} + \mathcal{O}(r^{-1})$ for $r \rightarrow \infty$. Consequently, the fall-off conditions for the spatial metric, lapse and shift are
\begin{align}
	\label{fall-off conditions for spatial metric, lapse and shift}
	q_{ab} \sim \delta_{ab} + \mathcal{O}(r^{-1}) 
	\, , \quad 
	N \sim N_{\infty}(t) + \mathcal{O}(r^{-1})
	\, , \quad
	N^{a} \sim \mathcal{O}(r^{-1}) 
	\, . 
\end{align}
The asymptotic lapse $N_{\infty} = \lim_{r\mapsto\infty} N$ depends on out choice of foliation, e.g. for $g_{00}=1$ will be $N_{\infty} = 1$, and represents the proper time element on the asymptotic boundary. The metric determinant is $\closedsqrt{q} \sim 1 + \mathcal{O}(r^{-1})$, same for $\closedsqrt{b}$, and thus $\closedsqrt{b} d^2 l \sim r^2$. The boundary normal is $s_{a} \sim \mathcal{O}(1)$. From the finiteness of action, $S \sim \mathcal{O}(1)$, we can derive the fall-off condition for the canonical momenta: $p^{ab} \sim \mathcal{O}(r^{-2})$. Applying these fall-off condition to the 4 boundary terms (\ref{asymptotic boundary terms - NR})-(\ref{asymptotic boundary terms - qNdp}) shows their asymptotic behaviour:
\begin{align*}
	B_{R}^{(N)} &\sim \mathcal{O}(r^{-1}) \overset{r \rightarrow \infty}{\longrightarrow} 0 
	\, , \quad
	B_{R}^{(q)} \sim \mathcal{O}(1)
	\\
	B_{N\nabla p}^{(p)} &\sim \mathcal{O}(r^{-1}) \overset{r \rightarrow \infty}{\longrightarrow} 0 
	\, , \quad
	B_{N\nabla p}^{(h)} \sim \mathcal{O}(r^{-2}) \overset{r \rightarrow \infty}{\longrightarrow} 0 
	\, . 
\end{align*}
All the terms with the exception of (\ref{asymptotic boundary terms - qR}) asymptotically vanish. Since our metric is assumed to be asymptotically flat, in the limit $r \rightarrow \infty$ we introduce asymptotic Cartesian coordinates $\mathrm{x} = \mathrm{x}^a$. The spatial covariant derivative then behaves as ${^{(\parallel)}\!}\nabla_{b} \sim \partial_{\mathrm{x}^b}$. Denoting $\closedsqrt{b} d^2 l \sim dS$ an surface element of the boundary, the non-vanishing term $B_{R}^{(q)}$ can be obtained as a variation ot the ADM energy
\begin{align}
	\label{ADM energy}
	E_{ADM}[q_{ab}](t)
	\equiv \lim_{r \rightarrow \infty} \, \frac{1}{2 \kappa} 
	\int_{\partial\Sigma} \Big( s^{m} \delta^{an} - s^{a} \delta^{mn} \Big) 
	\frac{\partial q_{mn}}{\partial \mathrm{x}^b} \, dS
	\, .
\end{align}
ADM energy is both a function of time and a functional of the spatial metric. The asymptotic behaviuor of boundary term (\ref{asymptotic boundary terms - qR}) can be then expressed as $B_{R}^{(h)} \sim - N_{\infty}(t) \, \delta E_{ADM}$. There are now several ways of how to handle the non-vanishing term. We will be brief, but more detailed analysis is provided in \cite{kuchar}. One option is to demand the lapse to have a fixed form at $r \rightarrow \infty$, which results in $\delta N_{\infty}(t) = 0$.
The boundary contribution to the action will the be
\begin{align}
	\label{ADM energy action contribution}
	S_{\infty} = - \int_{\mathbb{R}} N_{\infty} \, E_{ADM} \, dt
	\, .
\end{align}
Another approach is to parametrize $N_{\infty}(t) = \bdot{\tau}_{\infty}$ by time derivative of some function $\tau_{\infty}(t)$, and add this function to the canonical variables on the boundary. The associated canonical momentum $P_{\infty} = \delta_{\bdot{\tau}_{\infty}} S_{\infty} = - E_{ADM}$ is the ADM energy. This momentum does not depend on $\bdot{\tau}_{\infty}$, and cannot be inverted, giving us a primary constraint $\mathcal{C}_{\infty} \equiv P_{\infty} + E_{ADM} \approx 0$. The boundary contribution to the action is then 
\begin{align}
	\label{parametrized ADM energy action contribution}
	S_{\infty}[\tau_{\infty},\, P_{\infty};\, N]
	= \int_{\mathbb{R}} \left(P_{\infty} \bdot{\tau}_{\infty} - N \mathcal{C}_{\infty} \right) dt
	\, .
\end{align}
Variation of $S_{\infty}$ with respect to $P_{\infty}$ gives back the parametrization of the boundary lapse. Variation with respect to the full lapse $N$ gives the constraint $C_{\infty} \approx 0$, and variation with respect to $\tau_{\infty}$ yields $\bdot{P}_{\infty} = 0$, which implies conservation of the ADM energy: $\bdot{E}_{ADM} = 0$. This also means that the constraint is consistent, i.e. $\bdot{\mathcal{C}}_{\infty} = 0$, and the description is complete.

\section{Inclusion of Scalar Fields}
\label{Inclusion of Massless Scalar Field}

Dynamics of a scalar field $\phi(x)$ with potential $V$ on a space-time manifold $\mathcal{M}$ is naturally described by Lagrangian density
\begin{align}
	\label{scalar field lagrangian}
	\mathcal{L}_{\phi}
	= - \closedsqrt{-g} \left(\frac{\varepsilon}{2} \, g^{\alpha \beta} \nabla_{\alpha} \phi \, \nabla_{\beta} \phi + V(\phi) \right)
	\, ,
\end{align}
where the possible additional scalar field potential $V(\phi)$ depends only on the scalar field $\phi$ (or other parameters, but not on metric), for example, $V(\phi) = \frac{1}{2} m^2 \phi^2$ would be the mass term of a scalar field $\phi$. The $\varepsilon = \pm 1$ determines sign of the Lagrangian kinetic term. For normal ("real") fields, $\varepsilon = 1$, and for phantom ("ghost") fields $\varepsilon = -1$.
Since $\phi$ is a scalar field, the covariant derivatives of the field reduce to normal derivatives: $\nabla_{\mu} \phi = \partial_{\mu} \phi$. 
The corresponding stress-energy tensor is obtained by varying the Lagrangian density with respect to the metric:
\begin{align}
	\label{scalar field stress-energy tensor}
	T_{\mu\nu}
	= \frac{2}{\closedsqrt{-g}} \frac{\delta \mathcal{L}_{M}}{\delta g^{\mu\nu}}
	= \varepsilon \, \nabla_{\mu} \phi \, \nabla_{\nu} \phi 
	- \frac{\varepsilon}{2} \, g_{\mu \nu} g^{\alpha \beta} \, \nabla_{\alpha} \phi \, \nabla_{\beta} \phi
	-  g_{\mu \nu} V(\phi)
	\, .
\end{align}
Taking trace of Einstein field equations with the stress-energy tensor (\ref{scalar field stress-energy tensor}) as the source and substuting the result back to the Einstein equations gives us their simplified version
\begin{align}
	\label{simplified Einstein equations for scalar field source}
	R_{\mu\nu}
	= \kappa \, \varepsilon \, \nabla_{\mu} \phi \, \nabla_{\nu} \phi 
    + \kappa \, g_{\mu \nu} V(\phi)
	\, .
\end{align}
Conservation laws for $\phi$ are recovered from the contracted Bianchi identities, which imply $\nabla_{\mu} T^{\mu\nu} = 0$. Taking the divergence of (\ref{scalar field stress-energy tensor}) produces the "wave" equation 
\begin{align}
	\label{scalar field conservation equation}
	\varepsilon \, g^{\mu\nu} \, \nabla_{\mu} \nabla_{\nu} \phi - \partial_{\phi} V(\phi) = 0
\end{align}
for the scalar field.
Switching to adapted coordinates $x = (t, x^i)$, denoting $\bdot{\phi} = \partial_{t} \phi$, and substituting the decomposed metric (\ref{inverse metric matrix in adapted coordinates}) and its determinant, the scalar field Lagrangian density becomes
\begin{align}
	\label{massless scalar field lagrangian in adapted coordinates}
	\mathcal{L}_{\phi}
	= \frac{\varepsilon \closedsqrt{q}}{2 N} \Big[ \left( \bdot{\phi} - N^a \partial_{a} \phi \right)^2 
	- N^2 q^{ab} \partial_{a} \phi \, \partial_{b} \phi \Big]
	- N \closedsqrt{q} \, V(\phi)
	\, .
\end{align}
The associated scalar field momentum $p_{\phi}$ from the Lagrangian $L_{\phi} = \int_{\Sigma} \mathcal{L}_{\phi} \, d^3 y$ in the usual way
\begin{align}
	\label{scalar field momentum}
	p_{\phi}
	= \frac{\delta L_{\phi}}{\delta \bdot{\phi}} 
	= \frac{\partial \mathcal{L}_{\phi}}{\partial \bdot{\phi}} 
	= \frac{\varepsilon \closedsqrt{q}}{N} \left( \bdot{\phi} - N^a \, \partial_{a} \phi \right)
	\, .
\end{align}
The momentum depends on $\bdot{\phi}$ only linearly and can be thus easily inverted with respect to it, substituted into $\mathcal{L}_{\phi}$, and derive the corresponding Hamiltonian density
\begin{align}
	\label{scalar field Hamiltonian density}
	\mathcal{H}^{(\phi)}
	= p_{\phi} \, \bdot{\phi} - \mathcal{L}_{\phi}
	= \frac{\varepsilon N}{2 \closedsqrt{q}} \, p_{\phi}^{2} 
	+ p_{\phi} N^a \partial_{a} \phi 
	+ \frac{\varepsilon N \closedsqrt{q}}{2} \, q^{ab} \partial_{a} \phi \, \partial_{b} \phi 
	+ N \closedsqrt{q} \, V(\phi)
	\, .
\end{align}
This Hamiltonian density depends on the lapse $N$ and shift $N^{a}$ linearly. The terms proportional to lapse thus contribute to the Hamiltonian constraint (and as such to the super-Hamiltonian), while the terms proportional to the shift contribute to the momentum constraint. The Hamiltonian density can be then split into the linear combination $\mathcal{H}^{(\phi)} = N \mathcal{H}^{(\phi)}_{\perp} + N^a \mathcal{H}^{(\phi)}_{a}$ with scalar field super-Hamiltonian and super-momentum
\begin{align}
	\label{scalar field super-Hamiltonian}
	\mathcal{H}^{(\phi)}_{\perp} 
	&= \frac{\varepsilon \, p_{\phi}^{2}}{2 \closedsqrt{q}} 
	+ \frac{\varepsilon \closedsqrt{q}}{2} \, q^{ab} \partial_{a} \phi \, \partial_{b} \phi 
	+ \closedsqrt{q} \, V(\phi) 
	\, , 
	\\
	\label{scalar field super-momentum}
	\mathcal{H}^{(\phi)}_{a} 
	&= p_{\phi} \, \partial_{a} \phi 
	\, .
\end{align} 
The overall Hamiltonian for the system is obtained by adding these scalar field contributions to the vacuum case super-Hamiltonian $\mathcal{H}^{(G)}_{\perp}$ and super-momentum $\mathcal{H}^{(G)}_{a}$ from above in (\ref{vacuum super-Hamiltonian}) and (\ref{vacuum super-momentum}) respectively, and introducing new, non-vacuum super-Hamiltonian and super-momentum
\begin{align}
	\label{super-Hamiltonian and super-momentum with massless scalar field - schematic}
	\mathcal{H}^{(G,\,\phi)}_{\perp} \equiv \mathcal{H}^{(G)}_{\perp} + \mathcal{H}^{(\phi)}_{\perp}
	\, , \quad
	\mathcal{H}^{(G,\,\phi)}_{a} \equiv \mathcal{H}^{(G)}_{a} + \mathcal{H}^{(\phi)}_{a} 
	\, .
\end{align}
All together, this Hamiltonian is then a functional
\begin{align}
	\label{Hamiltonian with massless scalar field}
	H^{(G,\,\phi)} 
	= \int_{\Sigma} \left( N \, \mathcal{H}^{(G,\,\phi)}_{\perp} 
	+ N^a \, \mathcal{H}^{(G,\,\phi)}_{a} \right) d^3 y 
	\, ,
\end{align}
with phase space coordinates $(q_{ab},\, p^{ab},\, \phi ,\, p_{\phi})$. Evolution of functions/functionals on this phase space is prescribed by the Poisson brackets $\bdot{F} = \{F,\, H^{(G,\,\phi)}\}$ with the additional (and the only non-zero) canonical bracket 
\begin{align}
	\label{scalar field caninical Poisson brackets}
	\{ \phi(\bm{x}),\, p_{\phi}(\bm{y}) \} = {\delta}^{(3)}(\bm{x} - \bm{y})
\end{align}
for the scalar field and its momentum. The total action of the system is
\begin{align}
	\label{action with massless scalar field}
	S^{(G,\,\phi)}
	= \int_{\mathbb{R}} \int_{\Sigma} \left( p^{ab} \, \bdot{q}_{ab} + p_{\phi} \, \bdot{\phi} 
	- N \, \mathcal{H}^{(G,\,\phi)}_{\perp} - N^a \, \mathcal{H}^{(G,\,\phi)}_{a} \right) dt \, d^3 y 
	\, . 
\end{align}

The new super-Hamiltonian and super-momentum (\ref{Hamiltonian with massless scalar field}) satisfy the same constraint algebra as in the vacuum case above, namely
\begin{align}
	\label{constraint algebra with scalar field - 2 super-Hamiltonians}
	\left\{ \mathcal{H}^{(G,\,\phi)}_{\perp}(\bm{x}) ,\, \mathcal{H}^{(G,\,\phi)}_{\perp}(\bm{y}) \right\} 
	&= \left[q^{ab}(\bm{x}) \mathcal{H}^{(G,\,\phi)}_{a}(\bm{x})   
	+ q^{ab}(\bm{y}) \mathcal{H}^{(G,\,\phi)}_{a}(\bm{y}) \right] \partial_{b} {\delta}^{(3)}(\bm{x} - \bm{y})
	\\
	\label{constraint algebra with scalar field - super-Hamiltonian and super-momentum}
	\left\{ \mathcal{H}^{(G,\,\phi)}_{a}(\bm{x}) ,\, \mathcal{H}^{(G,\,\phi)}_{\perp}(\bm{y}) \right\} 
	&= \mathcal{H}^{(G,\,\phi)}_{\perp}(\bm{x}) \partial_{a} {\delta}^{(3)}(\bm{x} - \bm{y}) 
	\\
	\label{constraint algebra with scalar field - 2 super-momenta}
	\left\{ \mathcal{H}^{(G,\,\phi)}_{a}(\bm{x}) ,\, \mathcal{H}^{(G,\,\phi)}_{b}(\bm{y}) \right\} 
	&= \mathcal{H}^{(G,\,\phi)}_{a}(\bm{y}) \partial_{b} {\delta}^{(3)}(\bm{x} - \bm{y}) 
	+ \mathcal{H}^{(G,\,\phi)}_{b}(\bm{x}) \partial_{a} {\delta}^{(3)}(\bm{x} - \bm{y})
	\, .
\end{align}
where all the derivatives of Dirac delta functions are taken with respect to $\bm{x}$. The full verification is provided in Appendix \ref{Constraint Algebra with Scalar Field}. The constraint algebra hence remains preserved even after inclusion of the scalar field.

Variation of the scalar-field part of (\ref{action with massless scalar field}) with respect to the scalar field produces 2 additional boundary terms:
\begin{align*}
	B_{\phi}^{(N)} 
	= - \int_{\partial\Sigma} N s^{a} \left(\partial_{a} \phi \right) \delta\phi 
	\closedsqrt{b} \, d^2 l 
	\, , \quad
	B_{\phi}^{(p)} 
	= - \int_{\partial\Sigma} \frac{s_{a} N^{a}}{\closedsqrt{q}} p_{\phi} \delta \phi \, \closedsqrt{b} \, d^2 l 
	\, , 
\end{align*}
Both of these terms are directly proportional to $\delta\phi$, and can be then easily eliminated by demanding that the scalar field does not change at $\partial\Sigma$, i.e. $\delta\phi|_{\partial\Sigma} = 0$, or by some appropriate fall-off conditions for $\phi$ in case of asymptotic boundary $r \rightarrow \infty$.

On quantum level, states of the system shall be represented with wave functional $\Psi[q_{ij},\, \phi]$. Canonical commutators of $\widehat{\phi}(y)$ and $\widehat{p}_{\phi}(y)$ are 
\begin{align*}
	[\widehat{\phi}(\bm{x}), \, \widehat{p}_{\phi}(\bm{y})] 
	= i \hbar \, {\delta}^{(3)}(\bm{x} - \bm{y})
	\, ,
\end{align*}
where the rest of the commutators involving the scalar field and its momentum are zero. Similarly to the vacuum case, in metric representation, operators $\widehat{\phi}(y)$ and $\widehat{p}_{\phi}(y)$ act on the wave functionals $\Psi[q_{ij},\, \phi]$ as
\begin{align*} 
	\widehat{\phi} \, \Psi = \phi \, \Psi
	\, , \quad
	\widehat{p}_{\phi} \, \Psi = - \, i \hbar \, \frac{\delta \Psi}{\delta \phi}
	\, .
\end{align*}
The Hamiltonian and momentum constraints are $\widehat{\mathcal{H}}^{\,(G,\,\phi)}_{\perp} \, \Psi = 0$ and $\widehat{\mathcal{H}}^{\,(G,\,\phi)}_{a} \, \Psi = 0$ respectively, and finally, the Wheeler-DeWitt equations (in the conventional qp-ordering): 
\begin{align}
	\label{Wheeler-DeWitt equations with scalar field} 
	- 2 \kappa \hbar^2 \, G_{abcd} \, \frac{\delta^2 \Psi}{\delta q_{ab} \, \delta q_{cd}} 
	- \, \frac{\varepsilon \hbar^2}{2 \closedsqrt{q}} \, \frac{\delta^2 \Psi}{\delta \phi^2} 
	+ \frac{\closedsqrt{q}}{2\kappa} \left( \varepsilon \kappa \, q^{ab} \partial_{a} \phi \, \partial_{b} \phi 
    + 2 \kappa V(\phi) - {^{(\parallel)}\!}R\right) \Psi  
	&= 0
	\, ,
	\\
	\label{skalární zdroj: Wheelerovy-DeWittovy rovnice se skalárním polem - hybnostní vazba}
	2 \, i \hbar \, {^{(\parallel)}\!}\nabla_{c} \left( q_{ab} \, \frac{\delta \Psi}{\delta q_{cb}} \right) 
	- i \hbar \left(\partial_{a} \phi\right) \frac{\delta \Psi}{\delta \phi} 
	&= 0
	\, .
\end{align}

\section{Spherically Symmetric Spacetimes}
\label{Spherically Symmetric Spacetimes}

In the text, our main focus are spherically symmetric spacetimes. For our manifold $\mathcal{M} = \bigcup_{t \in \mathbb{R}} \Sigma(t)$ we thus choose a general spherically symmetric spatial metric $q_{ab}$ on the foliation hypersurfaces $\Sigma$. In spherical coordinates $y^{a} = (r, \theta, \varphi)$, such metric can be written in the form 
\begin{align}
	\label{spherically symmetric spacial metric}
	d\sigma^2 = A^2(t, r) dr^2 + B^2(t, r) d\Omega^2
	\, ,
\end{align}
with $d\Omega^2 \equiv d\theta^2 + \sin^2\theta \, d\varphi^2$. The shift vector $N^a$ and the lapse function $N$ both have to respect spherical symmetry of the spacetime which means that neither $N^a$ nor $N$ can depend on any of the angular coordinates $\theta$ and $\varphi$. Also, the angular components $N^{\theta}$ and $N^{\varphi}$ of the shift vector have to be zero. The only form of the shift vector and the lapse function that respects spherical symmetry is $N^{a} = (N^r ,\, 0,\, 0)$, where $N^r = N^r(t, r)$, and $N=N(t, r)$ respectfully. The full spacetime metric $g_{\mu\nu}$ on $\mathcal{M}$ from (\ref{square of lenght in adapted coordinates}) is then
\begin{align}
	\label{spherically symmetric stacetime metric}
	ds^{2} = \left( - N^{2} + A^2 (N^{r})^2 \right)dt^{2} + 2 \, A^2 N^{r} dt \, dr + d\sigma^2
	\, .
\end{align}
For functions like $F(t, r)$, the derivatives with respect to the foliation time will be denoted as $\bdot{F} \equiv \partial_t F$, and the derivatives with respect to the radial coordinate as $F\bm{'} \equiv \partial_r F$. The spatial metric (\ref{spherically symmetric spacial metric}) is diagonal with only 3 non-zero components: $q_{rr} = A^2$, $q_{\theta\theta} = B^2$, and $q_{\varphi\varphi} = B^2 \sin^2\theta$, and contains 2 independent functions $A(t, r)$ and $B(t, r)$. The spatial metric density is $\closedsqrt{q} = A B^2 \sin\theta$. The only non-zero Christoffel symbols are
\begin{equation*}
\begin{gathered}
	\Gamma\indices{^{r}_{rr}} = \frac{A\bm{'}}{A} 
	\, , \quad 
	\Gamma\indices{^{r}_{\theta\theta}} = - \frac{B B\bm{'}}{A^2}
	\, , \quad
	\Gamma\indices{^{r}_{\varphi\varphi}} = \Gamma\indices{^{r}_{\theta\theta}} \sin^2 \theta
	\, , 
	\\
	\Gamma\indices{^{\theta}_{\theta r}} = \Gamma\indices{^{\varphi}_{\varphi r}} = \frac{B\bm{'}}{B}
	\, , \quad
	\Gamma\indices{^{\theta}_{\varphi\varphi}} = - \sin\theta \cos\theta
	\, , \quad
	\Gamma\indices{^{\varphi}_{\varphi\theta}} = \frac{\cos\theta}{\sin\theta}
	\, .
\end{gathered}
\end{equation*}
The only non-zero components of the spatial Riemann tensor are:
\begin{align*}
	R\indices{^{r}_{\theta r \theta}} 
	= \frac{B}{A^3}\left(A\bm{'} B\bm{'} - A B\bm{''}\right)
	\, , \quad
	R\indices{^{r}_{\varphi r \varphi}} = R\indices{^{r}_{\theta r \theta}} \sin^2(\theta)
	\, , \quad
	R\indices{^{\theta}_{\varphi \theta \varphi}} 
	= \left( 1 - \frac{B\bm{'}^{\,2}}{A^2} \right) \sin^2\theta
	\, ,
\end{align*}
and for the Kretschmann scalar:
\begin{align}
	\label{Kretschmann scalar of shperically symmetric spatial metric}
	R^{abcd} R_{abcd} 
	= \frac{8}{A^4} \left( \frac{A\bm{'} B\bm{'}}{A B} - \frac{B\bm{''}}{B} \right)^2
	- \frac{4}{B^4} \left( 1 - \frac{B\bm{'}^{\,2}}{A^2} \right)^2
	\, .
\end{align}
The spatial Ricci tensor is
\begin{equation*}
\begin{gathered}
	R_{rr} 
	= 2\left(\frac{A\bm{'} B\bm{'}}{AB} - \frac{B\bm{''}}{B}\right)
	\, , \quad
	R_{\theta\theta} 
	= 1 + \frac{B A\bm{'} B\bm{'}}{A^3} - \frac{B\bm{'}^{\,2}}{A^2} 
	- \frac{B B\bm{''}}{A^2}
	\, , \quad
	R_{\varphi\varphi} = R_{\theta\theta} \sin^2\theta
	\, .
\end{gathered}
\end{equation*}
And finally, for the spatial scalar curvature we have
\begin{align}
	\label{scalar curvature of of shperically symmetric spatial metric}
	{^{(\parallel)}\!}R 
	= \frac{2}{B^2} + \frac{4 A\bm{'} B\bm{'}}{A^3 B}
	- \frac{2 B\bm{'}^{\,2}}{A^2 B^2} - \frac{4 B\bm{''}}{A^2 B}
	\, .
\end{align}
The exterior curvature (\ref{extrinsic curvature in adapted coordinates}) has components
\begin{equation}
	\label{extrinsic curvature in spherical symmetry}
	K_{rr} 
	= \frac{A}{N} \left( \bdot{A} - (AN^r)\bm{'} \right)
	\, , \quad
	K_{\theta\theta} 
	= \frac{B}{N} \left( \bdot{B} - B^{\bm\prime} N^r \right)
	\, , \quad
	K_{\varphi\varphi} = K_{\theta\theta} \sin^2 \theta
	\, .
\end{equation}
The vacuum Lagrangian density
\begin{align*}
	\mathcal{L}_{G}
	\overset{\text{int.}}{=} 
	\frac{N \closedsqrt{q}}{2\kappa} \left({^{(\parallel)}\!}R + K_{ab} K^{ab} - K^{2} \right)
	= \frac{N A B^2}{2\kappa} \left( {^{(\parallel)}\!}R 
	- \frac{4  K_{rr}  K_{\theta\theta}}{A^2 B^2}  
	- \frac{2  K_{\theta\theta}^2}{B^4} \right) \sin\theta 
\end{align*}
depends on the angular coordinates $\theta$ and $\varphi$ only trivially. The corresponding Lagrangian $L_{G} = \int_{0}^{\infty} \int_{0}^{\pi} \int_{0}^{2\pi} \mathcal{L}_{G} \, dr d\theta d\varphi$ can be then easily integrated over the angular coordinates, reducing it effectively to
\begin{align*}
	L_{G} \overset{\text{int.}}{=} \int_{0}^{\infty} \frac{8\pi}{\kappa} N A B^2 
	\left[ \frac{{^{(\parallel)}\!}R}{4} - \frac{K_{\theta\theta}}{B^2} \left( \frac{K_{rr}}{A^2} + \frac{K_{\theta\theta}}{2 B^2} \right) \right] dr 
	\, .
\end{align*}
The extrinsic curvature, and therefore the Lagrangian, contains only 2 independent "velocities" $\bdot{A}$ and $\bdot{B}$. There are hence only 2 independent momenta 
\begin{align}
	\label{spherically symmetric momenta}
	P_{\!A} \equiv \frac{\delta L_G}{\delta \bdot{A}}
	= - \frac{4\pi}{\kappa} \frac{K_{\theta\theta}}{A}
	\, , \quad
	P_{\!B} \equiv \frac{\delta L_G}{\delta \bdot{B}} 
	= - \frac{4\pi}{\kappa} A \left( \frac{K_{rr}}{A^2} + \frac{K_{\theta\theta}}{B^2} \right)
	\, .
\end{align}
To properly utilize general relations derived in the previous sections, we note that because of out parametrization of the metric (\ref{spherically symmetric spacial metric}) and the integration over angular coordinates, the general metric momenta $p^{rr}$ and $p^{\theta\theta}$ are related to $P_{\!A}$ and $P_{\!B}$ as
\begin{align*}
	p^{rr} = \frac{\sin\theta}{8 \pi A} \, P_{\!A}
	\, , \quad
	p^{\theta\theta} = \frac{\sin\theta}{8 \pi B} \, P_{\!B}
	\, ,
\end{align*} 
Assume that generally $\bdot{A} \neq 0$ and $\bdot{B} \neq 0$ it is possible invert the momenta calculate the ADM Hamiltonian
\begin{align*}
	H_{G} 
	= \int_{0}^{\infty} \left( N \mathcal{H}^{(G)}_{\perp} +  N^r \mathcal{H}^{(G)}_{r} \right) dr  
	\, ,
\end{align*} 
with the vacuum super-Hamiltonian and super-momentum
\begin{align}
	\label{spherically symmetric vacuum super-Hamiltonian and super-momentum}
	\mathcal{H}^{(G)}_{\perp}
	= \frac{\kappa}{8\pi} \, G_{IJ} \, p^I p^J 
	- \frac{2\pi}{\kappa} A B^2  {^{(\parallel)}\!}R
	\, , \quad
	\mathcal{H}^{(G)}_{r}
	= -  A  P^{\bm{\prime}}_{\! A}  +  B\bm{'} P_{\! B}
	\, ,
\end{align}
where the part of super-Hamiltonian containing products of the momenta
\begin{align}
	\label{effective inverse DeWitt supermetric contraction with momenta}
	G_{IJ} \, p^I p^J
	= \frac{A}{2 B^2} P_{\!A}^2 - \frac{1}{B} P_{\!A} P_{\!B}
\end{align}
has been simplified with the effective, inverse DeWitt (super)metric
\begin{align}
	\label{effective inverse DeWitt supermetric and effective momenta}
	G_{IJ}
	\equiv \left(\begin{matrix}
		\frac{A}{2 B^2} & - \frac{1}{2 B} \\[5pt]
		- \frac{1}{2 B} & 0 \\
	\end{matrix}\right)
	\, ,
\end{align}
and $p^I \equiv (P_{\!A},\, P_{\!B})$. The corresponding effective DeWitt supermetric is obtained as the inverse to $G_{IJ}$, where (unlike the theoretical case above) the two effective metrics are inverse to each other in the standard way $G^{IK} G_{KJ} = \delta^{I}_{J}$. Explicit for of the DeWitt supermetric is thus
\begin{align}
	\label{effective DeWitt supermetric}
	G^{IJ}
	= \left(\begin{matrix}
		0 & - 2 B \\
		- 2 B & - 2 A \\
	\end{matrix}\right)
    \, .
\end{align} 

Looking back at (\ref{spherically symmetric vacuum super-Hamiltonian and super-momentum}), in order to calculate value of this radial component of the super-momentum in the ADM Hamiltonian we had to integrate by parts, giving rise to the boundary term $\left[ A P_{\! A} N^r \right]_{r=0}^{\infty}$. It follows from the fall-off conditions for asymptotically flat metrics that at $r\rightarrow\infty$ the boundary terms behaves as $A P_{\! A} N^r \sim \mathcal{O}(r^{-3})$, and so $A P_{\! A} N^r \overset{r\rightarrow\infty}{\longrightarrow} 0$. The remaining part $\left( A P_{\! A} N^r \right)|_{r=0}$ has to be then eliminated by appropriate boundary conditions in the radial coordinate origin. One such option is to require $B|_{r=0} = 0$, implication of which is also that $P_{\! A}|_{r=0} = 0$, and the remaining part of the boundary term vanishes. We will therefore demand the function $B$ to satisfy the condition
\begin{align}
	\label{condition B=0}
	B|_{r=0}  =  0
	\, .
\end{align}

Let us now include a scalar field source to the system. We consider a scalar field $\phi$ that obeys spherical symmetry. Such scalar field has to be a function $\phi = \phi(t, r)$ when expressed in spherical coordinates. Following the formalism detailed in section \ref{Inclusion of Massless Scalar Field}, and integrating over angular coordinates, we obtain the scalar-field super-Hamiltonian and super-momentum 
\begin{align}
	\label{spherically symmetric scalar-field super-Hamiltonian and super-momentum}
	\mathcal{H}^{(\phi)}_{\perp} 
	= \frac{\varepsilon \, P_{\!\phi}^2}{8 \pi A B^2} 
    + \frac{2 \pi \varepsilon B^2}{A} \phi\bm{'}^{\,2}
    + 4 \pi A B^2 V(\phi)
	\, , \quad
	\mathcal{H}^{(\phi)}_{r} = \phi\bm{'} P_{\!\phi}
	\, .
\end{align}
The full (ADM) action of the system is 
\begin{align}
	\label{spherically symmetric AMD action with scalar field}
	&S^{(G, \phi)}[A, B, \phi, P_{\!A}, P_{\!B}, P_{\!\phi}; N, N^{r}]
	= \nonumber
	\\
	& \hspace{+4em} 
	=  \int_{\mathbb{R}} \int_{0}^{\infty} 
	\left( P_{\!A} \bdot{A} + P_{\!B} \bdot{B} + P_{\!\phi}  \bdot{\phi} 
	- N  \mathcal{H}^{(G, \phi)}_{\perp} - N^r  \mathcal{H}^{(G, \phi)}_{r} \right) dr \, dt
	\, .
\end{align}
with the super-Hamiltonian and super-momentum sums (\ref{super-Hamiltonian and super-momentum with massless scalar field - schematic}) of the vacuum contributions (\ref{spherically symmetric vacuum super-Hamiltonian and super-momentum}) and the scalar-field contributions (\ref{spherically symmetric scalar-field super-Hamiltonian and super-momentum}):
\begin{align}
	\label{spherically symmetric super-Hamiltonian - full version}
	\mathcal{H}^{(G, \phi)}_{\perp} 
	&= \frac{\kappa}{8\pi} \, Q_{IJ} \, p^I p^J
	-  \frac{2\pi}{\kappa}  A B^2  {^{(\parallel)}\!}R + \frac{2 \pi \varepsilon B^2}{A} \phi\bm{'}^{\,2} 
    + 4 \pi A B^2 V(\phi)
	\, ,
	\\
	\label{spherically symmetric super-momentum - full version}
	\mathcal{H}^{(G, \phi)}_{r}
	&= -  A  P\bm{'}_{\!A} +  B\bm{'} P_{\!B} + \phi\bm{'} P_{\!\phi}
	\, ,
\end{align}
where we have denoted the whole momentum part 
\begin{align}
	\label{spherically symmetric super-Hamiltonian - full version}
	\frac{\kappa}{8\pi} \, Q_{IJ} \, p^I p^J 
    \equiv \frac{\kappa}{8\pi} \left( \frac{A P_{\!A}^2}{2 B^2} - \frac{P_{\!A} P_{\!B}}{B} \right)
	+ \frac{\varepsilon \, P_{\!\phi}^2}{8 \pi A B^2}
\end{align}
with the canonical momenta $p^I \equiv (P_{\!A},\, P_{\!B},\, P_{\!\phi})$ and the "canonical supermetric" $Q^{IJ}$ and its inverse $Q_{IJ}$ ($Q^{IK} Q_{KJ} = \delta^I_J$):
\begin{align}
	\label{effective inverse DeWitt supermetric and effective momenta}
	Q^{IJ}
	\equiv 
    \left(\begin{matrix}
    0 & - 2 B & 0 \\
    - 2 B & -2 A & 0 \\
    0 & 0  & \varepsilon \kappa A B^2 \\
    \end{matrix}\right)
	\, , \quad
    Q_{IJ}
	\equiv 
    \left(\begin{matrix}
    \frac{A}{2 B^2} & - \frac{1}{2 B} & 0 \\
    - \frac{1}{2 B} & 0 & 0 \\
    0 & 0  & \frac{\varepsilon}{\kappa A B^2} \\
    \end{matrix}\right)
    \, .
\end{align}

A (formally finite) spherically symmetric boundary $\partial\Sigma = \{ (r, \theta, \varphi); r = r_{max}\}$ located at the radius $r_{max} = const.$ is given by the constraint $r - r_{max} = 0$. Its external normal $s_{\mu}$ is proportional to $\partial_{\mu}(r - r_{max}) = \delta^{r}_{\mu}$, and normalized as $q^{\mu\nu} \delta^{r}_{\mu} \delta^{r}_{\nu} = q^{rr} = A^2$. The unit normal to the boundary is hence $s_{\mu} = A \delta^{r}_{\mu}$. The induced boundary metric is the angular part $d\omega^2 = B^2 d\Omega^2$ of the spatial metric (\ref{spherically symmetric spacial metric}), with boundary metric density $\closedsqrt{b} = B^2 \sin\theta$. The boundary extrinsic curvature trace is equal to
\begin{align*}
	{^{(\partial\Sigma)}\!}K 
	= q^{ab}  {^{(\parallel)}\!}\nabla_{a} s_{b}
	= \frac{2 B\bm{'}}{A B^2}
	\, .
\end{align*}
The boundary contribution (\ref{finite boundary Hamiltonian contribution}) for a spherical boundary is equal to 
\begin{align}
	\label{spherical finite boundary action contribution}
	H_{\partial\Sigma}[N, \, N^a]
	= \left(N \frac{8\pi}{\kappa} \frac{B B\bm{'}}{A} 
	- N^{r} A P_{\!A} \right)\Bigg|_{r=r_{max}}
	\, , 
\end{align}
where the lapse $N|_{r=r_{max}}$ and the shift $N^{r}|_{r=r_{max}}$ are parameters on the boundary, independent of the lapse and shift from the interior of the area.

Another important quantity is the ADM energy. On an asymptotically Cartesian boundary it holds for asymptotically flat metrics that $q_{ab} \sim \delta_{ab}$. To calculate asymptotic Cartesian derivatives of a metric, the metric needs to be transformed from spherical coordinates $(r, \theta, \varphi)$ to Cartesian coordinates $\mathrm{x}^a = (x, y, z)$. The derivative (gradient) of a radial distance $r = \closedsqrt{\mathrm{x}^a\mathrm{x}_a}$ is equal to and will be denoted as $r_a \equiv \partial_{\mathrm{x}^a} r = r^{-1} \mathrm{x}_a$. A normal $s_a$ to the spherically symmetric boundary $r =const.$ is therefore $s_a = r_a$. Integrating (\ref{ADM energy}) over the angular coordinates gives us a starting formula for the ADM energy  
\begin{align}
	\label{ADM energy in spherical symmetry}
	E_{ADM} 
	= \lim_{r\rightarrow\infty} \frac{2\pi}{\kappa}  r^2 
	\Big( r^{m}  \delta^{an}  -  r^{a}  \delta^{mn} \Big) 
	\frac{\partial q_{mn}(\mathrm{x})}{\partial \mathrm{x}^a} 
	\, . 
\end{align}
An element of the radial distance transform to Cartesian coordinates as  $dr^2 = r_a r_b \, d\mathrm{x}^a d\mathrm{x}^b$. An angular element is then obtained simply from the Euclidean metric: $d\Omega^2 = r^{-2}\left(ds^2 - dr^2\right) = r^{-2}\left(\delta_{ab} - r_a r_b\right) d\mathrm{x}^a d\mathrm{x}^b$. The components of the new metric expressed in terms of the the old one are 
\begin{align}
	\label{sférická symetrie: sféricky symetrická metrika v asymptoticky kartézských souřadnicích}
	q_{ab}(\mathrm{x}) 
	= q_{rr} r_a r_b  +  \frac{q_{\theta\theta}}{r^2}\left(\delta_{ab} - r_a r_b\right)
	\, .
\end{align}
Next we need to calculate the Cartesian derivatives of the metric components. Derivatives of a function $F(t, r)$ on foliation hypersurfaces $\Sigma(t=const.)$ are $\frac{\partial F}{\partial \mathrm{x}^a} = \frac{\partial F}{\partial r} \frac{\partial r}{\partial \mathrm{x}^a} = F\bm{'} r_a$.  
A Cartesian derivative of the radial gradient $r_a$ is equal to $\frac{\partial r_a}{\partial \mathrm{x}^b} = r^{-1}\left(\delta_{ab} - r_a r_b\right)$. Cartesian derivatives of the metric components are then given as
\begin{align*}
	\frac{\partial q_{mn}(\mathrm{x})}{\partial \mathrm{x}^a}
	&= q_{rr}^{\bm\prime}  r_a r_m r_n 
	+  \left(\frac{q_{rr}}{r} - \frac{q_{\theta\theta}}{r^3}\right)
	\left(r_n\delta_{am} + r_m\delta_{an} - 2 r_a r_m r_n \right)  + 
	\nonumber
	\\
	&\quad\,\, +  \left( \frac{q_{\theta\theta}\bm{'}}{r^2} - \frac{2 q_{\theta\theta}}{r^3} \right) 
	\left(r_a \delta_{mn} - r_a r_m r_n\right) 
	\, .
\end{align*}
The contraction of two radial gradients is $r_a r^a = \frac{\mathrm{x}_a \mathrm{x}^a}{r^2} = 1$. We can now finally express the ADM energy explicitly in terms of the old metric: 
\begin{align}
	\label{ADM energy in spherical symmetry - final formula}
	E_{ADM} 
	= \lim_{r\rightarrow\infty}  \frac{8\pi}{\kappa}  \frac{r}{2} 
	\left( q_{rr} - \frac{q_{\theta\theta}\bm{'}}{r} + \frac{q_{\theta\theta}}{r^2} \right) 
	\, . 
\end{align}
Asymptotically flat metrics should satisfy fall-off condition $q_{ab}(\mathrm{x}) \sim \delta_{ab} + \mathcal{O}(r^{-1})$ which for the components $q_{rr}(t, r)$ and $q_{\theta\theta}(t, r)$ included in $q_{ab}(\mathrm{x})$, see (\ref{sférická symetrie: sféricky symetrická metrika v asymptoticky kartézských souřadnicích}), implies general asymptotic behaviour 
\begin{equation}
\begin{aligned}
	\label{fall-off conditions for spherical metric} 
    q_{rr}  
    &\sim 1 + \frac{2 \mu(t)}{r} + \mathcal{O}(r^{-1-\varepsilon})
    \, ,
    \\
    q_{\theta\theta}   
    &\sim r^2 + r \rho_{1}(t) + \rho_{0}(t) + \mathcal{O}(r^{-\varepsilon})  
    \, , 
\end{aligned}
\end{equation}
with $\mu(t)$, $\rho_{0}(t)$, $\rho_{1}(t)$ being some, yet unspecified functions. 
Finally, by substituting these fall-off conditions into the formula (\ref{ADM energy in spherical symmetry - final formula}) above, we get the following the value for the ADM energy of an asymptotically flat, spherically symmetric metric:
\begin{align}
	\label{ADM energy in spherical symmetry - fall-off formula}
	E_{ADM} = \frac{8\pi}{\kappa}  \mu(t)
	\, . 
\end{align}
The fall-off conditions for the metric functions $A$ and $B$ translate to
\begin{equation}
\begin{aligned}
	\label{fall-off conditions for parametrized spherical metric} 
	A &\sim 1 + \frac{\mu(t)}{r} + \mathcal{O}(r^{-1-\varepsilon}) 
	\, ,
	\\
	B &\sim r + \beta_{0}(t) - \frac{\beta_{-1}(t)}{r} + \mathcal{O}(r^{-1-\varepsilon})  
	\, .
\end{aligned}
\end{equation}
with some functions $\beta_{0}(t)$ and $\beta_{-1}(t)$.

On the classical level, the phase space\footnote{
    Strictly speaking, we are working on a reduced phase space of only the non-trivial dynamical canonical variables.
} 
or our spherically symmetrical system with a scalar field is described by canonical variables $q_I = (A,\, B,\, \phi)$ and the corresponding canonical momenta $p^I = (P_{\!A},\, P_{\!B},\, P_{\!\phi})$, with the only non-zero canonical Poisson brackets
\begin{align}
	\label{spherically symmetrical canonical Poisson brackets}
	\{ q_I(\bm{x}),\, p^J(\bm{y}) \} 
	= \delta^{I}_{J} \, {\delta}^{(3)}(\bm{x} - \bm{y})
	\, .
\end{align}
On the quantum level, we assign operators $\widehat{q}_I = (\widehat{A},\, \widehat{B},\, \widehat{\phi})$ to each canonical variable, and $\widehat{p}^{\,I} = (\widehat{P}_{\!A},\, \widehat{P}_{\!B},\, \widehat{P}_{\!\phi})$ to each canonical momenta, and define canonical commutators 
\begin{align}
	\label{spherically symmetrical canonical commutators}
	[\widehat{q}_I(\bm{x}),\, \widehat{p}^{\,J}(\bm{y})]
	= i \hbar \, \delta^{I}_{J} \, {\delta}^{(3)}(\bm{x} - \bm{y})
	\, ,
\end{align}
In metric representation, quantum states the system are described by wave functionals $\Psi[A,\, B,\, \phi]$, and the canonical operators are prescribed to act of such wave functionals as 
\begin{align} 
	\label{spherically symmetrical canonical operators act on wave functionls}
	\widehat{q}_I \, \Psi = q_I \, \Psi
	\, , \quad
	\widehat{p}^{\,I} \, \Psi = - \, i \hbar \, \frac{\delta \Psi}{\delta q_I}
	\, .
\end{align}
The Hamiltonian and momentum constraints are $\widehat{\mathcal{H}}^{\,(G,\,\phi)}_{\perp} \, \Psi = 0$ and $\widehat{\mathcal{H}}^{\,(G,\,\phi)}_{a} \, \Psi = 0$ respectively, and the Wheeler-DeWitt equations (in the conventional qp-ordering): 
\begin{align}
	\label{WDW Hamiltonian constraint in AB parameterization}
	\frac{\kappa \hbar^2}{8\pi} \, Q_{IJ} \, \frac{\delta^2 \Psi}{\delta q_I \delta q_J} 
    + \left( \frac{2 \pi \varepsilon B^2 \phi\bm{'}^{\,2}}{A} + 4 \pi A B^2 V(\phi) 
    - \frac{2\pi}{\kappa}  A B^2  {^{(\parallel)}\!}R\right) \Psi 
	&= 0
	\, ,
	\\
	\label{WDW momentum constraint in AB parameterization}
	B\bm{'} \, \frac{\delta \Psi}{\delta B} + \phi\bm{'} \, \frac{\delta \Psi}{\delta \phi}
	-  A \left[ \frac{\delta \Psi}{\delta A} \right]^{\bm{\prime}} 
	&= 0
	\, .
\end{align}

\section{Roberts and special Janis-Newman-Winicour Metric}
\label{Roberts and special Janis-Newman-Winicour Metric}

One of the spherically symmetric solutions to the Einstein field equations is a spacetime which we will from now on call Roberts spacetime (the expression (6.2) in \cite{tahamtan-svitek-2}; for more information and origin of the name see \cite{roberts}) that is described by the metric
\begin{align}
	\label{Roberts metric}
	ds^2 = - dT^2 + dR^2 + \left( R^2 - \chi^2 \right) d\Omega^2
	\, ,
\end{align}
where $\chi$ is a function  
\begin{align}
	\label{Roberts metric chi function}
	\chi(T,\, R) = \chi_0 - \frac{\chi_0}{C}\left( T - R \right)
\end{align}
and $\chi_0, C \geq 0$ are constant parameters. 
In this Roberts metric, the coordinate $R$ has the physical meaning of the radial distance, while the spherical curvature radius is given by the function $\closedsqrt{R^2 - \chi^2}$. There is a singularity located on the radius $R = \chi$. The radial coordinate $R$ ranges from $R \geq \chi$. A more detailed analysis of the metric (\ref{Roberts metric}) can be found in \cite{tahamtan-svitek-2}. The non-zero components of the Ricci tensor of the Roberts metric are
\begin{align}
	\label{Roberts metric Ricci tensor}
	R_{TT} = \frac{2 \chi_0^2  R^2}{C^2 \left( R^2 - \chi^2 \right)^2} 
	\, , \quad 
	R_{RR} = \frac{2 \chi_0^2 \left( 1 - \frac{T}{C} \right)^2}{\left( R^2 - \chi^2 \right)^2}
	\, .
\end{align}
The massless scalar field source $\phi$ in the Roberts metric  (\ref{Roberts metric}) has the explicit form 
\begin{align}
	\label{Roberts meric scalar source field}
	\phi(T,\, R) = \frac{1}{\closedsqrt{2 \kappa}} \ln\left(\frac{R - \chi}{R + \chi}\right)
	\, .
\end{align}
To utilise the general ADM formalism established above without enforcing any particular foliation we have to express the original Roberts coordinates $x^{\mu} = \left( T, R, \theta, \varphi\right)$ in terms of adapted spherical coordinates $y^{\mu} = \left( t, r, \theta, \varphi\right)$. Because of the spherical symmetry of our spacetime, we could choose spherical coordinates as an general frame of reference, adapted to the space-like hypersurfaces $\Sigma$ that foliate our spacetime, while keeping the foliation of our spacetime general\footnote{Although the foliation should be kept general, due to the nature of the studied spacetimes it also has to satisfy spherical symmetry, i.e. the space-like hypersurfaces of the foliation have to be spherically symmetric.}. The spherical symmetry requirement forces the original Roberts coordinates $T$ and $R$ to depend only on $t$ and $r$ as $T(t, r)$ and $R(t, r)$. The Roberts metric (\ref{Roberts metric}) then has the form
\begin{equation}
	\begin{aligned}
		\label{Roberts metric in spherical coordinates}
		ds^2 
        = 
		& - \left( \bdot{T}^2 - \bdot{R}^2 \right) dt^2 
		- 2 \left( \bdot{T} T\bm{'} - \bdot{R} R\bm{'} \right) dt dr 
		\, + 
		\\
		& + \left( R\bm{'}^{\,2} - T\bm{'}^{\,2} \right) dr^2 
		+ \left( R^2  -  \chi^2 \right) d\Omega^2
		\, .
	\end{aligned}
\end{equation} 
Compared to the 3+1 metric deconposition (\ref{square of lenght in adapted coordinates}), the spatial part $g_{ab} = q_{ab}$ of the Roberts metric has, in adapted coordinates, a diagonal form $\text{diag}(q_{rr}, q_{\theta\theta}, q_{\varphi\varphi})$ which corresponds to a spherically symmetric metric (\ref{spherically symmetric spacial metric}) where we identify
\begin{align}
	\label{Roberts a sJNW: prostorová část spec. JNW metriky}
	q_{rr} =  R\bm{'}^{\,2} - T\bm{'}^{\,2} = A^2 
	\, , \quad
	q_{\theta\theta} = R^2 - \chi^2 = B^2
    \, , \quad 
    q_{\varphi\varphi} = q_{\theta\theta} \sin^2 \theta
	\, .
\end{align}  
The covariant form of the shift vector $N_a$ has only its radial component $N_r = \bdot{R}R\bm{'} - \bdot{R}R\bm{'}$ non-zero. The shift vector $N^a = q^{ab} N_b$ then has (as a consequence of spherical symmetry) also non-zero only its radial component 
\begin{align}
	\label{Roberts metric shift vektor}
	N^r = \frac{\bdot{R} R\bm{'} - \bdot{T} T\bm{'}}{R\bm{'}^{\,2} - T\bm{'}^{\,2}}
	\, .
\end{align}
The lapse function $N$ can be obtained from the relation $g_{tt} = N^r N_r - N^2$. After some algebraic manipulations one gets
\begin{align}
	\label{Roberts metric lapse function}
	N = \frac{\bdot{T} R\bm{'} - \bdot{R} T\bm{'}}{\closedsqrt{R\bm{'}^{\,2} - T\bm{'}^{\,2}}}
	\, .
\end{align}
The parameters $R$ and $\chi$ in the Roberts metric and in the source scalar field can be reconstructed in the AB parameterisation (\ref{spherically symmetric spacial metric}) of the metric with the help of the scalar field $\phi$:
\begin{align*}
	B^2 
    = R^2 - \chi^2 
	= \left( R - \chi \right) \left( R + \chi \right)
	\, , \quad
	\exp\left( \closedsqrt{2\kappa} \phi \right) 
	= \frac{\left( R - \chi \right)}{\left( R + \chi \right)}
	\, ,
\end{align*}
from where one can easily obtain
\begin{align}
	\label{Roberts metric parameters R and chi via B and phi}
	R = B \cosh\left( \closedsqrt{\frac{\kappa}{2}}  \phi \right) 
	\, , \quad 
	\chi = B \sinh\left( \closedsqrt{\frac{\kappa}{2}}  \phi \right)
	\, .
\end{align}
The Roberts metric is not asymptotically flat: 
\begin{align}
	\label{Roberts metric is not asymptotically flat}
	\frac{q_{\theta\theta}}{R^2} 
	\xrightarrow{R\rightarrow\infty}  
	\left( 1 - \frac{\chi_0^2}{C^2} \right)
	\, .
\end{align}
The formula (\ref{Roberts metric}) is written in the special foliation $R = r$ and $T = t$, which corresponds to the fixation of the lapse to $N = 1$ and the shift to $N^{r} = 0$. In this foliation, the boundary term (\ref{spherical finite boundary action contribution}) for the Roberts metric,
\begin{align}
	\label{Roberts metric boundary term}
	H_{\partial\Sigma}[1,\, 0] 
	=  - \frac{8\pi}{\kappa} \left(1 - \frac{\chi_0^2}{C^2}\right) R_{max} 
    \, ,
\end{align}
diverges on the asymptotic boundary $R_{max}\rightarrow\infty$. Although we can, for example, eliminate the first term in (\ref{spherical finite boundary action contribution}) by a different choice of foliation (all acceptable foliations have to satisfy $N \neq 0$, however), the second term will remain, causing the same problems with divergences as before.
The foliation $(N, N^{r}) = (1, 0)$  corresponds to the spherical Minkowski metric and has a well understood, physical interpretation. For a Minkowski metric it is $\overline{H}_{\partial\Sigma}[1, 0] = - \frac{8\pi}{\kappa} R_{max}$. The divergent boundary term (\ref{Roberts metric boundary term}) of the Roberts metric therefore cannot be fully normalized with respect to the Minkowski metric. The reason for this is that the spherical curvature radius of the Roberts metric is creating an angular deficit: the surface area of a sphere in some fixed radial distance is always smaller than it would be in the flat Minkowski spacetime, which, because of the Roberts metric is not asymptotically flat, will not vanish even on the asymptotic boundary. The spherical surface deficit of the Roberts metric will then diverge in radial infinity. 
A fixation of the laps and the shift on the boundary would solve the problem with divergences but it would inevitably restrict our choice of a foliation which is undesirable. Another way to deal with the divergences is to consider the lapse and the shift on the boundary to be independent variables, parameterize the boundary term (\ref{spherical finite boundary action contribution}) and formulate another set of phase space constraints, valid only on the boundary.

In the limit $C \rightarrow \infty$ is $\chi \rightarrow \chi_0$. The Roberts metric (\ref{Roberts metric}) then becomes, so called, special Janis-Newman-Winicour (sJNW) metric\footnote{
    The special case of Janis-Newman-Winicour metric from (6.9) in \cite{tahamtan-svitek-2}, where we set $A\rightarrow \infty$.} 
(see the expression (6.9) in \cite{tahamtan-svitek-2})
\begin{align}
	\label{special Janis-Newman-Winicour metric}
	ds^2 = - dT^2 + dR^2 + \left( R^2 - \chi_0^2 \right) d\Omega^2
	\, .
\end{align}
The scalar-field source of the sJNW metric is simply 
\begin{align}
	\label{sJNW metris scalar source field}
	\phi(R) = \frac{1}{\closedsqrt{2 \kappa}} \ln\left(\frac{R - \chi_0}{R + \chi_0}\right)
	\, .
\end{align}
The sJNW metric contains a naked time-like singularity which is located on the constant radius $R = \chi_0$ for $\chi_0 > 0$ or on $R = - \chi_0$ for $\chi_0 < 0$. The coordinate $R$ is still interpreted as the radial distance measured from the location of the singularity, that is, from the constant radius $\chi_0$. The spherical curvature radius is $B = \closedsqrt{R^2 - \chi_0^2}$. 
If one prefers the curvature radius over the proper radial distance, they can replace the coordinate $R$ with $B$ since both quantities $R$ and $B$ have good physical interpretation. In terms of the curvature radius $B$, the sJNW metric has the form
\begin{align}
	\label{sJNW metric with B}
	ds^2 =  - dT^2 + \frac{dB^2}{\left(1 + \frac{\chi_0^2}{B^2}\right)} + B^2  d\Omega^2
	\, .
\end{align}
For $R = \chi_0$ is $B = 0$. In general, spherically symmetric foliation is $R(t, r)$. The condition $B|_{r=0} = 0$ which we demanded in (\ref{condition B=0}) is naturally satisfied iff $R|_{r=0} = \chi_0$, i.e. iff the proper radial distance $R$ is measured from the location of the singularity.

Unlike the Roberts metric, the sJNW metric is asymptotically flat. This can be easily seen in (\ref{Roberts metric is not asymptotically flat}) for $C \rightarrow \infty$. We can therefore calculate a finite value for the ADM energy of sJNW with respect to the Minkowski spacetime. The ADM energy does not depend on a particular choice of a metric for a given spacetime which allows us to simply compare the fall-off conditions (\ref{fall-off conditions for spherical metric}) for the sJNW metric components. We see that $q_{RR} = 1$ and so $\mu(T) = 0$. Alternatively, from
\begin{align*}
    q_{BB} = \left(1 + \frac{\chi_0^2}{B^2}\right)^{-1} \sim 1 - \frac{\chi_0^2}{B^2} + \mathcal{O}(B^{-4})
\end{align*}
we again get $\mu(T) = 0$. Both ways lead to the same result: the ADM energy of the sJNW spacetime is zero. The asymptotic boundary term (\ref{ADM energy action contribution}) thus vanishes. The limit $C \rightarrow \infty$ in (\ref{Roberts metric is not asymptotically flat}) produces the same result as for the Minkowski metric. It is therefore possible to normalize the boundary term with respect to the Minkowski metric, so that the relative ADM energy between the sJNW and the Minkowski metrics is zero.

Because of the problematic behaviour of the Roberts metric at the asymptotic boundary we will, for now, choose to quantize its static version, the sJNW metric.

\section{Canonical Quantization of the sJNW Metric}
\label{Canonical Quantization of the sJNW Metric}

In the case of a spherically symmetrical metric (\ref{spherically symmetric spacial metric}) with a scalar field $\phi$ the corresponding quantum system is described by a wave functional $\Psi = \Psi[A, B, \phi]$. Since the sJNW spacetime includes a real massless scalar field, we set $\varepsilon = 1$ and $V(\phi) = 0$ in the formulas derived above.
The Wheeler-DeWitt equations have, in a qp-ordering (i.e. all the momenta act first, before any of the coordinates) the form
\begin{align}
	\label{WDW Hamiltonian constraint in AB parameterization}
	\frac{\kappa \hbar^2}{8\pi} 
    \Bigg( 
    \frac{A}{2 B^2}  \frac{\delta^2 \Psi}{\delta A^2} 
	-  \frac{1}{B}  \frac{\delta^2 \Psi}{\delta B  \delta A} 
	+  \frac{1}{\kappa A B^2}  \frac{\delta^2 \Psi}{\delta \phi^2} 
    \Bigg)
	+ \frac{2\pi}{\kappa} 
    \Bigg( 
    \frac{\kappa B^2 \phi\bm{'}^{\,2}}{A} 
    -  A B^2  {^{(\parallel)}\!}R 
    \Bigg) \Psi 
	&= 0
	\, ,
	\\
	\label{WDW momentum constraint in AB parameterization}
	B\bm{'} \, \frac{\delta \Psi}{\delta B} + \phi\bm{'} \, \frac{\delta \Psi}{\delta \phi}
	-  A \left[ \frac{\delta \Psi}{\delta A} \right]^{\bm{\prime}} 
	&= 0
	\, .
\end{align}
In the following sections we will focus on solving these Wheeler-DeWitt equation.

\subsection{Momentum Constraint}
\label{Momentum Constraint}

Let us first focus on the momentum constraint 
\begin{align}
	\label{quantum momentum constraint: definition}
	B\bm{'} \frac{\delta Z}{\delta B} + \phi\bm{'} \frac{\delta Z}{\delta \phi}
	- A \left[ \frac{\delta Z}{\delta A} \right]^{\bm{\prime}} 
	= 0
	\, ,
\end{align}
as a functional equation for some functional $Z[A,\, B,\, \phi]$. 

The obvious solution to the momentum constraint (\ref{quantum momentum constraint: definition}) is a constant functional $Z_{const.}$ which is, in general, a function of some physical or other constants and numerical parameters of the spacetime, independent of any dynamical fields, so
\begin{align}
	\label{quantum momentum constraint trivial solution Z_const}
	Z_{const.} 
	= Z_{const.}\left(\chi_0, C^{-1}, G, c, \hbar, \pi, \ldots \right)
	\, .
\end{align}
Another obvious solution to (\ref{quantum momentum constraint: definition}) is a functional $Z_{0}[A, B, \phi]$ satisfying 
\begin{align}
	\label{quantum momentum constraint solution 0}
	\frac{\delta Z_{0}}{\delta A} = 0
	\, , \quad 
	\frac{\delta Z_{0}}{\delta B} = 0
	\, , \quad
	\frac{\delta Z_{0}}{\delta \phi} = 0
	\, .
\end{align}
Let us then focus first on the simple functional differential equation 
\begin{align}
	\label{quantum momentum constraint: deltaZ_0/deltaX=0}
	\frac{\delta Z_{0}}{\delta X} = 0
\end{align}
for some functional $Z_{0}[X]$ parametrized with a general spacetime field $X(t, r)$, where on any (space-like) foliation hypersurface $\Sigma(t)$ for given $t = const.$ is $X = X(r)$.  
The functional $Z_{0}$ is assumed to be in the most general form
\begin{align}
	\label{quantum momentum constraint: Z_0[X] ansatz}
	Z_{0}[X] = \int\limits_{r_{min}}^{r_{max}} f\left( X , X\bm{'} \right) dr
	\, ,
\end{align}
with some yet undetermined function $f$. From the perspective of the task  (\ref{quantum momentum constraint: deltaZ_0/deltaX=0}) itself the constant limits $r_{min}$ and $r_{max}$ of the integral can be arbitrary. In our case of the momentum constraint, however, specific values those limits (position of the endpoints) depend on an interpretation of the radial coordinate $r$ or more specifically on the coordinate origin from which the radial distance is measured. So , for example, the lower limit (endpoint) $r_{min}$ can be $0$ or $\chi_0$ and the upper limit $r_{max}$ is almost always set to $\infty$. By varying $Z_{0}$ one gets
\begin{align}
	\label{quantum momentum constraint: Z_0[X] variation}
	\delta Z_{0}[X]
	= \int\limits_{r_{min}}^{r_{max}}
	\Bigg( \frac{\partial f}{\partial X}
	- \left[ \frac{\partial f}{\partial X\bm{'}} \right]^{\bm{\prime}} \Bigg) 
	\delta X dr 
	+ \left[ \frac{\partial f}{\partial X\bm{'}} \delta X \right]_{r_{min}}^{r_{max}}
	\, .
\end{align}
The problem we are trying to solve is a Cauchy problem, and so the value of the field $X$ is prescribed in the endpoints $r_{min}$ and $r_{max}$. The field $X$ is thus fixed at these endpoints and its variation there is zero: $\delta X |_{r_{min}} = 0$ and $\delta X|_{r_{max}} = 0$. The last term on the right side of  \ref{quantum momentum constraint: Z_0[X] variation}) is therefore zero. The Functional derivative of $Z_{0}[X]$ with respect to $X$ is then
\begin{align}
	\label{quantum momentum constraint: Z_0[X] functional derivative}
	\frac{\delta Z_{0}}{\delta X}
	= \frac{\partial f}{\partial X}
	- \left[ \frac{\partial f}{\partial X\bm{'}} \right]^{\bm{\prime}}
	\, .
\end{align} 
For the functionals of the form (\ref{quantum momentum constraint: Z_0[X] ansatz}) then the original functional differential equation (\ref{quantum momentum constraint: deltaZ_0/deltaX=0}) converts to the partial differential equation
\begin{align}
	\label{quantum momentum constraint: Z_0[X] functional derivative equation}
	\frac{\partial f}{\partial X} - \left[ \frac{\partial f}{\partial X\bm{'}} \right]^{\bm{\prime}} = 0
	\, .
\end{align}
This partial differential equation is solved by a function of the form $f = X\bm{'} \, \widetilde{f}(X)$, where $\widetilde{f}(X)$ is an arbitrary differentiable function of the field $X$. The field $X\bm{'}$ appears in this solution only linearly. If we attempt to generalize the solution to the multiplicative form $\widetilde{f}(X) v(X\bm{'})$ with some function $v(X\bm{'})$ of the field $X\bm{'}$, then by substituting the multiplicative term into (\ref{quantum momentum constraint: Z_0[X] functional derivative equation}) we get a differential equation for $v(X\bm{'})$ which, after some manipulations, simplifies to
\begin{align*}
	\frac{d \ln\widetilde{f}}{d X} \left( v - \frac{d v}{d X\bm{'}} X\bm{'} \right) 
	= \left[ \frac{d v}{d X\bm{'}} \right]^{\bm{\prime}}
\end{align*}
The right side of this equation does not depend on $X$ at all, while the left side can generally depend on $X$. Since the functions $f$ and $v$ are required to depend on hypersurface coordinates (here on $r$) only indirectly through dynamical fields (here $X$), the only way to satisfy such equation is that its right side has to be zero and thus also the expression in the brackets on its left side has to vanish, i.e.
\begin{align*}
	v = \frac{d v}{d X\bm{'}}  X\bm{'} 
	\quad\quad \& \quad\quad
	\left[ \frac{d v}{d X\bm{'}} \right]^{\bm{\prime}}
	= 0
	\, .
\end{align*}
The only non-trivial solution to those two ordinary differential equations is (up to multiplicative constant which can be included into $\widetilde{f}$) of the form: $v(X\bm{'}) = X\bm{'}$, that is, the function $v$ has to an identity. The single solution $f = X\bm{'} \widetilde{f}(X)$ cannot be further (non-trivially) generalized in the argument $X\bm{'}$. The equation (\ref{quantum momentum constraint: Z_0[X] functional derivative equation}) is linear in the derivatives and is therefore solved by an arbitrary linear combination of the terms $X\bm{'} \widetilde{f}(X)$. The solution can be then generalized to a series $f = X\bm{'} \sum_{n\in\mathbb{Z}}\widetilde{f}_n(X) k_n$ with some numerical coefficients $k_n$ that are constant with respect to $X$ and $X\bm{'}$. The general solution (of a separate type, i.e. which is in the form of a product of a function of $X$ and a function of $X\bm{'}$) to the partial differential equation (\ref{quantum momentum constraint: Z_0[X] functional derivative equation}) thus is
\begin{align}
	\label{quantum momentum constraint: general solution f(X,X')}
	f\left( X, X\bm{'} \right) = X\bm{'} \widetilde{f}(X)
	\, ,
\end{align}
where $\widetilde{f}(X)$ is an arbitrary differentiable function of $X$.
The functional $Z_{0}[A, B, \phi]$ that solves the momentum constraint  (\ref{quantum momentum constraint: definition}) through the expressions (\ref{quantum momentum constraint solution 0}) is
\begin{align}
	\label{quantum momentum constraint: Z_0[A,B,phi]}
	Z_0[A, B, \phi]
	= \int\limits_{r_{min}}^{\infty}
	f\left(A, B, \phi ; A\bm{'}, B\bm{'}, \phi\bm{'} \right) dr
	\, ,
\end{align}
with
\begin{align}
	\label{quantum momentum constraint: f=f_A+f_B+f_phi}
	f\left(A, B, \phi ; A\bm{'}, B\bm{'}, \phi\bm{'} \right)
	= f_{A}\left(A, A\bm{'}\right) 
	+ f_{B}\left(B, B\bm{'}\right) 
	+ f_{\phi}\left(\phi, \phi\bm{'} \right)
	\, ,
\end{align}
where the functions $f_{A}$, $f_{B}$ and $f_{\phi}$ are, for the fields $X = A, B, \phi$ given by the expression (\ref{quantum momentum constraint: general solution f(X,X')}) as $f_{X}\left( X, X\bm{'} \right) = f\left( X, X\bm{'} \right)$. The solution in the form of the functional $Z_0$ then contains three arbitrary differentiable functions $\widetilde{f}_{A}(A)$, $\widetilde{f}_{B}(B)$ and $\widetilde{f}_{\phi}(\phi)$.

Next we will examine the part of the momentum constraint that is symmetric with respect to $B$ and $\phi$, that is, the part 
\begin{align}
	\label{quantum momentum constraint: B,phi part of the constraint}
	B\bm{'} \frac{\delta Z}{\delta B} + \phi\bm{'} \frac{\delta Z}{\delta \phi} =  0
	\, .
\end{align}
and we are looking for such functional $Z = Z_{g}[B, \phi]$ that would satisfy $\frac{\delta Z_{g}}{\delta B} \neq 0$ and $\frac{\delta Z_{g}}{\delta \phi} \neq 0$ but still solve the equation (\ref{quantum momentum constraint: B,phi part of the constraint}). The initial form of $Z_{g}$ is assumed to be
\begin{align}
	\label{quantum momentum constraint: Z_g[B,phi]}
	Z_{g}[B, \phi] = \int\limits_{r_{min}}^{\infty} g\left(B, \phi ; B\bm{'}, \phi\bm{'} \right) dr
	\, ,
\end{align}
with some function $g$, so that
\begin{align*}
	\frac{\delta Z_{g}}{\delta B} 
	\propto \phi\bm{'} \widetilde{g}(B, \phi)
	\, , \quad 
	\frac{\delta Z_{g}}{\delta \phi} 
	\propto - B\bm{'} \widetilde{g}(B, \phi)
	\, .
\end{align*}
All these conditions are satisfied by the function   
\begin{align}
	\label{hybnostní vazba: funkce g = g_B + g_phi}
	g\left(B, \phi ; B\bm{'}, \phi\bm{'} \right)
	= B\bm{'} g_{B}(B, \phi) + \phi\bm{'} g_{\phi}(B, \phi)
	\, ,
\end{align}
where $g_{B}(B, \phi)$ and $g_{\phi}(B, \phi)$ are arbitrary differentiable functions of both $B$ and $\phi$.
Denoting 
\begin{align}
	\label{quantum momentum constraint: g^tilde = g^tilde_B - g^tilde_phi}
	\widetilde{g} \equiv \frac{\partial g_{B}}{\partial \phi} - \frac{\partial g_{\phi}}{\partial B}
	\, ,
\end{align}
it holds that 
\begin{align}
	\label{quantum momentum constraint: functional derivatives of Z_g[B,phi]}
	\frac{\delta Z_{g}}{\delta B} = - \phi\bm{'} \widetilde{g}(B, \phi)
	\, , \quad 
	\frac{\delta Z_{g}}{\delta \phi} = B\bm{'} \widetilde{g}(B, \phi)
	\, .
\end{align}
The functional $Z_{g}$ thus solves the functional differential equation  (\ref{quantum momentum constraint: B,phi part of the constraint}) and noticing that $Z_{g}[B, \phi]$ does not contain the field $A$, the functional $Z_{g}$ hence also solves the entire momentum constraint (\ref{quantum momentum constraint: definition}).

In the vacuum case, when is $\phi = 0$ and $P_{\!\phi} = 0$ the momentum constraint has (in qp-ordering), after some adjustments, the alernative form
\begin{align}
	\label{quantum momentum constraint: vacuum constraint}
	\frac{B\bm{'}}{A} \frac{\delta Z}{\delta B} - \left[ \frac{\delta Z}{\delta A} \right]^{\bm{\prime}} = 0
	\, .
\end{align}
We again want to find a functional $Z_{q}[A, B]$ for which is $\frac{\delta Z_{q}}{\delta A} \neq 0$ and $\frac{\delta Z_{q}}{\delta B} \neq 0$ but it solves the vacuum momentum constraint (\ref{quantum momentum constraint: vacuum constraint}). We will again search for $Z_{q}$ by assuming
\begin{align}
	\label{quantum momentum constraint: Z_h[A,B] - ansatz with a(A)}
	Z_{q}[A, B]
	= \int\limits_{r_{min}}^{\infty}
	a(A) \cdot q_{B}\left(B, \frac{B\bm{'}}{A} \right) dr
	\, .
\end{align}
The functional derivatives of $Z_{q}[A, B]$ with respect to $A$ and $B$ are
\begin{align}
	\label{quantum momentum constraint: functional derivatives of Z[A,B]}
	\frac{\delta Z_{q}}{\delta A}
	= q_{B} \frac{d a}{d A} - \frac{a}{A} \frac{B\bm{'}}{A} \frac{\partial q_{B}}{\partial \frac{B\bm{'}}{A}}  
	\, , \quad
	\frac{\delta Z_{q}}{\delta B}
	= a \frac{\partial q_{B}}{\partial B} 
	- \left[\frac{a}{A} \frac{\partial q_{B}}{\partial \frac{B\bm{'}}{A}} \right]^{\bm{\prime}} 
	\, .
\end{align} 
Substituting these into the vacuum momentum constraint (\ref{quantum momentum constraint: vacuum constraint}) produces a differential equation containing terms proportional to the derivatives $A\bm{'}$ and $B\bm{''}$ which, however, are not present in any of the functions $a$ or $q_{B}$. The term proportional to $A\bm{'}$ and $B\bm{''}$ therefore have to be zero. These vanishing terms are in fact three differential equations that can be further reduced to two equations
\begin{align*}
	\frac{d a}{d A} = \frac{a}{A} 
	\quad \& \quad
	\frac{d^2 a}{dA^2} =  0
	\, .
\end{align*}
Up to a multiplicative constant, the only solution to this system of equations is the function $a(A) = A$. But we also see that the functional $Z_{q}$, defined in (\ref{quantum momentum constraint: Z_h[A,B]}) with $a=A$ satisfies the vacuum momentum constraint independently of the form of the function $q_{B}\left(B, \frac{B\bm{'}}{A} \right)$.  
The solution to the vacuum momentum constraint (\ref{quantum momentum constraint: vacuum constraint}) is therefore given by the functional 
\begin{align}
	\label{quantum momentum constraint: Z_h[A,B]}
	Z_{q}[A, B]
	= \int\limits_{r_{min}}^{\infty}
	A \cdot q_{B}\left(B, \frac{B\bm{'}}{A} \right) dr
\end{align}
with an arbitrary differentiable function $q_{B}\left(B, \frac{B\bm{'}}{A} \right)$. 

Due to the fact that $B$ and $\phi$ appear in the full momentum constraint (\ref{quantum momentum constraint: definition}) symmetrically, a solution in the form of the functional $Z_{q}$ can be expanded for the scalar field $\phi$ in a very straightforward way. The general non-trivial solution to the full momentum constraint (\ref{quantum momentum constraint: definition}) is then 
\begin{align}
	\label{hybnostní vazba: funkcionál Z_h[A,B,phi]}
	Z_{q}[A, B, \phi]
	= \int\limits_{r_{min}}^{\infty}
	A \cdot h\left(B, \phi ; \frac{B\bm{'}}{A}, \frac{\phi\bm{'}}{A} \right) dr
	\, ,
\end{align}
with an arbitrary differentiable function $h\left(B, \phi ; \frac{B\bm{'}}{A}, \frac{\phi\bm{'}}{A} \right)$.  

Assembling all the solutions together, the general solution to the full version (\ref{quantum momentum constraint: definition}) of the momentum constraint is
\begin{align}
	\label{quantum momentum constraint: Z = Z_0 + Z_g + Z_h + Z_konst}
	Z[A, B, \phi]
	= Z_{0}[A, B, \phi] 
	+ Z_{g}[B, \phi] 
	+ Z_{q}[A, B,\, \phi] 
	+ Z_{const.}\left(\chi_0, C^{-1}\right) 
	\, ,
\end{align}
where the individual functionals $Z_{0}$, $Z_{g}$ and $Z_{q}$ are defined above in (\ref{quantum momentum constraint: Z_0[A,B,phi]}), (\ref{quantum momentum constraint: Z_g[B,phi]}) and (\ref{hybnostní vazba: funkcionál Z_h[A,B,phi]}).
Functional derivatives of $Z$ are explicitly 
\begin{align}
	\label{quantum momentum constraint: functional derivative of Z with respect to A}
	\frac{\delta Z}{\delta A}
	&= h - \frac{B\bm{'}}{A} \frac{\partial h}{\partial \frac{B\bm{'}}{A}} 
	- \frac{\phi\bm{'}}{A} \frac{\partial h}{\partial \frac{\phi\bm{'}}{A}}
	\, ,
	\\
	\label{quantum momentum constraint: functional derivative of Z with respect to B}
	\frac{\delta Z}{\delta B}
	&= - \phi\bm{'} \widetilde{g}  
	+ A \frac{\partial h}{\partial B} 
	- \left[ \frac{\partial h}{\partial \frac{B\bm{'}}{A}} \right]^{\bm{\prime}}
	\, ,
	\\
	\label{quantum momentum constraint: functional derivative of Z with respect to phi}
	\frac{\delta Z}{\delta \phi}
	&= B\bm{'} \widetilde{g}  
	+ A \frac{\partial h}{\partial \phi} 
	- \left[ \frac{\partial h}{\partial \frac{\phi\bm{'}}{A}} \right]^{\bm{\prime}}
	\, .
\end{align}

The wave functional $\Psi$ can be now expressed \cite{kenmoku-kubotani-takasugi-yamazaki} as a wave function of $Z$:
\begin{align}
	\label{quantum momentum constraint: wave functional as function of Z}
	\Psi = \psi\left(Z\right) 
	\, .
\end{align}
All functional derivatives of $\Psi$ with respect to the canonical fields then translate into ordinary derivatives of $\psi$ with respect to $Z$:
\begin{align}
	\label{quantum momentum constraint: functional derivatives of the wave functional}
	\frac{\delta \Psi}{\delta X}
	= \frac{\delta Z}{\delta X} \frac{d \psi}{d Z}
	\, , \quad 
	\frac{\delta^2 \Psi}{\delta X \delta Y}
	= \frac{\delta^2 Z}{\delta X \delta Y} \frac{d \psi}{d Z} 
	+ \frac{\delta Z}{\delta X} \frac{\delta Z}{\delta Y} \frac{d^2 \psi}{d Z^2}
	\, ,
\end{align}
with $X, Y \in \{A, B, \phi\}$. From the definition of $Z$ as a definite integral over $r$, it holds
\begin{align}
	\label{quantum momentum constraint: Z'=0}
	Z\bm{'} = 0 
	\, .
\end{align}
An important aspect of the prescription (\ref{quantum momentum constraint: wave functional as function of Z}) is the fact that, due to the relations (\ref{quantum momentum constraint: functional derivatives of the wave functional}) and (\ref{quantum momentum constraint: Z'=0}) if the functional $Z$ satisfies the momentum constraint, the constraint will then be also solved by any differentiable function of $Z$, so by any wave function $\psi(Z)$. The wave functional $\Psi = \psi(Z)$ therefore satisfies the momentum constraint completely independently of any specific dependence of the wave function $\psi$ on $Z$:
\begin{align*}
	B\bm{'} \frac{\delta \Psi}{\delta B} 
	+ \phi\bm{'} \frac{\delta \Psi}{\delta \phi} 
	- A \left[ \frac{\delta \Psi}{\delta A} \right]^{\bm{\prime}} 
	= \left( B\bm{'} \frac{\delta Z}{\delta B} 
	+ \phi\bm{'} \frac{\delta Z}{\delta \phi} 
	- A \left[ \frac{\delta Z}{\delta A} \right]^{\bm{\prime}} \right) \frac{d \psi}{d Z}
	= 0
	\, .
\end{align*}
The final form of the solution to the momentum constraint is thus (\ref{quantum momentum constraint: wave functional as function of Z}).

\subsection{Hamiltonian Constraint}
\label{Hamiltonian Constraint}

The Hamiltonian constraint (\ref{WDW Hamiltonian constraint in AB parameterization}) can be rewritten using the expression $\Psi = \psi(Z)$ in the form of a partial (or ordinary) differential equation for the wave function $\psi$. If we also rescale the scalar field $\phi$ to absorb the Einstein gravitation constant $\kappa$ as $\closedsqrt{\kappa} \phi \rightarrow \phi$, we get
\begin{align}
	\label{Hamiltonian constraint: Hamiltonian constraint via wave function psi(Z)}
    &\frac{\kappa \hbar^2}{8\pi} \left[ 
    \frac{A}{2 B^2} \left(\frac{\delta Z}{\delta A}\right)^{\! 2} 
	- \frac{1}{B} \frac{\delta Z}{\delta B}  \frac{\delta Z}{\delta A}
	+ \frac{1}{A B^2} \left(\frac{\delta Z}{\delta \phi}\right)^{\! 2} 
	\right] \frac{d^2 \psi}{dZ^2}
	\, +
	\nonumber
	\\
	&+ \frac{\kappa \hbar^2}{8\pi} \left[
    \frac{A}{2 B^2} \frac{\delta^2 Z}{\delta A^2} 
	- \frac{1}{B} \frac{\delta^2 Z}{\delta B \delta A}
	+ \frac{1}{A B^2} \frac{\delta^2 Z}{\delta \phi^2}  
	\right] \frac{d \psi}{d Z}
	\, +
	\nonumber
	\\
	&+ \frac{2\pi}{\kappa} \left[ 
    \frac{B^2}{A} \phi\bm{'}^{\,2}
    - A B^2 {^{(\parallel)}\!}R \right] \psi 
	= 0
	\, .
\end{align}
The resulting differential equation is still complicated enough in order to search for the solution directly. It is also not entirely clear in which operator ordering it should be formulated. Moreover, we haven't yet found any particular ordering of its operators that would simplify the constraint and help solving it. 

The Wheele-DeWitt equation (\ref{Hamiltonian constraint: Hamiltonian constraint via wave function psi(Z)}) holds for general dynamical spherically symmetric spacetimes. If we focus first on quantization of static spacetimes as is, for example, sJNW spacetime, the Wheele-DeWitt equation above will simplify considerably. Thus in the next section we will explore some specifics and properties of static spacetimes and we will finally  proceed to the quantization of the sJNW spacetime.

\subsection{Static Spherically Symmetric Spacetimes}
\label{Static Spherically Symmetric Spacetimes}

Static spacetimes are time-independent spacetimes that do not permit stationary rotations. Locally, stationary spacetimes have the structure $\mathbb{R}\times\Sigma$ where $\Sigma$ are space-like hypersurfaces. We consider a globally hyperbolic spacetime $\mathcal{M} = \bigcup_{t \in \mathbb{R}} \Sigma(t)$ foliated into Cauchy hypersurfaces $\Sigma(t)$ that are parametrized by global foliation time $t$, and describe it with an adapted coordinate system $y=(t, y^{a})$, $y^{a} = \bm{y} \in \Sigma$. As a source, we consider a massless static scalar field $\phi(\bm{y})$ described by (\ref{scalar field lagrangian}) where we set $\varepsilon = 1$ and $V = 0$. A metric of such spacetimes has the form 
\begin{align}
	\label{static spherical symmetry: general metric}
	ds^2 = - N^2(\bm{y}) dt^2 + q_{ab}(\bm{y}) dy^{a} dy^{b}
	\, .
\end{align}
The shift vector for static spacetimes expressed in adapted coordinates vanishes: $N^{a} = 0$. The extrinsic curvature $K_{ab}$ of the space-like hypersurfaces $\Sigma$ is thus zero: 
\begin{align*}
	K_{ab}(y) 
	= \frac{1}{2 N} \Big( \underbrace{\partial_{t}  q_{ab}}_{0} 
	- \partial_{a} \underbrace{N_{b}}_{0}  -  \partial_{b} \underbrace{N_{a}}_{0} 
	+ 2 \Gamma_{c b a} \underbrace{N^{c}}_{0} \Big)
	= 0
	\, .
\end{align*}
The scalar curvature is then equal (up to the boundary terms) to the spatial scalar curvature: $R \overset{\text{int}}{=} {^{(\parallel)}\!}R$. As a result of $K_{ab} = 0$, the momenta $p^{ab}$ associated with the spatial metric $q_{ab}$ vanish. The momentum $p_{\phi}$ of the source scalar field $\phi(\bm{y})$ is also zero. All the momenta constitute a set of primary constraints 
\begin{align}
	\label{static spherical symmetry: momenta = 0}
	p^{ab} = 0
	\, , \quad
	p_{\phi} = 0
\end{align}
on the phase space. In the absence of a shift vector, the momentum constraint $\mathcal{C}_{a}$ has to be added to the system via some arbitrary multipliers $\lambda^{a}$. The relations (\ref{static spherical symmetry: momenta = 0}) then ensure that $\mathcal{C}_{a} \approx 0$ because the hypersurface given by the constraints (\ref{static spherical symmetry: momenta = 0}) is a subspace of the hypersurface given by the momentum constraint. The momentum constraint is then satisfied identically and it is not necessary to add it to the system via some other additional multipliers $\lambda^{a}\mathcal{C}_{a}$. The Hamiltonian constraint reduces to
\begin{align}
	\label{static spherical symmetry: static Hamiltonian constraint}
	\mathcal{C}^{(G,\,\phi)}_{stat.}
	\equiv  \mathcal{C}^{(G, \phi)}_{\perp} \Big| {}_{\substack{p^{ab}=0 \\ p_{\phi}=0}}
	= \frac{\closedsqrt{q}}{2\kappa} \left(\kappa \, q^{ab} \partial_{a} \phi \, \partial_{b} \phi 
	- {^{(\parallel)}\!}R \right)
    \approx  0
	\, .
\end{align}
This Hamiltonian constraint reproduces the contracted Einstein field equations (\ref{simplified Einstein equations for scalar field source}). The ADM Hamiltonian is
\begin{align}
	\label{static spherical symmetry: Hamiltonian}
	H^{(G, \phi)}_{stat.} 
	=  \int_{\Sigma} \Big(N \mathcal{C}^{(G, \phi)}_{stat.} + \lambda_{ab} p^{ap} 
	+ \lambda^{\phi}  p_{\phi} \Big) d^3 y 
	\, . 
\end{align}
The constraint algebra is generally no longer conserved. It is thus necessary to satisfy consistency conditions of the constraints (\ref{static spherical symmetry: momenta = 0}). These consistency conditions demand $\bdot{p}^{\,ab} \approx 0$ and $\bdot{p}_{\phi} \approx 0$, which is explicitly:
\begin{align}
	\label{static spherical symmetry: consistency condition for metric momenta}
	\bdot{p}^{\,ab}
	&= - \frac{N \closedsqrt{q}}{2\kappa} \left( {^{(\parallel)}\!}R^{ab} - \frac{1}{2} q^{ab} {^{(\parallel)}\!}R \right) 
	\, + 
	\nonumber
	\\
	& \quad\,\, + \frac{N \closedsqrt{q}}{2} \left( q^{ai} q^{bj} - \frac{1}{2} q^{ab} q^{ij} \right)
	\partial_{i} \phi \partial_{j} \phi 
	\, +
	\nonumber
	\\
	& \quad\,\, + \frac{\closedsqrt{q}}{2\kappa} \Big( q^{ai} q^{bj} - q^{ab} q^{ij} \Big)
	\left( {^{(\parallel)}\!}\nabla_{i} {^{(\parallel)}\!}\nabla_{j} N \right)  
	\overset{!}{\approx} 0
	\, , 
	\\
	\label{static spherical symmetry: consistency condition for scalar field momentum}
	\bdot{p}_{\phi} 
	&= - N \closedsqrt{q} q^{ab} \, {^{(\parallel)}\!}\nabla_{a} {^{(\parallel)}\!}\nabla_{b} \phi  
	- \closedsqrt{q} q^{ab} \partial_{a} \phi \, \partial_{b} N 
	\overset{!}{\approx} 0
	\, .
\end{align}
First, we examine the second of these conditions; the consistency condition for the scalar-field momentum. By expanding the conservation laws (\ref{scalar field conservation equation}) for the scalar field $\phi$ we get: 
\begin{align*}
	g^{\alpha \beta}  \nabla_{\alpha} \nabla_{\beta} \phi 
	&= q^{ab} \nabla_{a} \nabla_{b} \phi 
	+ g^{tt} \nabla_{t} \nabla_{t} \phi 
	= q^{ab} \, {^{(\parallel)}\!}\nabla_{a} {^{(\parallel)}\!}\nabla_{b} \phi 
	- \frac{1}{N^2}  \Gamma\indices{^{a}_{tt}} \partial_{a} \phi
	= \nonumber
	\\
	&= q^{ab} \, {^{(\parallel)}\!}\nabla_{a} {^{(\parallel)}\!}\nabla_{b} \phi 
	+ \frac{1}{N} q^{ab} \partial_{b} N \, \partial_{a} \phi 
	\, .
\end{align*}
Substituting this expression into the second condition (\ref{static spherical symmetry: consistency condition for scalar field momentum}) gives us the conservations law for the scalar field, see (\ref{scalar field conservation equation}):
\begin{align}
	\label{static spherical symmetry: scalar field wave equation}
	\bdot{p}_{\phi} 
    = - N \closedsqrt{q} g^{\alpha \beta} \nabla_{\alpha} \nabla_{\beta} \phi 
    = 0
\end{align}
Second, the first condition for consistency of the spatial-metric momentum constraint (\ref{static spherical symmetry: scalar field wave equation}) can be further simplified by substituting the Hamiltonian constraint (\ref{static spherical symmetry: static Hamiltonian constraint}) producing 
\begin{align}
	\label{static spherical symmetry: simplified consistency condition for metric momenta}
	\bdot{p}^{\,ab}
	&= \frac{N \closedsqrt{q}}{2}  q^{ai}  q^{bj}  \partial_{i} \phi  \partial_{j} \phi 
	-  \frac{N \closedsqrt{q}}{2\kappa}  {^{(\parallel)}\!}R^{ab}  
	\, + \nonumber
	\\
	& \quad\, + \frac{\closedsqrt{q}}{2\kappa} \left( q^{ai} q^{bj} - q^{ab} q^{ij} \right)
	{^{(\parallel)}\!}\nabla_{i} {^{(\parallel)}\!}\nabla_{j} N   
	\overset{!}{\approx}  0
	\, .
\end{align} 
This condition is, in fact, a condition for any lapse function $N$ in case of static spacetimes. Satisfying (\ref{static spherical symmetry: simplified consistency condition for metric momenta}) then ensures that the used foliation (e.g. the choice of a lapse $N$) is consistent with the staticity, which should be viewed, in this context, as demanding additional constraint for all the momenta (\ref{static spherical symmetry: momenta = 0}). Assuming the consistency condition (\ref{static spherical symmetry: simplified consistency condition for metric momenta}) is satisfied and using the static Hamiltonian constraint, we get for the lapse $N$ the relation 
\begin{align}
	\label{static spherical symmetry: contracted consistency conditions for metric momenta}
	q_{ab} \bdot{p}^{\,ab}
	=  - \kappa \, q^{ij} {^{(\parallel)}\!}\nabla_{i} {^{(\parallel)}\!}\nabla_{j} N   
	=  0
	\, .
\end{align} 
However, this contracted condition, together with the static Hamiltonian constraint does not generally compensate for the six consistency conditions (\ref{static spherical symmetry: simplified consistency condition for metric momenta}). On the classical level, we can further simplify the relations (\ref{static spherical symmetry: simplified consistency condition for metric momenta}) by using the Einstein field equations (\ref{simplified Einstein equations for scalar field source}), which should be equivalent to the Hamiltonian formalism. Rewriting the Einstein field equations into coordinates adapted to the foliation hypersurfaces\footnote{
    In case of $K_{\mu\nu} = 0$, the 3+1 decomposition of Ricci tensor is $R_{\mu\nu} = {^{(\parallel)}\!}R_{\mu\nu} - {^{(\parallel)}\!}\nabla_{\nu} a_{\mu} - a_{\mu} a_{\nu}$.}
and utilizing relations $K_{ab} = 0$, $\partial_{t} N = 0$ and $\partial_{t} \phi = 0$ valid for the static spacetimes yields, in static case, the equation 
\begin{align}
	\label{static spherical symmetry: adapted Einstein equations} 
	\kappa \, \delta_{a}^{\mu} \delta_{b}^{\nu} \, q^{ai} q^{bj} \partial_{i} \phi \, \partial_{j} \phi 
	&=  \delta_{t}^{\mu} \delta_{t}^{\nu} N^{-2} \, q^{ij} 
	\left( {^{(\parallel)}\!}\nabla_{j} a_{i} + a_{j} a_{i} \right) 
	\, + \nonumber
	\\
	&\quad\,\, + \delta_{a}^{\mu} \delta_{b}^{\nu} \Big[ {^{(\parallel)}\!}R^{ab} 
	- q^{ai} q^{bj} \left( {^{(\parallel)}\!}\nabla_{j} a_{i} + a_{j} a_{i} \right) \Big]
	\, ,
\end{align}
where we have taken advantage of the fact that in the static case $N^i = 0$ is $n^{\mu} = N^{-1} \delta_{t}^{\mu}$. Acceleration $a_{\mu}$ is defined as $a_{\mu} = n^{\nu}  \nabla_{\nu} n_{\mu}$ with $n_{\mu} = - N \delta^{t}_{\mu}$, explicitly: 
\begin{align*}
	a_{\mu}  
	=  n^{\nu}  \nabla_{\nu} n_{\mu}  
	=  - \frac{1}{N}  \nabla_{t} \left(N  \delta_{\mu}^{t} \right) 
	=  \Gamma\indices{^{t}_{t\mu}}
	=  \delta_{\mu}^{a}  \frac{\partial_{a} N}{N}
	\, ,
\end{align*}
from where it can be easily seen that $a_{i} = N^{-1} \partial_{i} N$. The spatial covariant derivative of the acceleration hence is\footnote{
    Because lapse is a scalar function: $\partial_{i} N = {^{(\parallel)}\!}\nabla_{i} N$.} 
\begin{align*}
	{^{(\parallel)}\!}\nabla_{j} a_{i}  
	=  {^{(\parallel)}\!}\nabla_{j} \left( \frac{\partial_{i} N}{N} \right)   
	=  \frac{1}{N}  {^{(\parallel)}\!}\nabla_{j} {^{(\parallel)}\!}\nabla_{i} N  
	-  \frac{1}{N^2} \partial_{j} N \partial_{i} N
	\, .
\end{align*}
Because $a_{j} a_{i} = N^{-2} \partial_{i} N \partial_{j} N$ it holds  
\begin{align*}
	{^{(\parallel)}\!}\nabla_{j} a_{i} + a_{j} a_{i} 
	= \frac{1}{N} {^{(\parallel)}\!}\nabla_{j} {^{(\parallel)}\!}\nabla_{i} N
	\, .
\end{align*}
From the adapted static Einstein field equations (\ref{static spherical symmetry: adapted Einstein equations}) we have  
\begin{itemize}
	\item
	for $\mu = \nu = t$:
	\begin{align}
		\label{static spherical symmetry: tt component of adapted Einstein equations}
		\frac{1}{N^2} q^{ij} \left( {^{(\parallel)}\!}\nabla_{j} a_{i} + a_{j} a_{i} \right) 
		= \frac{1}{N^3} q^{ij} {^{(\parallel)}\!}\nabla_{j} {^{(\parallel)}\!}\nabla_{i} N
		= 0
		\, ,
	\end{align}
	
	\item
	and for $\mu = a$, $\nu = b$:
	\begin{align}
		\label{static spherical symmetry: ab component of adapted Einstein equations}
		\kappa \,  q^{ai}  q^{bj} \partial_{i} \phi \, \partial_{j} \phi 
		&=  {^{(\parallel)}\!}R^{ab} 
		-  q^{ai} q^{bj} \left( {^{(\parallel)}\!}\nabla_{j} a_{i} + a_{j} a_{i} \right)
		= 
		\nonumber
		\\[5pt] 
		&=  {^{(\parallel)}\!}R^{ab} 
		- \frac{1}{N} q^{ai} q^{bj} {^{(\parallel)}\!}\nabla_{j} {^{(\parallel)}\!}\nabla_{i} N 
		\, .
	\end{align}
\end{itemize}
Substituting the spatial component (\ref{static spherical symmetry: ab component of adapted Einstein equations}) of the adapted static Einstein field equations into the condition (\ref{static spherical symmetry: simplified consistency condition for metric momenta}) for consistency of spatial-metric momentum constraint gives us an equivalent to the contracted consistency conditions  (\ref{static spherical symmetry: contracted consistency conditions for metric momenta}). On the space of all static metrics that satisfy the Einstein field equations with a scalar field, the original consistency conditions (\ref{static spherical symmetry: consistency condition for metric momenta}) for the primary constraints (\ref{static spherical symmetry: momenta = 0}) are reduced to one single differential equation for the lapse function $N$: 
\begin{align}
	\label{static spherical symmetry: lapse equation}
	q^{ij}  {^{(\parallel)}\!}\nabla_{i} {^{(\parallel)}\!}\nabla_{j} N  
	=  q^{ij} \partial_{i} \partial_{j} N - q^{ij} \Gamma\indices{^{k}_{ij}} \partial_{k} N  
	=  0
	\, .
\end{align}
The space of all the metrics satisfying Einstein field equations (\ref{static spherical symmetry: adapted Einstein equations}) does certainly not  constitute the whole representation space $\mathscr{F}$. If we include the functions that form the consistency conditions (\ref{static spherical symmetry: consistency condition for metric momenta}) for the primary constraints of staticity into the Hamiltonian as a secondary constraints, the functions would, on the quantum level in metric representation, appear among the Wheeler-DeWitt equations. Considering that the consistency conditions for the primary constraints of staticity do not contain any momenta, the equations obtained from their quantization in metric representation would be investigated on a classical level where, as we know, these consistency conditions are satisfied. The space of all the metrics satisfying Einstein field equations would be, in that case the space $\mathscr{F}_{WDW}$.

On the quantum level we have the following constraints for $\Psi = \Psi[q_{ij}, \phi]$; the Hamiltonian constraint  
\begin{align}
	\label{static spherical symmetry: static Hamiltonian constraint}
	\widehat{\mathcal{C}}^{\, (G, \phi)}_{stat.}  \Psi
	= \frac{\closedsqrt{q}}{2\kappa} \left( \kappa \, q^{ab} \partial_{a} \phi \, \partial_{b} \phi 
	- {^{(\parallel)}\!}R \right) \Psi
	= 0
	\, ,
\end{align}
and also the quantum versions of the constraints (\ref{static spherical symmetry: momenta = 0}):
\begin{align}
		\label{static spherical symmetry: constraint for zero momenta}
		\widehat{p}^{\, ab} \Psi = - i \hbar \frac{\delta \Psi}{\delta q_{ab}} = 0
		\, , \quad
		\widehat{p}_{\phi} \Psi = - i \hbar \frac{\delta \Psi}{\delta \phi} = 0
		\, .
\end{align}
The only relations that restrict the dependence of the wave functional $\Psi$ on the fields $q_{ij}$ and $\phi$ are thus the equations 
\begin{align}
	\label{static spherical symmetry: constraint for zero momenta ver.2}
	\frac{\delta \Psi}{\delta q_{ab}} = 0
	\, , \qquad
	\frac{\delta \Psi}{\delta \phi} = 0
	\, .
\end{align}

We will be now more specific and restrict ourselves to static spherically symmetric spacetimes. The spatial metric of a static spherically symmetric spacetime has the form
\begin{align}
	\label{static spherical symmetry: general metric}
	d\sigma^2 
	=  q_{ab}(\bm{y})  dy^{a} dy^{b}
	=  A^2(r)  dr^2  +  B^2(r)  d\Omega^2
	\, .
\end{align}
In addition, we also have a massless static spherically symmetric scalar field $\phi = \phi(r)$. The primary constraints (of staticity) for momenta and the Hamiltonian constraint are
\begin{align}
	\label{static spherical symmetry: momenta = 0 v AB parametrizaci}
	P_{\!A} = 0
	\, , \quad
	P_{\!B} = 0
	\, , \quad
	P_{\!\phi} = 0
	\, ,
	\\
	\label{static spherical symmetry: Hamiltonian constraint in AB parametrization}
	A B^2  {^{(\parallel)}\!}R -  \frac{\kappa B^2 \phi\bm{'}^{\,2}}{A} =  0
	\, .
\end{align}
The consistency conditions for the primary constraints (\ref{static spherical symmetry: momenta = 0 v AB parametrizaci}) for momenta are (after their simplification with the Hamiltonian constraint):
\begin{align}
	\label{static spherical symmetry: consistency conditions for P_A}
	\bdot{P}_{\!A}
	&=  \frac{4\pi A^2 B^2}{\kappa} \left[ 
    N \left( \frac{\kappa \phi\bm{'}^{\,2}}{A^4} - {^{(\parallel)}\!}R^{rr} \right) 
    - \frac{B\bm{'} N\bm{'}}{A^4 B} \right] 
    \overset{!}{\approx} 0
	\, , 
	\\
	\label{static spherical symmetry: consistency conditions for P_B}
	\bdot{P}_{\!B}
	&= - \frac{4\pi A B^3}{\kappa} \left[ N  {^{(\parallel)}\!}R^{\theta\theta}  
	+ \frac{1}{A^2 B^2} \left( N\bm{''} - \frac{A\bm{'} N\bm{'}}{A} 
	+ \frac{B\bm{'} N\bm{'}}{B} \right) \right] 
	\overset{!}{\approx} 0
	\, , 
	\\
	\label{static spherical symmetry: consistency conditions for P_phi}
	\bdot{P}_{\!\phi} 
	&= - \frac{4\pi B^2 N}{A} \left[ \phi\bm{''} - \left(\frac{A\bm{'}}{A} 
	- \frac{2 B\bm{'}}{B} - \frac{N\bm{'}}{N}\right) \phi\bm{'} \right] 
	\overset{!}{\approx} 0
	\, .
\end{align}
Spatial scalar curvature is explicitly 
\begin{align*}
	{^{(\parallel)}\!}R 
	= A^2 \, {^{(\parallel)}\!}R^{rr} + 2 B^2 \,{^{(\parallel)}\!}R^{\theta\theta}
	\, . 
\end{align*}
If we substitute the Hamiltonian constraint (\ref{static spherical symmetry: Hamiltonian constraint in AB parametrization}) into the expression $A \bdot{P}_{\! A} + 2 B \bdot{P}_{\! B}$, we get 
\begin{align}
	\label{static spherical symmetry: contracted consistency conditions for lapse in AB parametrization}
	A \bdot{P}_{\!A} + 2 B \bdot{P}_{\!B} 
    = - \frac{8\pi B^2}{\kappa A} \left( N\bm{''} - \frac{A\bm{'} N\bm{'}}{A} 
	+ \frac{2 B\bm{'} N\bm{'}}{B} \right) 
	\overset{!}{\approx}  0 
	\, ,
\end{align}
which (up to multiplicative numerical constant $8 \pi$) exactly corresponds to the contracted conditions (\ref{static spherical symmetry: contracted consistency conditions for metric momenta}) and it is therefore equivalent to the equation (\ref{static spherical symmetry: lapse equation}) for the lapse $N$:
\begin{align}
	\label{static spherical symmetry: lapse equation v AB parametrizaci}
	q^{ij}  {^{(\parallel)}\!}\nabla_{i} {^{(\parallel)}\!}\nabla_{j} N 
	=  \frac{1}{A^2} \left( N\bm{''} - \frac{A\bm{'} N\bm{'}}{A} \right) 
	+  \frac{2 B\bm{'} N\bm{'}}{A^2 B} 
	=  0
	\, .
\end{align}
Rearranging the equation to a more appropriate form
\begin{align}
	\label{static spherical symmetry: lapse equation v AB parametrizaci ver.2}
	N\bm{''} - \left(\frac{A\bm{'}}{A} - \frac{2 B\bm{'}}{B} \right) N\bm{'} 
	=  \frac{2}{B} \left(B N\bm{'}\right)^{\bm{\prime}} - \frac{1}{A} \left(A N\bm{'}\right)^{\bm{\prime}}
	=  0
	\, ,
\end{align}
its apparent solution is a constant lapse function $N = const.$ which is also our case of sJNW metric. Another possible solutions is, for example, Schwarzschild's  $N = \closedsqrt{1 - 2M r^{-1}}$, $A = N^{-1}$, $B = r$ with $M$ constant. 
Knowing the specific form of the functions $A(r)$ and $B(r)$ with the assumption that $N\bm{'} \neq 0$ allows us to separate variables in the equation (\ref{static spherical symmetry: lapse equation v AB parametrizaci ver.2}) with respect to $N\bm{'}$ giving us the compact expression
\begin{align}
	\label{static spherical symmetry: variable separation in lapse equation}
	\left( \ln\left( N\bm{'} \right) \right)^{\bm{\prime}}
	=  \frac{A\bm{'}(r)}{A(r)} - \frac{2 B\bm{'}(r)}{B(r)} 
	\, . 
\end{align}
The consistency conditions for the primary constraints (\ref{static spherical symmetry: momenta = 0}) in the case of the Schwarzshild spacetime are discussed in more detail in \cite{bojowald}, chapter \textit{4.3.1 Schwarzschild solution}. 
The equation (\ref{static spherical symmetry: lapse equation v AB parametrizaci ver.2}) can be regarded as a relation for the lapse function, which ensures a consistency of the foliation (calibration $N$, $N^{r} = 0$) with the requirement of staticity of the spacetime, i.e. the vanishment of all the momenta. The explanation provided by \cite{bojowald} is that these consistency conditions restrict freedom of choice of the lapse to such $N$ for which the vector field tangent to the hypersurface $\mathcal{C}^{(G, \phi)}_{stat.}[N]$ of the Hamiltonian constraint is also simultaneously tangent to the hypersurface of the primary constraints for the vanishing momenta, and that this state remains unchanged throughout the whole gauge orbit.

On a quantum level, we have the following set of equations for the wave functional $\Psi[A, B, \phi]$:  
\begin{align}
	\label{static spherical symmetry: spherically symmetrical Hamiltonian constraint}
	\left( A B^2  {^{(\parallel)}\!}R - \frac{\kappa B^2 \phi\bm{'}^{\,2}}{A} \right) \Psi = 0
	\, ,
	\\
	\label{static spherical symmetry: spherically symmetrical momentum constraints}
	\frac{\delta \Psi}{\delta A} = 0
	\, , \quad
	\frac{\delta \Psi}{\delta B} = 0
	\, , \quad
	\frac{\delta \Psi}{\delta \phi} = 0
	\, .
\end{align} 
The relationship between the canonical variables $A$, $B$ and $\phi$ is described by the Hamiltonian constraint (\ref{static spherical symmetry: Hamiltonian constraint in AB parametrization}) in the from of an equation
\begin{align}
	\label{statická sférická symetrie: vztahy mezi A, B a phi dané hamiltonovskou vazbou}
	\frac{4 \pi G}{c^4} \phi\bm{'}^{\,2} - \frac{A^2}{2} {^{(\parallel)}\!}R 
	=  \frac{4 \pi G}{c^4} \phi\bm{'}^{\,2} - \frac{2 A\bm{'} B\bm{'}}{A B} + \frac{2 B\bm{''}}{B}  + \frac{B\bm{'}^{\,2}}{B^2} - \frac{A^2}{B^2}
	=  0
	\, ,
\end{align}
which is automatically solved by all the static spacetimes with a scalar field that satisfies the Einstein field equations. It is thus also satisfied by the sJNW spacetime. That concludes our analysis of the static Hamiltonian constraint.

As we already know from the section \ref{Momentum Constraint}, the solution to the equations (\ref{static spherical symmetry: spherically symmetrical momentum constraints}) is the functional 
\begin{align}
	\label{static spherical symmetry: functional Z=Z_0+Z_const}
	Z[A, B, \phi] =  Z_{0}[A, B, \phi] + Z_{const.} 
	\, ,
\end{align}
where $Z_{const.}$ is a constant (or a function of physical constants and other parameters of the spacetime) and the funkcional $Z_{0}$ is
\begin{align}
	\label{static spherical symmetry: static solution as Z_0 - integral}
	Z_{0}[A, B, \phi]
	=  \int_{r_{min}}^{\infty}
	\Big( A\bm{'} f_{A}\left(A\right)  +  B\bm{'} f_{B}\left(B\right) 
	+  \phi\bm{'} f_{\phi}\left(\phi\right) \Big) dr
	\, ,
\end{align}
with arbitrary (differentiable) functions $f_{A}$, $f_{B}$ and $f_{\phi}$.
To these functions, let us define their primitive functions $F_{A}$, $F_{B}$ and $F_{\phi}$, collectively for $X = A, B, \phi$ as
\begin{align}
	\label{static spherical symmetry: static solution as Z_0 - F(X)}
	F_{X}(X) \equiv \int f_{X}(X) dX
	\, .
\end{align}
It then holds that  
\begin{align}
	\label{static spherical symmetry: static solution as Z_0 - differention equation for F(X)}
	F\bm{'}_{X}(X) = X\bm{'} f_{X}(X)
	\, .
\end{align}
For the functional $Z_0$ we then get the expression
\begin{align}
	\label{static spherical symmetry: static solution as Z_0 - integrated solution}
	Z_{0}[A,\, B,\, \phi]
	= \Big[ F_{A}\left(A(r)\right) + F_{B}\left(B(r)\right) 
	+ F_{\phi}\left(\phi(r)\right) \Big]_{r=r_{min}}^{r=\infty} 
	\, .
\end{align}
Naturally, we want the functional $Z_0$ to be well defined. To be specific we require that $|Z_0| < \infty$. The functions $X\bm{'} f_{X}$ then have to be  (Lebesgue) integrable on the interval $(r_{min},\, \infty)$. For a sJNW metric in the foliation (\ref{special Janis-Newman-Winicour metric}) is $r_{min}= R_{min} = \chi_0$ and in the foliation (\ref{sJNW metric with B}) is $r_{min}= B_{min} = 0$. 

In the special foliation $R = r$ of a sJNW metric, that favours a proper radial distance $R$ (over a spherical curvature radius) we get following explicit values of the metric functions: $A = 1$, $B = \closedsqrt{r^2 - \chi_0^2}$. 
The consistent lapse function\footnote{
    With the assumption that $N\bm{'} \neq 0$ we get for this foliation, from the equation (\ref{static spherical symmetry: variable separation in lapse equation}) the relation $N\bm{'} = c_1 B^2$, and hence $N(r) = c_1 \left( \frac{1}{3} r^3 - \chi_0^2 r\right) + c_2$ with integration constants $c_1$ and $c_2$. In accordance with the original assumption $N\bm{'} \neq 0$ we then set  $c_2 = 0$.} 
is $N = 1$ which corresponds to $T = t$. 
On the other hand, in the foliation $R = \closedsqrt{r^2 + \chi_0^2}$ of a sJNW metric, favouring a spherical curvature radius $B$, we get: $A = \left(1 + \chi_0^2 r^{-2}\right)^{-1/2}$, $B = r$. 
The consistent lapse function\footnote{
    Assuming $N\bm{'} \neq 0$ allows us again to find a consistent lapse function from the equation (\ref{static spherical symmetry: variable separation in lapse equation}), which has for this choice of foliation a general solution $N(r) = \frac{c_1}{\chi_0}\left[ \ln(r) - \ln(\chi_0 \closedsqrt{r^2 + \chi_0^2} + \chi_0^2)\right] + c_2$ with integration constants $c_1$ and $c_2$. Then, in accordance with $N\bm{'} \neq 0$ it has to hold that $c_2 = 0$.} 
is then given as $N = 1$, which corresponds to $T = t$.  
For both mentioned foliations we have the same scalar field $\phi = \frac{c^2}{4 \closedsqrt{\pi G}} \left[ \ln(r - \chi_0) - \ln(r + \chi_0) \right]$. Functionals $Z_{0}^{(R)}$ and $Z_{0}^{(B)}$, corresponding to these two foliations have a form of the functions
\begin{align}
	\label{static spherical symmetry: static solution as Z_0 - R function}
	Z_{0}^{(R)}(\chi_0)
	&=  \int_{\chi_0}^{\infty}
	\left( \frac{r  f_{B}(B(r))}{\closedsqrt{r^2 - \chi_0^2}}  
	+ \frac{1}{\closedsqrt{2\kappa}} \frac{\chi_0 f_{\phi}(\phi(r))}{\left(r^2 - \chi_0^2\right)} \right) dr 
	\, , \quad
	\\
	\label{static spherical symmetry: static solution as Z_0 - B function}
	Z_{0}^{(B)}(\chi_0)
	&=  \int_{0}^{\infty}
	\left( - \frac{\chi_0^2 f_{A}(A(r))}{\left(r^2 + \chi_0^2\right)^{\frac{3}{2}}} 
	+ f_{B}(r) 
	+ \frac{1}{\closedsqrt{2\kappa}} \frac{\chi_0  f_{\phi}(\phi(r))}{\left(r^2 - \chi_0^2\right)} \right) dr 
	\, .
\end{align}
The wave functional $\Psi$ of this system can be expressed as a wave function $\Psi = \psi\left(Z_0 + Z_{const.}\right)$. The value of the constant functional $Z_{const.}$ can be determined if we know the behaviour of the wave function $\psi(Z)$ and the functional $Z_0$ (and, in a general dynamical case also the baviour of $Z_g$ and $Z_h$), and if we place some additional conditions restricting any possibly undesirable behaviour of the wave function $\psi(Z)$. Depending on the specific behaviour of $Z_0$, these additional conditions might help to determine $Z_{const.}$. In our case for the above-described choices of a foliation, however, we don't know the specific form of the wave function $\psi(Z)$. The wave functional $\Psi = \psi(Z)$ is thus an undetermined wave function 
\begin{align}
	\label{static spherical symmetry: wave function psi(chi0)}
	\Psi  =  \psi(Z)  =  \psi\left(Z_0(\chi_0) + Z_{const.}\right)  =  \Psi(\chi_0)
	\, .
\end{align}
So far, we have only demanded $Z_0$ to be well defined (finite) which meant that all the functions $A\bm{'} f_{A}(A)$, $B\bm{'} f_{B}(B)$ and $\phi\bm{'} f_{\phi}(\phi)$, for both abode-discussed choices of foliation are required to be integrable on the intervals $(\chi_0, \infty)$ or $(0, \infty)$, depending on the specific foliation. Although these conditions do restrict a possible choice of $f_{A}$, $f_{B}$ and $f_{\phi}$, they are not sufficient enough to determine them. Without any additional conditions moderating the behaviour of the wave functions, the solutions (\ref{static spherical symmetry: static solution as Z_0 - integrated solution}) or (\ref{static spherical symmetry: static solution as Z_0 - R function}) and (\ref{static spherical symmetry: static solution as Z_0 - B function}) for any preferred foliation of our problem of quantization of the sJNW spacetime will contain a triple (pair) of insufficiently specified functions. The presence of arbitrary functions is typical for solutions to  Wheeler-DeWitt equations. Because of the absence of momenta in the static Hamiltonian constraint one cannot find the specific dependence of the wave function $\psi$ on the functional $Z_0$. This "unfortunate" lack of knowledge does not allow us to, for example, precisely investigate the nature and fate of singularities, detailed quantum effect in their proximity or even whether or not their are allowed to exist (on a quantum level) in the studied spacetimes.

Let us now investigate our solution (\ref{static spherical symmetry: wave function psi(chi0)}) from a qualitative perspective. 
The natural interpretation of the wave function $\Psi(\chi_0)$ is that it represents a superposition of naked time-like sJNW singularities located on various different radii $\pm\chi_0$. The states $\Psi(\chi_0)$ of the sJNW geometry are static and retain the same value on all hypersurfaces for any given foliation $\{A(r), B(r)\}$ at any time $t \in \mathbb{R}$. Since we now work with some wave functions $\Psi(\chi_0)$ and not with functionals, we may attempt to define a scalar product in the naive form
\begin{align}
	\label{static spherical symmetry: scalar product naive definition}
	\braket{\Psi_{1}|\Psi_{2}}_{w}
	= \int\limits_{-\infty}^{\infty} \Psi^{*}_{1}(\chi_0) \Psi_{2}(\chi_0) w(\chi_0) d\chi_0
	\, ,
\end{align}
with some weight function $w$. 
The scalar product is well defined if we require the wave functions to be square-integrable on $\mathbb{R}$ with respect to some, yet to be determined weight $w$. The associated Hilbert space of all the states $\Psi(\chi_0)$ is then a space $L^2_{w}(\mathbb{R})$ of all $w$-weighted square-integrable functions. With a scalar product defined we can furthermore demand the normalization of (physical) wave functions to be $\braket{\Psi|\Psi}_{w} = 1$. The quantity $\Psi^{*}(\chi_0) \Psi(\chi_0) w(\chi_0)$ will be then interpreted as a probability density to find the system in the sJNW geometry of spacetime with the singularity located on the radius $\chi_0$. Unfortunately, this brief analysis comprises all the main properties of the wave functions $\Psi(\chi_0)$ and their behaviour we are currently able to uncover.

\section{Conclusion}

We investigated canonical quantization of spherically symmetric (midisuperspace) spacetimes with a massless scalar field via the Wheeler-DeWitt equations. These equations were solved in case of some simpler minisuperspace cosmological models (Hartle, Hawking \cite{hartle-hawking}, and others). However, in case of midisuperspace model the situation is rather complicated. Despite that, reduced phase space quantization was successfully performed for vacuum spherically symmetric spacetimes, namely for Schwarzschild primordial black holes (Kucha\v r \cite{kuchar}). Attempts to generalize Kucha\v r's approach also to non-vacuum spherically symmetric spacetimes with scalar-field source have been unsuccessful so far. The problem is that in this framework no proper definition of time have been found yet. For those reasons we have decided to use the methods leading to Wheeler-DeWitt equations. 

For a midisuperspace model we initially chose the Roberts spacetime (expression (6.2) in \cite{tahamtan-svitek-2}) which is a dynamical spherically symmetrical spacetime with a scalar field. However, we discovered that the Roberts spacetime does not behave well in the asymptotic region. The Roberts metric is not asymptotically flat and its respective boundary terms diverge. Furthermore, the Roberts metric cannot be normalized to a Minkowski metric because some of the divergences persist even after its re-normalization to a flat space. Because of this problematic behaviour of the Roberts metric we decided to work with its limit instead. The limit leads to asymptotically flat special Janis-Newnam-Winicour (sJNW) spacetime. Apparently, the sJNW spacetime is also spherically symmetric non-vacuum midisuperspace model with scalar field but unlike Roberts, sJNW  is static and contains a naked time-like singularity. Furthermore, the boundary term  of the sJNW spacetime can be normalized to Minkowski spacetime in respect to which the sJNW has zero ADM energy. We therefore proceeded with quantization of the sJNW spacetime. 

In case of general spherically symmetric midisuperspace models with a scalar field we have found a general, separate form, solution (section \ref{Momentum Constraint}) of the quantum momentum constraint in metrical representation. The form $Z = Z_0 + Z_g + Z_h + Z_{const.}$ of the solution is described in section \ref{Momentum Constraint}. Unfortunately, we were not able to found any non-trivial solution to the quantum Hamiltonian constraint. Considering staticity of the sJNW metric, we further focused on the specifics of static 3-metrics, on the additional conditions in the ADM formalist, and on their quantization. 

Lastly, we investigated the WDW equations for static spherically symmetric midisuperspace spacetimes with scalar field and we have found their solution. This solution contains free, insufficiently specified functions and hence cannot be quantitatively examined to the full extend. Applying the solution to the sJNW metric yields a specific solution $\Psi = \Psi(\chi_0)$ which we interpret as a superposition of naked time-like sJNW singularities located on various radii $\pm\chi_0$. States $\Psi(\chi_0)$ of the spacetime are static and hence remain constant for all hypersurfaces (of a given foliation) independently of the choice of $t \in \mathbb{R}$. Karel Kuchař found a similar solution $\Psi = \Psi(m)$ for the Schwarzschild spacetime (which is also spherically symmetric midisuperspace model and, unlike our spacetime, it is vacuum). This Kuchař's wave function $\Psi(m)$ was interpreted analogously to our case as a superposition of primordial black holes with different masses $m$. Wheeler-DeWitt equations for Schwarzschild spacetime were solved by Masakatsu Kenmoku, Hiroto Kubotani, Eiichi Takasugi and Yuki Yamazaki in their article \cite{kenmoku-kubotani-takasugi-yamazaki}. The key point of their approach was to use the mass function $M$ (originally found by Kuchař as a reconstructed mass of a Schwarzschild black hole in canonical variables), which is a conserved quantity, to express the Hamiltonian constraint as a linear combination of $M$ and the momentum constraint. Instead of quantizing rather complex Hamiltonian constraint it is then possible to quantize considerably simpler mass function and momentum constraint instead. Unfortunately, in our case of sJNW spacetime with a scalar field we haven't found any analogue to Kuchař's conservant mass function. The reason for this might be the vanishing ADM mass of sJNW because in case of Schwarzschield the ADM energy is exactly the the mass of a Schwartzschild black hole.

The results of this work can be further extended in the future. One of the possible remaining tasks is to study the properties of Roberts spacetime in more depth, especially its metric that has a bad asymptotic behaviour in the space infinity $i^0$ but it behaves reasonably in the future null (light) infinity $\mathcal{I}^{+}$. The Bondi mass of the Roberts metric is finite (see \cite{tahamtan-svitek-2}). This reveals a possibility of foliating the Roberts spacetime with null hypersurfaces, in which case it will be necessary however to deal with number of both technical and physical  difficulties like ensuring causality etc.

\renewcommand{\thesubsection}{\Alph{subsection}}
\appendix
\section*{Appendix}
\label{Appendix}

\subsection{Remarks on Dirac Delta Function}
\label{Remarks on Dirac Delta Function}

Let us summarize some useful properties of Dirac delta function (distribution) $\delta(x-y)$. 
Dirac delta function is symmetric: $\delta(-x) = \delta(x)$. Its simple derivatives can be then manipulated as 
\begin{align*}
	\partial_{x} \delta(x-y) 
	&= \partial_{(x-y)} \delta(x-y) = \partial_{x} \delta(y-x) = 
	\nonumber
	\\
	&= - \partial_{(y-x)} \delta(y-x) = - \partial_y \delta(y-x) = - \partial_y \delta(x-y)
	\, ,
\end{align*}
and for multiple derivatives of delta function:
\begin{align*}
	\partial_{x} \partial_{x} \delta(x-y) 
	= - \partial_{x} \partial_{y} \delta(x-y)
	= \partial_{y} \partial_{y} \delta(x-y)
	\, .
\end{align*}
One of the defining features of Dirac delta functions is 
\begin{align*}
	\int_{\Omega} \varphi(x) \delta(x-y) \, dx = \varphi(y)
\end{align*}
for any testing function $\varphi(x)$ that is smooth and with compact support on $\Omega$, i.e. $\varphi$ vanishes at the boundary: $\varphi |_{\partial\Omega} = 0$. Consider a function $F(x)$. Distribution $F(x) \delta(x-y)$ is then required to act on testing functions $\varphi$ as
\begin{align*}
	\int_{\Omega} \varphi(x) F(x) \delta(x-y) \, dx
	&= \varphi(y) F(y)
	= F(y) \int_{\Omega} \varphi(x) \delta(x-y) \, dx
	=
	\\
	&= \int_{\Omega} \varphi(x) F(y) \delta(x-y) \, dx \, .
\end{align*}
which can be then interpreted as
\begin{align}
	\label{Constraint algebra in vacuum: F*delta}
	F(x) \delta(x-y) = F(y) \delta(x-y) \, .
\end{align}
As a consequence, we can write expressions that contain functions evaluated in different points but are multiplied by delta function as if they were taken in the same point:
\begin{align*} 
	F(x)G(y)\delta(x-y) &= F(y)G(y)\delta(x-y) = F(y)G(x)\delta(x-y) =  
	\\
	&= F(x)G(x)\delta(x-y) \equiv F G \delta(x-y)
\end{align*}
Above relations are useful when dealing with expressions containing derivative of Dirac delta function.  
Distribution $F(x) \partial_{x} \delta(x-y)$ acts on testing functions $\varphi$, $\varphi |_{\partial\Omega} = 0$, as 
\begin{align*}
	&\int_{\Omega} \varphi(x) F(x) \partial_{x} \delta(x-y) \, dx
	=
	\\
	& \quad\quad\quad =
	- \int_{\Omega} dx \, \Big[ \varphi(x) \partial_{x} F(x) 
	+ F(x) \partial_{x} \varphi(x) \Big] \delta(x-y) \, dx
	= 
	\\ 
	& \quad\quad\quad =
	- \varphi(y) \partial_{y} F(y) - F(y) \partial_{y} \varphi(y) 
	=
	\\
	& \quad\quad\quad =
	- \partial_{y} F(y) \int_{\Omega} \varphi(x) \delta(x-y) 
	- F(y) \int_{\Omega} dx \, \partial_{x} \varphi(x) \, \delta(x-y) \, dx
	=
	\\
	& \quad\quad\quad =
	\int_{\Omega} dx \, \varphi(x) \Big[- \partial_{y} F(y) \, \delta(x-y) 
	+ F(y) \partial_{x} \delta(x-y) \Big] \, dx
	\, ,
\end{align*}
implying that effectively
\begin{align}
	\label{Constraint algebra in vacuum: F*ddelta}
	F(x) \partial_{x} \delta(x-y) - F(y) \partial_{x} \delta(x-y)
	= - \partial_{y} F(y) \, \delta(x-y)
	\, .
\end{align}
This result can be also formally obtained by differentiating identity (\ref{Constraint algebra in vacuum: F*delta}): 
\begin{align*}
	F(y) \delta(x-y) &= F(x) \delta(x-y) \,\, \big/^{\partial_x} 
	\\
	F(y) \partial_{x} \delta(x-y) &= F(x) \partial_{x} \delta(x-y) + \partial_{x} F(x) \, \delta(x-y)
	\, .
\end{align*}
which is precisely (\ref{Constraint algebra in vacuum: F*ddelta}). In the following text we will often use identities 
\begin{align}
	\label{Constraint algebra in vacuum: F*G*ddelta}
	F(x) G(y) \partial_{x} \delta(x-y) 
	= 
	\begin{cases}
		F(x) G(x) \partial_{x} \delta(x-y) + F(x) \left[\partial_{x} G(x)\right] \delta(x-y) 
		\\[5pt]
		F(y) G(y) \partial_{x} \delta(x-y) - \left[\partial_{y} F(y)\right] G(y) \delta(x-y)
	\end{cases}
	\, .
\end{align}
Both these branches are indeed equivalent as can be easily proven using (\ref{Constraint algebra in vacuum: F*ddelta}):
\begin{align*}
	&F(y) G(y) \partial_{x} \delta(x-y) - \left[\partial_{y} F(y)\right] G(y) \delta(x-y) 
	= 
	\\
	&\qquad= 
	F(x) G(x) \partial_{x} \delta(x-y) + \partial_{x} \left[\cancel{F(x)} G(x)\right] \delta(x-y) 
	- \cancel{\left[\partial_{x} F(x)\right] G(x) \delta(x-y)}
	\, .
\end{align*}
Another useful identity that can be derived form (\ref{Constraint algebra in vacuum: F*ddelta}) is 
\begin{align}
	\label{Constraint algebra in vacuum: F*G*ddelta}
	&F(x) G(y) \partial_{x} \delta(x-y) + F(y) G(x) \partial_{x} \delta(x-y) 
	= \nonumber
	\\
	&\qquad=  
	F(x) G(x) \partial_{x} \delta(x-y) + \cancel{F(x) \left[\partial_{x} G(x)\right] \delta(x-y)} 
	+ \nonumber 
	\\
	&\qquad\quad\, +
	F(y) G(y) \partial_{x} \delta(x-y) - \cancel{F(y) \left[\partial_{y} G(y)\right] \delta(x-y)}
	= \nonumber
	\\[5pt]
	&\qquad=
	F(x) G(x) \partial_{x} \delta(x-y) + F(y) G(y) \partial_{x} \delta(x-y)
	\, .
\end{align}
Distribution $F(x) \partial_{x} \partial_{x} \delta(x-y)$, containing second derivatives of Dirac delta function can be obtained in similar way. The distribution acts on a testing function $\varphi(x)$ that satisfies boundary conditions $\varphi |_{\partial\Omega} = \partial_{x} \varphi(x) |_{\partial\Omega} = 0$ as
\begin{align*}
	&\int_{\Omega} \varphi(x) F(x) \partial_{x} \partial_{x} \delta(x-y) \, dx
	=
	\\
	& \quad =
	- \int_{\Omega} \Big[ \varphi(x) \partial_{x} F(x) 
	+ F(x) \partial_{x} \varphi(x) \Big] \partial_{x} \delta(x-y) \, dx
	=
	\\
	& \quad =
	\int_{\Omega} \Big[ \varphi(x) \partial_{x} \partial_{x} F(x) 
	+ 2 \partial_{x} \varphi(x) \, \partial_{x} F(x) + F(x) \partial_{x} \partial_{x} \varphi(x) \Big] \delta(x-y) \, dx
	= 
	\\ 
	& \quad =
	\varphi(y) \partial_{y} \partial_{y} F(y) + 2 \partial_{y} \varphi(y) \, \partial_{y} F(y) 
	+ F(y) \partial_{y} \partial_{y} \varphi(y) 
	=
	\\
	& \quad =
	\partial_{y} \partial_{y} F(y) \int_{\Omega} \varphi(x) \delta(x-y) \, dx
	+ 2 \partial_{y} F(y) \int_{\Omega} \partial_{x} \varphi(x) \delta(x-y) \, dx + 
	\\
	& \quad\quad + F(y) \int_{\Omega} \partial_{x} \partial_{x} \varphi(x) \, \delta(x-y) \, dx
	=
	\\
	& \quad =
	\int_{\Omega} \varphi(x) \Big[\partial_{y} \partial_{y} F(y) \, \delta(x-y) 
	- 2 \partial_{y} F(y) \, \partial_{x} \delta(x-y) 
	+ F(y) \partial_{x} \partial_{x} \delta(x-y) \Big] dx
	\, .
\end{align*}
Hence 
\begin{align}
	\label{Constraint algebra in vacuum: F*dddelta}
	&F(x) \partial_{x} \partial_{x} \delta(x-y) - F(y) \partial_{x} \partial_{x} \delta(x-y) = 
	\nonumber
	\\
	& \quad\quad\quad\quad\quad\quad = 
	\partial_{y} \partial_{y} F(y) \, \delta(x-y) - 2 \partial_{y} F(y) \, \partial_{x} \delta(x-y) 
	\, ,
\end{align}
which can be alternatively derived by differentiating (\ref{Constraint algebra in vacuum: F*ddelta}) twice.

\subsection{Poisson Brackets for Fields}
\label{Poisson brackets for fields}

Take a system described by canonical fields $q_{I}(x)$ and conjugated momenta $p^{I}(y)$, where the field indices denote all the different fields that describe our system (i.e. all fields that makes the configuration space of our system). Poisson brackets for functions or functionals $F[q_A,\, p^A]$, $G[q_A,\, p^A]$ etc. on phase space of fundamental variables $(q_I,\, p^J)$ with $q_{I}(x)$ and $p^{J}(x)$ being canonical fields and momenta are  
\begin{align}
	\label{Poisson brackets for fields: Poisson brackets for fields}
	\left\{ F, \, G \right\} 
	= 
	\sum\limits_A
	\int\limits_{\Sigma} 
	\bigg( \frac{\delta F}{\delta q_A(x)} \frac{\delta G}{\delta p^A(x)} 
	- \frac{\delta G}{\delta q_A(x)} \frac{\delta F}{\delta p^A(x)} \bigg) d x  
	\, .
\end{align}
Mutual (in)dependence of the canonical fields and momenta are expressed via their functional derivatives  
\begin{equation}
	\label{Poisson brackets for fields: canonical functional derivatives}
	\frac{\delta q_{A}(x)}{\delta q_{B}(y)} = \delta^{B}_{A} \delta(x - y) 
	\, , \quad
	\frac{\delta p^{A}(x)}{\delta p^{B}(y)} = \delta^{A}_{B} \delta(x - y)
	\, , \quad
	\frac{\delta q_{A}}{\delta p^{B}} = 0
	\, , \quad 
	\frac{\delta p^{A}}{\delta q_{B}} = 0 
	\, .
\end{equation}
The canonical Poisson brackets of fundamental fields $q_{A}(x)$ and conjugated momenta $p^{B}(y)$ are then
\begin{equation}
	\label{Poisson brackets for fields: canonical Poisson brackets}
	\Big\{ q_{A}(x) ,\, p^{B}(y) \Big\} = \delta^{B}_{A} \delta(x - y) 
	\, , \quad
	\Big\{ q_{A} ,\, q_{B} \Big\} = 0
	\, , \quad 
	\Big\{ p^{A} ,\, p^{B} \Big\} = 0 
	\, .
\end{equation}
Some basic properties of Poisson brackets are
\begin{itemize}
	\item Antisymmetry: $\left\{ G ,\, F \right\} = - \left\{ F ,\, G \right\} $
	\item Bilinearity: $\left\{ F_1 + F_2 ,\, G \right\} = \left\{ F_1 ,\, G \right\} + \left\{ F_2 ,\, G \right\}$, and $\left\{ c F ,\, G \right\} = c \left\{ F ,\, G \right\}$ for $c$ constant with respect to phase space variables. Linearity in the second argument follows from the antisymmetric property of the bracket.
	\item Leibnitz rule: $\left\{ F G ,\, H \right\} = F \left\{ G ,\, H \right\} + \left\{ F ,\, H \right\} G$, and the same for the second argument.
	\item Jacobi identity: $\left\{ F ,\, \left\{ G ,\, H \right\}\right\} + \left\{ H ,\, \left\{ F ,\, G \right\}\right\} + \left\{ G ,\, \left\{ H ,\, F \right\}\right\} = 0$
\end{itemize}
Poisson bracket of momentum and any (differentiable) functional $F[q_{A}]$ dependent solely on canonical coordinates is
\begin{align}
	\label{Poisson brackets for fields: Poisson bracket of F[q] and p}
	\left\{ F[q_{A}],\, p^{B}(y) \right\} 
	\nonumber
	&= 
	\int\limits_{\Sigma} 
	\bigg( \frac{\delta F[q_{A}]}{\delta q_{I}(z)} \frac{\delta p^{B}(y)}{\delta p^{I}(z)} 
	- \underbrace{\frac{\delta F[q_{A}]}{\delta p^{I}(z)}}_{0} 
	\underbrace{\frac{\delta p^{B}(y)}{\delta q_{I}(z)}}_{0} \bigg) d z 
	= \nonumber
	\\
	&=
	\int\limits_{\Sigma} 
	\frac{\delta F[q_{A}]}{\delta q_{I}(z)} 
	\delta^{B}_{I} \delta(y - z) d z
	= 
	\frac{\delta F[q_{A}]}{\delta q_{B}(y)} 
	\, .
\end{align}
If we have function $f\left(q_{A}(x)\right)$ instead of functional, we would get expression
\begin{align}
	\label{Poisson brackets for fields: Poisson bracket of f(q) and p}
	\left\{ f\left(q_{A}(x)\right),\, p^{B}(y) \right\} 
	&= \int\limits_{\Sigma} 
	\bigg( \frac{\delta f\left(q_{A}(x)\right)}{\delta q_{I}(z)} \frac{\delta p^{B}(y)}{\delta p^{I}(z)} 
	- \underbrace{\frac{\delta f\left(q_{A}(x)\right)}{\delta p^{I}(z)}}_{0} 
	\underbrace{\frac{\delta p^{B}(y)}{\delta q_{I}(z)}}_{0} \bigg) d z 
	= \nonumber
	\\
	&= \int\limits_{\Sigma} 
	\frac{\partial f\left(q_{A}(x)\right)}{\partial q_{J}(x)} \frac{\delta q_{J}(x)}{\delta q_{I}(z)} 
	\frac{\delta p^{B}(y)}{\delta p^{I}(z)}  d z
	= \nonumber
	\\
	&= \int\limits_{\Sigma} 
	\frac{\partial f\left(q_{A}(x)\right)}{\partial q_{J}(x)}
	\delta^{I}_{J} \delta(x - z) \delta^{B}_{I} \delta(y - z) d z
	= \nonumber
	\\
	&= \frac{\partial f\left(q_{A}(x)\right)}{\partial q_{B}(x)} \delta(x - y)
	\, .
\end{align}
Analogously for functionals and functions of momenta: 
\begin{align}
	\label{Poisson brackets for fields: Poisson brackets of F[p] and f(p) with q}
	\left\{ F[p^{A}],\, q_{B}(y) \right\} 
	= - \frac{\delta F[p^{A}]}{\delta p^{B}(y)} 
	\, , \quad 
	\left\{ f(p^{A}(x)),\, q_{B}(y) \right\}
	= - \frac{\partial f(p^{A}(x))}{\partial p^{B}(x)} \delta(x - y)
\end{align}
where the minus sign comes from reverse order od coordinates and momenta. Above derived relations allow us to reformulate Poisson bracket and instead of (\ref{Poisson brackets for fields: Poisson brackets for fields}) use
\begin{align}
	\label{Poisson brackets for fields: Poisson brackets for fields ver. 2}
	\left\{ F ,\, G \right\} 
	= \int\limits_{\Sigma} 
	\Big( \big\{ F ,\, q_A(x) \big\} \big\{ G ,\, p^A(x) \big\} 
	- \big\{ F ,\, p^A(x) \big\} \big\{ G ,\, q_A(x) \big\} \Big) d x  
	\, .
\end{align}
Because functional derivatives and variations commute with (partial or total) derivatives 
\begin{align*}
	\frac{\delta}{\delta \phi(y)} \partial_{x} f(\phi(x)) 
	= \partial_{x} \frac{\delta f(\phi(x))}{\delta \phi(y)} 
	\, , \quad 
	\delta \partial_{x} f(\phi(x)) 
	= \partial_{x} \delta f(\phi(x))
	\, ,
\end{align*}
we can write
\begin{align}
	\label{Poisson brackets for fields: Poisson bracket of dq and f(p)}
	\left\{ \partial_{x} q_{A}(x),\, f(p^{B}(y)) \right\} 
	&= \int\limits_{\Sigma} 
	\frac{\delta \partial_{x} q_{A}(x)}{\delta q_{I}(z)} \frac{\delta f(p^{B}(y))}{\delta p^{I}(z)} d z 
	= \nonumber
	\\
	&= \int\limits_{\Sigma} 
	\partial_{x} \frac{\delta q_{A}(x)}{\delta q_{I}(z)} \,  
	\frac{\partial f(p^{B}(y))}{\partial p^{J}(y)} \frac{\delta p^{J}(y)}{\delta p^{I}(z)}  d z 
	= \nonumber
	\\
	&= \int\limits_{\Sigma} 
	\delta^{A}_{I} \partial_{x} \delta(x - z) \, 
	\frac{\partial f(p^{B}(y))}{\partial p^{J}(y)} \delta^{J}_{I} \delta(y - z) d z 
	= \nonumber
	\\
	&= \frac{\partial f(p^{B}(y))}{\partial p^{A}(y)} \partial_{x} \delta(x - y)
	\, .
\end{align}
Analogously
\begin{align}
	\label{Poisson brackets for fields: Poisson bracket of f(q) and dp}
	\left\{ f(q_{A}(x)) ,\, \partial_{y} p^{B}(y) \right\} 
	=  \frac{\partial f(q_{A}(x))}{\partial q_{B}(x)} \partial_{y} \delta(x - y)
	\, .
\end{align}
And for both sides of the canonical bracket
\begin{align}
	\label{Poisson brackets for fields: Poisson bracket of dq and dp}
	\left\{ \partial_{x} q_{A}(x),\, \partial_{y} p^{B}(y) \right\} 
	&= \int\limits_{\Sigma} 
	\frac{\delta \partial_{x} q_{A}(x)}{\delta q_{I}(z)} \frac{\delta \partial_{y} p^{B}(y)}{\delta p^{I}(z)} d z 
	= \int\limits_{\Sigma} 
	\partial_{x} \frac{\delta q_{A}(x)}{\delta q_{I}(z)} \, \partial_{y} \frac{\delta p^{B}(y)}{\delta p^{I}(z)}  d z 
	= \nonumber
	\\
	&= \int\limits_{\Sigma} 
	\delta^{I}_{A} \partial_{x} \delta(x - z) \, \delta^{B}_{I} \partial_{y} \delta(y - z) \, d z 
	= \nonumber
	\\
	&= \partial_{x} \int\limits_{\Sigma} \delta^{B}_{A} \delta(x - z) \, \partial_{y} \delta(y - z) \, d z
	= \nonumber
	\\
	&= \delta^{B}_{A} \partial_{x} \partial_{y} \delta(y - x)
	\, .
\end{align}

In case of General relativity expressed in ADM formalism, we are using phase space of fundamental variables $( q_{ab},\, p^{ab})$ i. e. we have 3-metric $q_{A} = q_{ab}$ and its 3-momenta $p^{A} = p^{ab}$. 
 Using Hamilton's canonical equations, time (foliation) evolution of functionals $F[q_A,\, p^A]$ can be written in terms of Poisson bracket
\begin{align}
	\label{Poisson brackets for fields: time evolution of functionals}
	\bdot{F} 
	= \int\limits_{\Sigma} \bigg( \frac{\delta F}{\delta q_{ab}(x)} \bdot{p}^{ab}(x) 
	+ \frac{\delta F}{\delta p^{ab}(x)} \bdot{q}_{ab}(x)  \bigg) d^3 x  
	= \left\{ F, \, H_{G} \right\} 
	\, .
\end{align}
Canonical Poisson brackets for 3-metric and its 3-momentum are
\begin{equation}
	\label{Poisson brackets for fields: canonical Poisson brackets}
	\Big\{ q_{ab}(\bm{x}), \, p^{ij}(\bm{y}) \Big\} 
	= \delta^{i}_{(a} \delta^{j}_{b)} {\delta}^{(3)}(\bm{x} - \bm{y}) 
	\, , \quad 
	\Big\{ q_{ab}, \, q_{ij} \Big\} = 0
	\, , \quad
	\Big\{ p^{ab}, \, p^{ij} \Big\} = 0 
	\, .
\end{equation}
Poisson bracket of two functions (or functionals) both depending solely on 3-metric or solely on momenta are zero, i.e. 
\begin{align} 
	\label{Poisson brackets for fields: Poisson bracket of two functions depending solely on 3-metric or solely on momenta}
	\Big\{ f_1(q_{ab}(\bm{x})),\, f_2(q_{ij}(\bm{y})) \Big\} = 0 
	\, , \quad
	\Big\{ f_1(p^{ab}(\bm{x})),\, f_2(p^{ij}(\bm{y})) \Big\} = 0 
	\, .
\end{align}
In addition, consider a composite scalar function $f_{1}(q_{ab}) f_{2}(p^{ab})$ of a metric and its momenta, but not their derivatives. Then 
\begin{align*} 
	&\left\{ f_1(q_{ab}(\bm{x})) f_2(p^{ab}(\bm{x})) ,\, f_1(q_{ij}(\bm{y})) f_2(p^{ij}(\bm{y})) \right\} = 
	\\
	&= f_1(q_{ab}(\bm{x})) \left\{ f_2(p^{ab}(\bm{x})) ,\, f_1(q_{ij}(\bm{y})) \right\} f_2(p^{ij}(\bm{y})) + 
	\\
	&\hspace{125pt}\quad\, + f_1(q_{ij}(\bm{y})) \left\{ f_1(q_{ab}(\bm{x})) ,\, f_2(p^{ij}(\bm{y})) \right\} f_2(p^{ab}(\bm{x}))
	\, .
\end{align*}
Because, according to our assumption, $f_{1}(q_{ab}) f_{2}(p^{ab})$ does not depend on derivatives of the metric nor the momenta, their Poisson bracket is schematically 
\begin{align*} 
	&f_1(q_{ab}(\bm{x})) \left\{ f_2(p^{ab}(\bm{x})) ,\, f_1(q_{ij}(\bm{y})) \right\} f_2(p^{ij}(\bm{y})) = 
	\\ 
	&= \text{func.}\big(q_{ab}(\bm{x}),\, p^{ab}(\bm{x});\, q_{ij}(\bm{y}),\, p^{ij}(\bm{y})\big) {\delta}^{(3)}(\bm{x} - \bm{y})
	= \text{func.}(q_{kl},\, p^{kl}) |_{\bm{x}} \, {\delta}^{(3)}(\bm{x} - \bm{y})
	\, ,
\end{align*}
that is, the resulting function is multiplied by Dirac's delta distribution. The resulting function can be then taken at either the point $\bm{x}$, or $\bm{y}$ on $\Sigma$. Antisymmetry of Poisson brackets implies 
\begin{align*} 
	&f_1(q_{ab}(\bm{x})) \left\{ f_2(p^{ab}(\bm{x})) ,\, f_1(q_{ij}(\bm{y})) \right\} f_2(p^{ij}(\bm{y})) = 
	\\
	&\hspace{125pt}= - \, f_1(q_{ij}(\bm{y})) \left\{ f_1(q_{ab}(\bm{x})) ,\, f_2(p^{ij}(\bm{y})) \right\} f_2(p^{ab}(\bm{x}))
	\, ,
\end{align*} 
and so
\begin{align} 
	\label{Poisson brackets for fields: Poisson bracket of two composite scalar functions of metric and momenta but not their derivatives}
	\left\{ f_1(q_{ab}(\bm{x})) f_2(p^{ab}(\bm{x})) ,\, f_1(q_{ij}(\bm{y})) f_2(p^{ij}(\bm{y})) \right\} = 0
	\, .
\end{align}
Poisson bracket of 3-metric determinant $q(\bm{x}) \equiv \det\left(q_{ab}(\bm{x})\right)$ and the metric momenta is 
\begin{align} 
	\label{Poisson brackets for fields: Poisson bracket of determinant of 3-metric with momenta}
	\left\{ q(\bm{x}),\, p^{ij}(\bm{y}) \right\}
	= \frac{\partial q(\bm{x})}{\partial q_{ij}(\bm{x})} {\delta}^{(3)}(\bm{x} - \bm{y})
	= q(\bm{x}) q^{ij}(\bm{x}) {\delta}^{(3)}(\bm{x} - \bm{y})
	\, .
\end{align}
For Kronecker delta apparently is $\{ \delta^{a}_{b},\, p^{ij}(\bm{y}) \} = 0$, from where
\begin{align} 
\label{Poisson brackets for fields: Poisson bracket of kronecker delta and momenta}
	\left\{ \delta^{a}_{b},\, p^{ij}(\bm{y}) \right\}
	&= \left\{ q^{ak}(\bm{x}) q_{kb}(\bm{x}),\, p^{ij}(\bm{y}) \right\}
	= \nonumber
	\\
	&= \left\{ q^{ak}(\bm{x}),\, p^{ij}(\bm{y}) \right\} q_{kb}(\bm{x}) 
	+ q^{ak}(\bm{x}) \left\{q_{kb}(\bm{x}),\, p^{ij}(\bm{y}) \right\}
	= \nonumber
	\\
	&= \left\{ q^{ak}(\bm{x}),\, p^{ij}(\bm{y}) \right\} q_{kb}(\bm{x}) 
	+ q^{ak}(\bm{x}) \delta^{i}_{(k} \delta^{j}_{b)} {\delta}^{(3)}(\bm{x} - \bm{y})
	= 0
	\, .
\end{align}
Poisson bracket of inverse 3-metric and the momenta is thus
\begin{align} 
\label{Poisson brackets for fields: Poisson bracket of inverse 3-metric and momenta}
	\left\{ q^{ab}(\bm{x}),\, p^{ij}(\bm{y}) \right\}
	\nonumber
	&= - q^{ak}(\bm{x}) q^{bl}(\bm{x}) \delta^{i}_{(k} \delta^{j}_{l)} {\delta}^{(3)}(\bm{x} - \bm{y})
	= \nonumber
	\\
	&= - q^{a(i}(\bm{x}) q^{j)b}(\bm{x}) {\delta}^{(3)}(\bm{x} - \bm{y})
	\, .
\end{align}

\subsection{Constraint Algebra with Scalar Field}
\label{Constraint Algebra with Scalar Field}

In section \ref{Inclusion of Massless Scalar Field} above, we introduced a scalar field described by Lagrangian (\ref{scalar field lagrangian}) into out system, and computed its contributions (\ref{scalar field super-Hamiltonian}) and (\ref{scalar field super-momentum}) to the super-Hamiltonian and super-momentum. The overall super-Hamiltonian and super-momentum, defined in (\ref{super-Hamiltonian and super-momentum with massless scalar field - schematic}) are explicitly
\begin{align}
	\label{super-Hamiltonian with scalar field}
	\mathcal{H}^{(G,\,\phi)}_{\perp}
	&= 2 \kappa G_{abcd} p^{ab} p^{cd} - \frac{\closedsqrt{q}}{2 \kappa} {^{(\parallel)}\!}R 
	+ \frac{\varepsilon \, p_{\phi}^{2}}{2 \closedsqrt{q}} 
	+ \frac{\varepsilon \closedsqrt{q}}{2} q^{ab} \partial_{a} \phi \, \partial_{b} \phi 
	+ \closedsqrt{q} V(\phi)
	\, ,
	\\
	\label{supermomentum with scalar field}
	\mathcal{H}^{(G,\,\phi)}_{a}
	&= - 2 q_{ab} {^{(\parallel)}\!}\nabla_{c} p^{cb} + p_{\phi} \partial_{a} \phi
	\, .
\end{align}

We will now examine constraint algebra of the extended constraints and verify that the set of constrains is closed, and thus the overall super-Hamiltonian and super-momentum remain first class. We already know the brackets (\ref{constraint algebra - 2 super-Hamiltonians}), (\ref{constraint algebra - super-Hamiltonian and supermomentum}), and (\ref{constraint algebra - 2 supermomenta}) for vacuum case. 
Let us first compute the Poisson bracket of two non-vacuum momentum constraints. Observe that the scalar-field super-momentum $\mathcal{H}^{(\phi)}_{a} = p_{\phi} \partial_{a} \phi$ does not contain any spatial metric components nor any spatial-metric momenta. Its bracket with vacuum super-momentum therefore vanishes:
\begin{align*}
	\left\{ \mathcal{H}^{(G)}_{a}(\bm{x}) ,\, \mathcal{H}^{(\phi)}_{b}(\bm{y}) \right\} = 0
	\, , \quad
	\left\{ \mathcal{H}^{(\phi)}_{a}(\bm{x}) ,\, \mathcal{H}^{(G)}_{b}(\bm{y}) \right\} = 0
	\, .
\end{align*}
We are then left with  
\begin{align}
	\label{Constraint algebra with non-vacuum scalar field: Poisson bracket of two non-vacuum diffeomorphism constraints}
	\left\{ \mathcal{H}^{(G,\,\phi)}_{a}(\bm{x}) ,\, \mathcal{H}^{(G,\,\phi)}_{b}(\bm{y}) \right\} 
	&= \left\{ \mathcal{H}^{(G)}_{a}(\bm{x}) + \mathcal{H}^{(\phi)}_{a}(\bm{x}) ,\, 
	\mathcal{H}^{(G)}_{b}(\bm{y}) + \mathcal{H}^{(\phi)}_{b}(\bm{y}) \right\}
	= \nonumber
	\\
	&= \left\{ \mathcal{H}^{(G)}_{a}(\bm{x}) ,\, \mathcal{H}^{(G)}_{b}(\bm{y}) \right\} 
	+ \left\{ \mathcal{H}^{(\phi)}_{a}(\bm{x}) ,\, \mathcal{H}^{(\phi)}_{b}(\bm{y}) \right\}
	\, .
\end{align}
The first bracket in this expression is just Poisson bracket of two vacuum super-momenta and is equal to (\ref{constraint algebra - 2 supermomenta}). The second bracket, the Poisson bracket of two scalar-field super-momenta, represents a new contribution of the scalar field to the constraint algebra:  
\begin{align*}
	&\left\{ \mathcal{H}^{(\phi)}_{a}(\bm{x}) ,\, \mathcal{H}^{(\phi)}_{b}(\bm{y}) \right\}
	= \left\{ p_{\phi}(\bm{x}) \partial_{a} \phi(\bm{x}) ,\, p_{\phi}(\bm{y}) \partial_{b} \phi(\bm{y}) \right\} 
	= \nonumber
	\\
	&\qquad= 
	p_{\phi}(\bm{y}) \partial_{a} \phi(\bm{x}) \left\{ p_{\phi}(\bm{x}) ,\, \partial_{b} \phi(\bm{y}) \right\} 
	+ p_{\phi}(\bm{x}) \partial_{b} \phi(\bm{y}) \left\{ \partial_{a} \phi(\bm{x}) ,\, p_{\phi}(\bm{y}) \right\} 
	= \nonumber
	\\
	&\qquad= - p_{\phi}(\bm{y}) \partial_{a} \phi(\bm{x}) \, \partial_{y^{b}} {\delta}^{(3)}(\bm{x} - \bm{y})
	+ p_{\phi}(\bm{x}) \partial_{b} \phi(\bm{y}) \, \partial_{x^{a}} {\delta}^{(3)}(\bm{x} - \bm{y}) 
	= \nonumber
	\\
	&\qquad= p_{\phi}(\bm{y}) \partial_{a} \phi(\bm{y}) \, \partial_{x^{b}} {\delta}^{(3)}(\bm{x} - \bm{y})
	- \cancel{p_{\phi} \left(\partial_{b} \partial_{a} \phi \right) {\delta}^{(3)}(\bm{x} - \bm{y})}
	+ \nonumber 
	\\
	&\qquad\quad\, + p_{\phi}(\bm{x}) \partial_{b} \phi(\bm{x}) \, \partial_{x^{a}} {\delta}^{(3)}(\bm{x} - \bm{y}) 
	+ \cancel{p_{\phi} \left(\partial_{a} \partial_{b} \phi \right) {\delta}^{(3)}(\bm{x} - \bm{y})}
	= \nonumber
	\\
	&\qquad= p_{\phi}(\bm{y}) \partial_{a} \phi(\bm{y}) \, \partial_{x^{b}} {\delta}^{(3)}(\bm{x} - \bm{y}) 
	+ p_{\phi}(\bm{x}) \partial_{b} \phi(\bm{x}) \, \partial_{x^{a}} {\delta}^{(3)}(\bm{x} - \bm{y})
	= \nonumber
	\\
	&\qquad= \mathcal{H}^{(\phi)}_{a}(\bm{y}) \partial_{x^{b}} {\delta}^{(3)}(\bm{x} - \bm{y}) 
	+ \mathcal{H}^{(\phi)}_{b}(\bm{x}) \partial_{x^{a}} {\delta}^{(3)}(\bm{x} - \bm{y})
	\, .
\end{align*}
where we used relation (\ref{Constraint algebra in vacuum: F*G*ddelta}) in the next to last equality. 
We see that the bracket of two non-vacuum super-momenta precisely reproduces corresponding vacuum-case bracket:
\begin{align}
	\label{Poisson bracket of 2 non-vacuum super-momenta - result}
	\left\{ \mathcal{H}^{(G,\,\phi)}_{a}(\bm{x}) ,\, \mathcal{H}^{(G,\,\phi)}_{b}(\bm{y}) \right\} 
	= \mathcal{H}^{(G,\,\phi)}_{a}(\bm{y}) \partial_{x^{b}} {\delta}^{(3)}(\bm{x} - \bm{y}) 
	+ \mathcal{H}^{(G,\,\phi)}_{b}(\bm{x}) \partial_{x^{a}} {\delta}^{(3)}(\bm{x} - \bm{y})
	\, .
\end{align}
Secondly, we will compute the bracket of two non-vacuum super-Hamiltonians, 
\begin{align}
	\label{Poisson bracket of two non-vacuum Hamiltonian constraints}
	\left\{ \mathcal{H}^{(G,\,\phi)}_{\perp}(\bm{x}) ,\, 
	\mathcal{H}^{(G,\,\phi)}_{\perp}(\bm{y}) \right\} 
	&= \left\{ \mathcal{H}^{(G)}_{\perp}(\bm{x}) + \mathcal{H}^{(\phi)}_{\perp}(\bm{x}) ,\, 
	\mathcal{H}^{(G)}_{\perp}(\bm{y}) + \mathcal{H}^{(\phi)}_{\perp}(\bm{y}) \right\}
	= \nonumber
	\\
	&= \left\{ \mathcal{H}^{(G)}_{\perp}(\bm{x}) ,\, \mathcal{H}^{(G)}_{\perp}(\bm{y}) \right\} 
	+ \left\{ \mathcal{H}^{(\phi)}_{\perp}(\bm{x}) ,\, \mathcal{H}^{(\phi)}_{\perp}(\bm{y}) \right\}
	+ \nonumber
	\\
	&\quad\, + \left\{ \mathcal{H}^{(G)}_{\perp}(\bm{x}) ,\, 
	\mathcal{H}^{(\phi)}_{\perp}(\bm{y}) \right\}
	+ \left\{ \mathcal{H}^{(\phi)}_{\perp}(\bm{x}) ,\, \mathcal{H}^{(G)}_{\perp}(\bm{y}) \right\}
	\, .
\end{align}
Now, because the vacuum super-Hamiltonian $\mathcal{H}^{(G)}_{\perp}$  does not contain any derivatives of spatial-metric momenta nor does it contain any scar field canonical variables, the last two cross brackets in (\ref{Poisson bracket of two non-vacuum Hamiltonian constraints}) are proportional only to delta functions but not to their derivatives. And since on the classical level all functions commute, the sum of the two cross brackets vanishes 
\begin{align*}
	&\left\{ \mathcal{H}^{(G)}_{\perp}(\bm{x}) ,\, \mathcal{H}^{(\phi)}_{\perp}(\bm{y}) \right\}
	+ \left\{ \mathcal{H}^{(\phi)}_{\perp}(\bm{x}) ,\, \mathcal{H}^{(G)}_{\perp}(\bm{y}) \right\}
	= \nonumber
	\\
	&\qquad = F(\bm{x}) G(\bm{y}) {\delta}^{(3)}(\bm{x} - \bm{y}) - G(\bm{x}) F(\bm{y}) {\delta}^{(3)}(\bm{x} - \bm{y}) 
	= \nonumber
	\\
	&\qquad = \big( F G - G F \big) {\delta}^{(3)}(\bm{x} - \bm{y}) = 0
	\, .
\end{align*} 
The first bracket in (\ref{Poisson bracket of two non-vacuum Hamiltonian constraints}) is already known from vacuum-case constraint algebra. The only term that remains to be evaluated is bracket of two scalar-field super-Hamiltonians: 
\begin{align*}
	&\left\{ \mathcal{H}^{(\phi)}_{\perp}(\bm{x}) ,\, \mathcal{H}^{(\phi)}_{\perp}(\bm{y}) \right\}
	= \left\{ \left( \frac{\varepsilon \, p_{\phi}^{2}}{2 \closedsqrt{q}} 
	+ \frac{\varepsilon \closedsqrt{q}}{2} q^{ab} \partial_{a} \phi \, \partial_{b} \phi 
	+ \closedsqrt{q} V(\phi) \right)\bigg|_{\bm{x}} , \right.
	\\
	&\qquad\qquad\qquad\qquad\qquad \left. ,\, \left( 
	\frac{\varepsilon \, p_{\phi}^{2}}{2 \closedsqrt{q}} 
	+ \frac{\varepsilon \closedsqrt{q}}{2} q^{ab} \partial_{a} \phi \, \partial_{b} \phi 
	+ \closedsqrt{q} V(\phi) \right)\bigg|_{\bm{y}} \right\} 
	= \nonumber
	\\
	&\qquad = \underbrace{\left\{ 
	\frac{\varepsilon \, p_{\phi}^{2}}{2 \closedsqrt{q}}\bigg|_{\bm{x}} ,\, 
	\frac{\varepsilon \, p_{\phi}^{2}}{2 \closedsqrt{q}}\bigg|_{\bm{y}} \right\}}_{0} 
	+ \left\{ \frac{\varepsilon \, p_{\phi}^{2}}{2 \closedsqrt{q}}\bigg|_{\bm{x}} ,\, 
	\left(\frac{\varepsilon \closedsqrt{q}}{2} q^{ab} \partial_{a} \phi \, \partial_{b} \phi 
	+ \closedsqrt{q} V(\phi) \right)\bigg|_{\bm{y}} \right\}
	+ \nonumber
	\\
	&\qquad\quad\, + \left\{ 
	\left(\frac{\varepsilon \closedsqrt{q}}{2} q^{ab} \partial_{a} \phi \, \partial_{b} \phi 
	+ \closedsqrt{q} V(\phi)\right)\bigg|_{\bm{x}} ,\, 
	\frac{\varepsilon \, p_{\phi}^{2}}{2 \closedsqrt{q}}\bigg|_{\bm{y}} \right\}
	+ \nonumber
	\\
	&\qquad\quad\, + \underbrace{\left\{ 
	\left(\frac{\varepsilon \closedsqrt{q}}{2} q^{ab} \partial_{a} \phi \, \partial_{b} \phi 
	+ \closedsqrt{q} V(\phi)\right)\bigg|_{\bm{x}} ,\, 
	\left(\frac{\varepsilon \closedsqrt{q}}{2} q^{ab} \partial_{a} \phi \, \partial_{b} \phi 
	+ \closedsqrt{q} V(\phi) \right)\bigg|_{\bm{y}} \right\}}_{0} 
	\, .
\end{align*}
Continuing from above (and keeping in mind that $\varepsilon^2 = 1$):
\begin{align*}
	&\left\{ \mathcal{H}^{(\phi)}_{\perp}(\bm{x}) ,\, \mathcal{H}^{(\phi)}_{\perp}(\bm{y}) \right\}
	= \ldots =
	\\
	&\qquad = \left\{ \frac{\varepsilon \, p_{\phi}^{2}}{2 \closedsqrt{q}}\bigg|_{\bm{x}} ,\, \left(\frac{\varepsilon \closedsqrt{q}}{2} q^{ab} \partial_{a} \phi \, \partial_{b} \phi \right)\bigg|_{\bm{y}} \right\} 
	+ \left\{ \frac{\varepsilon \, p_{\phi}^{2}}{2 \closedsqrt{q}}\bigg|_{\bm{x}} ,\, \left(\closedsqrt{q} V(\phi) \right)\big|_{\bm{y}} \right\}
	+ \nonumber
	\\
	&\qquad\quad\, + \left\{ \left( \frac{\varepsilon \closedsqrt{q}}{2} q^{ab} \partial_{a} \phi \, \partial_{b} \phi \right)\bigg|_{\bm{x}} ,\, 
	\frac{\varepsilon \, p_{\phi}^{2}}{2 \closedsqrt{q}}\bigg|_{\bm{y}} \right\}
	+ \left\{ \left(\closedsqrt{q} V(\phi)\right)\big|_{\bm{x}} ,\, 
	\frac{\varepsilon \, p_{\phi}^{2}}{2 \closedsqrt{q}}\bigg|_{\bm{y}} \right\}  
	= \nonumber
	\\
	&\qquad = \frac{p_{\phi}^{2}}{\closedsqrt{q}}\bigg|_{\bm{x}} \left(\closedsqrt{q} q^{ab} \partial_{a}\phi \right)\Big|_{\bm{y}} \partial_{x^{b}} {\delta}^{(3)}(\bm{x} - \bm{y}) 
	- \cancel{\varepsilon \, p_{\phi} \frac{\partial V}{\partial \phi} {\delta}^{(3)}(\bm{x} - \bm{y})}
	+ \nonumber
	\\
	&\qquad\quad\, + \left(\closedsqrt{q} q^{ab} \partial_{a}\phi \right)\Big|_{\bm{x}} \frac{p_{\phi}^{2}}{\closedsqrt{q}}\bigg|_{\bm{y}} \partial_{x^{b}} {\delta}^{(3)}(\bm{x} - \bm{y})
	+ \cancel{\varepsilon \, p_{\phi} \frac{\partial V}{\partial \phi} {\delta}^{(3)}(\bm{x} - \bm{y})}
	= \nonumber
	\\
	&\qquad = \left( p_{\phi} q^{ab} \partial_{a}\phi \right)\Big|_{\bm{y}} 
	\partial_{x^{b}} {\delta}^{(3)}(\bm{x} - \bm{y}) 
	+ \bcancel{\frac{p_{\phi}^{2}}{\closedsqrt{q}} \partial_{b}\left(\closedsqrt{q} q^{ab} \partial_{a}\phi \right) 
		{\delta}^{(3)}(\bm{x} - \bm{y})}
	+ \nonumber
	\\
	&\qquad\quad\, + \left( p_{\phi} q^{ab} \partial_{a}\phi \right)\Big|_{\bm{x}} 
	\partial_{x^{b}} {\delta}^{(3)}(\bm{x} - \bm{y}) 
	- \bcancel{\frac{p_{\phi}^{2}}{\closedsqrt{q}} \partial_{b}\left(\closedsqrt{q} q^{ab} \partial_{a}\phi \right) 
		{\delta}^{(3)}(\bm{x} - \bm{y})}
	= \nonumber
	\\
	&\qquad = \left( q^{ab} p_{\phi} \partial_{a}\phi \right)\Big|_{\bm{y}} 
	\partial_{x^{b}} {\delta}^{(3)}(\bm{x} - \bm{y}) 
	+ \left( q^{ab} p_{\phi} \partial_{a}\phi \right)\Big|_{\bm{x}} 
	\partial_{x^{b}} {\delta}^{(3)}(\bm{x} - \bm{y})
	\, .
\end{align*}
The Poisson bracket of two non-vacuum super-Hamiltonians hence reproduces the corresponding vacuum-case bracket:
\begin{align}
	\label{Poisson bracket of 2 non-vacuum super-Hamiltonians - result}
	\left\{ \mathcal{H}^{(G,\,\phi)}_{\perp}(\bm{x}) ,\, \mathcal{H}^{(G,\,\phi)}_{\perp}(\bm{y}) \right\} 
	= \left[q^{ab}(\bm{x}) \mathcal{H}^{(G,\,\phi)}_{a}(\bm{x})   
	+ q^{ab}(\bm{y}) \mathcal{H}^{(G,\,\phi)}_{a}(\bm{y}) \right] \partial_{x^{b}} {\delta}^{(3)}(\bm{x} - \bm{y}) 
	\, .
\end{align}
The last Poisson bracket to be computed is that of the non-vacuum constraint algebra is that of super-momentum and super-Hamiltonian: 
\begin{align}
	\label{Poisson bracket of non-vacuum supermomentum and super-Hamiltonian}
	&\left\{ \mathcal{H}^{(G,\,\phi)}_{a}(\bm{x}) ,\, 
	\mathcal{H}^{(G,\,\phi)}_{\perp}(\bm{y}) \right\} 
	= \left\{ \mathcal{H}^{(G)}_{a}(\bm{x}) ,\, \mathcal{H}^{(G)}_{\perp}(\bm{y}) \right\} 
	+ \left\{ \mathcal{H}^{(G)}_{a}(\bm{x}) ,\, \mathcal{H}^{(\phi)}_{\perp}(\bm{y}) \right\}
	+ \nonumber
	\\
	&\qquad\qquad\qquad\qquad\qquad\,\quad\, 
	+ \left\{ \mathcal{H}^{(\phi)}_{a}(\bm{x}) ,\, 
	\mathcal{H}^{(\phi)}_{\perp}(\bm{y}) \right\} 
	+ \underbrace{\left\{ \mathcal{H}^{(\phi)}_{a}(\bm{x}) ,\, 
		\mathcal{H}^{(G)}_{\perp}(\bm{y}) \right\}}_{0}
	= \nonumber
	\\
	&\qquad = \left\{ \mathcal{H}^{(G)}_{a}(\bm{x}) ,\, \mathcal{H}^{(G)}_{\perp}(\bm{y}) \right\}
	- \left\{ \big(2 q_{ab} {^{(\parallel)}\!}\nabla_{c} p^{cb} \big)\big|_{\bm{x}} ,\, 
	\mathcal{H}^{(\phi)}_{\perp}(\bm{y}) \right\} 
	+ \left\{ \big(p_{\phi} \partial_{a}\phi\big)\big|_{\bm{x}} ,\, 
	\mathcal{H}^{(\phi)}_{\perp}(\bm{y}) \right\} 
	\, .
\end{align}
The first bracket in the last line is from vacuum constraint algebra and is already known. 
Before we proceed to compute the other two brackets we shall make some simplifying observations. 
The explicit form of the vacuum super-momentum (\ref{supermomentum with scalar field}) is 
\begin{align*}
	\mathcal{H}^{(G)}_{a} 
	&= 2 q_{ab} {^{(\parallel)}\!}\nabla_{c} p^{cb} 
	= 2 q_{ab} \left( \partial_{c} p^{cb} + \cancel{\Gamma^{c}_{cd} p^{db}} 
	+ \Gamma^{b}_{cd} p^{cd} - \cancel{\Gamma^{d}_{dc} p^{cb}}  \right)
	= \nonumber
	\\
	&= 2 q_{ab} \partial_{c} p^{cb} + 2 \Gamma_{acd} p^{cd}
	= 2 q_{ab} \partial_{c} p^{cb} 
	+ \left( 2 \partial_{c} q_{ad} - \partial_{a} q_{cd} \right) p^{cd}
	\, ,
\end{align*}
where the addition Christoffel symbol at the end of 1st first line comes from the fact that super-momentum is a tensor density (and not just an ordinary tensor). 
The bracket of vacuum momentum and scalar-field super-Hamiltonian is 
\begin{align}
	\label{Poisson bracket of vacuum momentum and scalar-field super-Hamiltonian}
	&\left\{ p^{cd}(\bm{x}) ,\, \mathcal{H}^{(\phi)}_{\perp}(\bm{y}) \right\} 
	= \left\{ p^{cd}(\bm{x}) ,\, \left( \frac{\varepsilon \, p_{\phi}^{2}}{2 \closedsqrt{q}} 
	+ \frac{\varepsilon \closedsqrt{q}}{2} q^{kl} \partial_{k} \phi \, \partial_{l} \phi 
	+ \closedsqrt{q} V(\phi)\right)\bigg|_{\bm{y}} \right\}
	= \nonumber
	\\
	&\qquad = \varepsilon \, p_{\phi}^{2}(\bm{y}) \left\{ p^{cd}(\bm{x}) ,\, 
	\frac{1}{2 \closedsqrt{q}}\bigg|_{\bm{y}} \right\}
	+ \varepsilon \left(q^{kl} \partial_{k}\phi \, \partial_{l}\phi\right)\Big|_{\bm{y}} 
	\left\{ p^{cd}(\bm{x}) ,\, \frac{\closedsqrt{q}}{2}\bigg|_{\bm{y}} \right\} 
	+ \nonumber
	\\
	&\qquad\quad\, + \varepsilon \left(\closedsqrt{q} \partial_{k}\phi \, \partial_{l}\phi\right)|_{\bm{y}} 
	\left\{ p^{cd}(\bm{x}) ,\, q^{kl}(\bm{y}) \right\}
	+ V(\phi)|_{\bm{y}}\left\{ p^{cd}(\bm{x}) ,\, \closedsqrt{q}|_{\bm{y}} \right\} 
	= \nonumber
	\\
	&\qquad = \left( \frac{\varepsilon \, p_{\phi}^{2} q^{cd}}{4 \closedsqrt{q}} 
	+ \frac{\varepsilon \closedsqrt{q}}{2} \left[q^{ck} q^{dl} - \frac{1}{2} q^{cd} q^{kl}\right] 
	\partial_{k} \phi \, \partial_{l} \phi 
	- \frac{\closedsqrt{q}}{2} q^{cd} V(\phi)\right)\bigg|_{\bm{y}}  
	{\delta}^{(3)}(\bm{x} - \bm{y})
	\, .
\end{align}
And the bracket of scalar-field super-momentum and super-Hamiltonian:
\begin{align}
	\label{Poisson bracket of scalar-field supermomentum and scalar-field super-Hamiltonian}
	&\left\{ \big(p_{\phi} \partial_{a}\phi\big)\big|_{\bm{x}} ,\, \mathcal{H}^{(\phi)}_{\perp}(\bm{y}) \right\} 
	= \nonumber
	\\
	&\qquad = \left\{ p_{\phi}(\bm{x}) \partial_{a}\phi(\bm{x}) ,\, 
	\left(\frac{\varepsilon \, p_{\phi}^{2}}{2 \closedsqrt{q}} 
	+ \frac{\varepsilon \closedsqrt{q}}{2} q^{kl} \partial_{k} \phi \, \partial_{l} \phi 
	+ \closedsqrt{q} V(\phi)\right)\bigg|_{\bm{y}} \right\} 
	= \nonumber
	\\
	&\qquad = p_{\phi}(\bm{x}) \frac{\varepsilon \, p_{\phi}}{\closedsqrt{q}}\bigg|_{\bm{y}} 
	\partial_{x^a} {\delta}^{(3)}(\bm{x} - \bm{y}) 
	+ \partial_{a}\phi(\bm{x}) \left(\varepsilon \closedsqrt{q} q^{kl} \partial_{k}\phi \right)\Big|_{\bm{y}} \partial_{x^l} {\delta}^{(3)}(\bm{x} - \bm{y}) 
	- \nonumber
	\\
	&\qquad\quad\, - \closedsqrt{q} \underbrace{\frac{d V}{d \phi} \partial_{a}\phi}_{\partial_{a} V(\phi)} \,  
	{\delta}^{(3)}(\bm{x} - \bm{y})
	\, .
\end{align}
Now we are able to explicitly write the bracket of non-vacuum supermomentum and scalar-field super-Hamiltonian:
\begin{align*}
	&\left\{ \mathcal{H}^{(G,\,\phi)}_{a}(\bm{x}) ,\, \mathcal{H}^{(\phi)}_{\perp}(\bm{y}) \right\} = 
	\\
	&\qquad = - 2 q_{ab}(\underset{\substack{\downarrow \\ \bm{y}}}{\bcancel{\bm{x}}}) 
	\left( \frac{\varepsilon \, p_{\phi}^{2} q^{cd}}{4 \closedsqrt{q}} 
	+ \frac{\varepsilon \closedsqrt{q}}{2} \left[q^{ck} q^{dl} - \frac{1}{2} q^{cd} q^{kl}\right] \partial_{k} \phi \, \partial_{l} \phi 
	- \right. \nonumber
	\\
	&\qquad\qquad\qquad\qquad \left. - \frac{\closedsqrt{q}}{2} q^{cd} V(\phi)\right)\bigg|_{\bm{y}} \partial_{x^c} {\delta}^{(3)}(\bm{x} - \bm{y}) + 
	\\
	&\qquad\quad\, + \left(\partial_{a}q_{cd} - \bcancel{2\partial_{c}q_{ad}}\right) 
	\left( \frac{\varepsilon \, p_{\phi}^{2} q^{cd}}{4 \closedsqrt{q}} 
	+ \frac{\varepsilon \closedsqrt{q}}{2} \left[q^{ck} q^{dl} 
	- \frac{1}{2} q^{cd} q^{kl}\right] \partial_{k} \phi \, \partial_{l} \phi 
	- \right. 
	\\
	&\qquad\qquad\qquad\qquad\qquad\qquad \left. 
	- \frac{\closedsqrt{q}}{2} q^{cd} V(\phi)\right) {\delta}^{(3)}(\bm{x} - \bm{y}) + 
	\\
	&\qquad\quad\, + \varepsilon \, p_{\phi}(\bm{x}) \frac{p_{\phi}}{\closedsqrt{q}}\bigg|_{\bm{y}} 
	\partial_{x^a} {\delta}^{(3)}(\bm{x} - \bm{y}) 
	- \closedsqrt{q}\, \partial_{a} V(\phi) \, {\delta}^{(3)}(\bm{x} - \bm{y}) + 
	\\
	&\qquad\quad\, + \varepsilon \, \partial_{a}\phi(\bm{x}) \left(\closedsqrt{q} q^{kl} \partial_{k}\phi \right)\Big|_{\bm{y}} 
	\partial_{x^l} {\delta}^{(3)}(\bm{x} - \bm{y}) 
	\, .
\end{align*}
Continuing from above:
\begin{align}
	\label{Poisson bracket of non-vacuum supermomentum and scalar-field super-Hamiltonian}
	&\left\{ \mathcal{H}^{(G,\,\phi)}_{a}(\bm{x}) ,\, \mathcal{H}^{(\phi)}_{\perp}(\bm{y}) \right\} 
	= \ldots = \nonumber
	\\
	&\qquad = - \frac{\varepsilon \, p_{\phi}^{2}}{2 \closedsqrt{q}}\Big|_{\bm{y}} \partial_{x^a} {\delta}^{(3)}(\bm{x} - \bm{y})
	- \left(\varepsilon \closedsqrt{q} q^{ck} \partial_{k}\phi \, \partial_{a}\phi \right)\big|_{\bm{y}} \partial_{x^c} {\delta}^{(3)}(\bm{x} - \bm{y}) 
	+ \nonumber
	\\
	&\qquad\quad\, + \left.\left(\frac{\varepsilon \closedsqrt{q}}{2} q^{kl} \partial_{k}\phi \, \partial_{a}\phi \right)\right|_{\bm{y}} \partial_{x^a} {\delta}^{(3)}(\bm{x} - \bm{y})
	+ \big(\closedsqrt{q} V(\phi) \big)\big|_{\bm{y}} \partial_{x^a} {\delta}^{(3)}(\bm{x} - \bm{y})
	+ \nonumber
	\\
	&\qquad\quad\, + \left(\partial_{a}q_{cd}\right) 
	\left( \frac{\varepsilon \, p_{\phi}^{2} q^{cd}}{4 \closedsqrt{q}} 
	+ \frac{\varepsilon \closedsqrt{q}}{2} \left[q^{ck} q^{dl} 
	- \frac{1}{2} q^{cd} q^{kl}\right] \partial_{k} \phi \, \partial_{l} \phi 
	- \right. \nonumber
	\\
	&\qquad\qquad\qquad\qquad \left. 
	- \frac{\closedsqrt{q}}{2} q^{cd} V(\phi)\right) {\delta}^{(3)}(\bm{x} - \bm{y})
	+ \nonumber
	\\
	&\qquad\quad\, + p_{\phi}(\bm{x}) \frac{\varepsilon \, p_{\phi}}{\closedsqrt{q}}\Big|_{\bm{y}} 
	\partial_{x^a} {\delta}^{(3)}(\bm{x} - \bm{y}) 
	- \closedsqrt{q}\, \partial_{a} V(\phi) \, {\delta}^{(3)}(\bm{x} - \bm{y}) 
	+ \nonumber
	\\
	&\qquad\quad\, + \partial_{a}\phi(\bm{x}) \left(\varepsilon \closedsqrt{q} q^{kl} \partial_{k}\phi \right)\big|_{\bm{y}} 
	\partial_{x^l} {\delta}^{(3)}(\bm{x} - \bm{y})
	\, .
\end{align}
In order to make calculations of (\ref{Poisson bracket of non-vacuum supermomentum and scalar-field super-Hamiltonian}) clear and simple we shall address certain groups of terms separately. Let us then focus first on terms containing scalar potential $V$. These are
\begin{align*}
	&\big(\closedsqrt{q} V(\phi) \big)\big|_{\bm{y}} \partial_{x^a} {\delta}^{(3)}(\bm{x} - \bm{y}) 
	- \left[\frac{\closedsqrt{q}}{2} q^{cd} V(\phi) + \closedsqrt{q}\, \partial_{a} V(\phi) \right] 
	{\delta}^{(3)}(\bm{x} - \bm{y}) 
	=
	\\
	&\qquad = \big(\closedsqrt{q} V(\phi) \big)\big|_{\bm{y}} \partial_{x^a} {\delta}^{(3)}(\bm{x} - \bm{y}) 
	- \partial_{a}\left[\closedsqrt{q} V(\phi)\right] {\delta}^{(3)}(\bm{x} - \bm{y})
	\\
	&\qquad = \big(\closedsqrt{q} V(\phi) \big)\big|_{\bm{x}} \partial_{x^a} {\delta}^{(3)}(\bm{x} - \bm{y})
	\, .
\end{align*}  
Terms of (\ref{Poisson bracket of non-vacuum supermomentum and scalar-field super-Hamiltonian}) containing scalar-field momenta are 
\begin{align*}
	& - \frac{\varepsilon \, p_{\phi}^{2}}{2 \closedsqrt{q}}\Big|_{\bm{y}} \partial_{x^a} {\delta}^{(3)}(\bm{x} - \bm{y}) 
	+ p_{\phi}(\bm{x}) \frac{\varepsilon \, p_{\phi}}{\closedsqrt{q}}\Big|_{\bm{y}} \partial_{x^a} {\delta}^{(3)}(\bm{x} - \bm{y}) 
	+ \frac{\varepsilon \, p_{\phi}^{2} q^{cd}}{4 \closedsqrt{q}} {\delta}^{(3)}(\bm{x} - \bm{y}) =
	\\
	&\qquad = - \frac{\varepsilon \, p_{\phi}^{2}}{2 \closedsqrt{q}}\Big|_{\bm{x}} \partial_{x^a} {\delta}^{(3)}(\bm{x} - \bm{y}) 
	+ \frac{\varepsilon \, p_{\phi}^{2}}{\closedsqrt{q}}\Big|_{\bm{x}} \partial_{x^a} {\delta}^{(3)}(\bm{x} - \bm{y}) - 
	\\
	&\qquad\quad\, - \varepsilon \underbrace{\left[p_{\phi}^{2} \partial_{a}\left(\frac{1}{2\closedsqrt{g}}\right) 
	+ \partial_{a}\left(\frac{p_{\phi}^{2}}{2\closedsqrt{g}}\right)
	- p_{\phi} \partial_{a}\left(\frac{p_{\phi}}{\closedsqrt{g}}\right) \right]}_{0} {\delta}^{(3)}(\bm{x} - \bm{y})
	\\
	&\qquad = \frac{\varepsilon \, p_{\phi}^{2}}{2 \closedsqrt{q}}\Big|_{\bm{x}} \partial_{x^a} {\delta}^{(3)}(\bm{x} - \bm{y})
	\, .
\end{align*}
Finally, the terms of (\ref{Poisson bracket of non-vacuum supermomentum and scalar-field super-Hamiltonian}) containing derivatives of scalar field are
\begin{align*}
	&\left.\left(\frac{\varepsilon \closedsqrt{q}}{2} q^{kl} \partial_{k}\phi \, \partial_{a}\phi \right)\right|_{\bm{y}} \partial_{x^a} {\delta}^{(3)}(\bm{x} - \bm{y})
	- \left(\varepsilon \closedsqrt{q} q^{ck} \partial_{k}\phi \, \partial_{a}\phi \right)\Big|_{\bm{y}} \partial_{x^c} {\delta}^{(3)}(\bm{x} - \bm{y}) + 
	\\
	& + \left[\frac{\varepsilon \closedsqrt{q}}{2} \left(\partial_{a}q_{cd}\right) q^{ck} q^{dl} \partial_{k}\phi \, \partial_{l}\phi
	- \frac{\varepsilon \closedsqrt{q}}{4} q^{cd} \left(\partial_{a}q_{cd}\right) q^{kl} \partial_{k}\phi \, \partial_{l}\phi \right] {\delta}^{(3)}(\bm{x} - \bm{y}) + 
	\\
	& + \partial_{a}\phi(\bm{x}) \left(\varepsilon \closedsqrt{q} q^{kl} \partial_{k}\phi \right)\Big|_{\bm{y}} \partial_{x^l} {\delta}^{(3)}(\bm{x} - \bm{y})
	\, ,
\end{align*}
which, when recast and simplified using Dirac delta function formulas, gives: 
\begin{align*}
	&\left.\left(\frac{\closedsqrt{q}}{2} q^{kl} \partial_{k}\phi \, \partial_{a}\phi \right)\right|_{\bm{x}} \partial_{x^a} {\delta}^{(3)}(\bm{x} - \bm{y}) 
	+ \partial_{a}\left(\frac{\closedsqrt{q}}{2} q^{kl} \partial_{k}\phi \, \partial_{a}\phi \right) 
	{\delta}^{(3)}(\bm{x} - \bm{y}) - 
	\\
	&\quad\, - \cancel{\left(\closedsqrt{q} q^{ck} \partial_{k}\phi \, \partial_{a}\phi \right)\big|_{\bm{y}} \partial_{x^c} {\delta}^{(3)}(\bm{x} - \bm{y})} - 
	\\
	&\quad\, 
	- \left[\frac{\closedsqrt{q}}{2} \left(\partial_{a}q^{kl}\right) \partial_{k}\phi \, \partial_{l}\phi
	+ \partial_{a}\left(\frac{\closedsqrt{q}}{2}\right) q^{kl} \partial_{k}\phi \, \partial_{l}\phi \right] {\delta}^{(3)}(\bm{x} - \bm{y}) + 
	\\
	&\quad\, + \cancel{\left(\closedsqrt{q} q^{kl} \partial_{k}\phi \, \partial_{a}\phi \right)\big|_{\bm{y}} 
		\partial_{x^l} {\delta}^{(3)}(\bm{x} - \bm{y})}
	- \closedsqrt{q} q^{kl} \partial_{k}\phi \, \partial_{l}\partial_{a}\phi \, {\delta}^{(3)}(\bm{x} - \bm{y}) =
	\\
	&= \left.\left(\frac{\closedsqrt{q}}{2} q^{kl} \partial_{k}\phi \, \partial_{a}\phi \right)\right|_{\bm{x}} \partial_{x^a} {\delta}^{(3)}(\bm{x} - \bm{y}) 
	+ \bcancel{\partial_{a}\left(\frac{\closedsqrt{q}}{2} q^{kl} \partial_{k}\phi \, \partial_{a}\phi \right) {\delta}^{(3)}(\bm{x} - \bm{y})} - 
	\\
	&\quad\, 
	- \bcancel{\left[ \partial_{a}\left(\frac{\closedsqrt{q}}{2} q^{kl}\right) \partial_{k}\phi \, \partial_{l}\phi  
	+ \frac{\closedsqrt{q}}{2} q^{kl} \partial_{a}\left(\partial_{k}\phi \, \partial_{l}\phi\right) \right] {\delta}^{(3)}(\bm{x} - \bm{y})} =
	\\
	&= \left.\left(\frac{\closedsqrt{q}}{2} q^{kl} \partial_{k}\phi \, \partial_{a}\phi \right)\right|_{\bm{x}} \partial_{x^a} {\delta}^{(3)}(\bm{x} - \bm{y})
	\, .
\end{align*}
Substituting for all of the three pieces from results above in (\ref{Poisson bracket of non-vacuum supermomentum and scalar-field super-Hamiltonian}) yields familiar expression for the bracket of non-vacuum super-momentum and scalar-field super-Hamiltonian:
\begin{align*}
	&\left\{ \mathcal{H}^{(G,\,\phi)}_{a}(\bm{x}) ,\, \mathcal{H}^{(\phi)}_{\perp}(\bm{y}) \right\} 
	= \nonumber
	\\
	&\qquad = \left.\left( \frac{\varepsilon \, p_{\phi}^{2}}{2 \closedsqrt{q}}  
	+ \frac{\varepsilon \closedsqrt{q}}{2} q^{kl} \partial_{k}\phi \, \partial_{a}\phi  
	+ \closedsqrt{q} V(\phi) \right)\right|_{\bm{x}} \partial_{x^a} {\delta}^{(3)}(\bm{x} - \bm{y}) 
	= \nonumber
	\\
	&\qquad = \mathcal{H}^{(\phi)}_{\perp}(\bm{x}) \partial_{x^a} {\delta}^{(3)}(\bm{x} - \bm{y}) 
	\, .
\end{align*}
The Poisson bracket of super-momentum and super-Hamiltonian is thus  
\begin{align}
	\label{Poisson bracket of non-vacuum supermomentum and super-Hamiltonian - result}
	\left\{ \mathcal{H}^{(G,\,\phi)}_{a}(\bm{x}) ,\, \mathcal{H}^{(G,\,\phi)}_{\perp}(\bm{y}) \right\} 
	= \mathcal{H}^{(G,\,\phi)}_{\perp}(\bm{x}) \partial_{x^a} {\delta}^{(3)}(\bm{x} - \bm{y})
	\, ,
\end{align}
which again reproduces the corresponding vacuum-case bracket. We had therefore verified the constraint algebra (\ref{constraint algebra with scalar field - 2 super-Hamiltonians})-(\ref{constraint algebra with scalar field - 2 super-momenta}) for non-vacuum case with a scalar field.

\end{document}